\documentclass[letterpaper,12pt,peerreviewca,draftcls]{IEEEtran}
\IEEEoverridecommandlockouts
\usepackage{csm16}
\usepackage[margin=1in]{geometry}
\usepackage{amsmath} 

\usepackage{mwe} 
\usepackage{etoolbox} 
\usepackage{url}
\usepackage{graphicx,xcolor}
\graphicspath{{./Figs/}}
\usepackage{verbatim}
\makeatletter
\newcommand{\verbatimfont}[1]{\def\verbatim@font{#1}}%
\makeatother
\verbatimfont{\ttfamily\small}
\usepackage[english]{babel} 
\usepackage{blindtext}

\newcommand{\sbsection}[2]{
\clearpage
\newpage
\section[#1]{Sidebar: #1}
\label{sb:#2}
}

\newcommand{\sbend}{
\newpage
\clearpage
}

\usepackage{listofitems}

\usepackage{tikz,pgfplots}
\pgfplotsset{compat=newest}
\usetikzlibrary{arrows,shapes,spy,calc,positioning,backgrounds,fit,decorations.markings}
\usepgfplotslibrary{groupplots,external,dateplot,units}
\tikzexternaldisable

\newcommand*{\customtikzsetnextfilename}[1]{%
	\refstepcounter{figure}\edef\tempa{figure\thefigure}%
	\expandafter\tikzsetnextfilename{\tempa_#1}%
	\addtocounter{figure}{-1}%
}

\newcommand{\includetikz}[1]{%
    \refstepcounter{figure}
    \includegraphics{figure\thefigure_#1.pdf}
    \addtocounter{figure}{-1}
}

\newlength\figureheight
\newlength\figurewidth

\newcommand{\includetikzfigure}[2]{%
    \begin{figure}[!h]
    \centering
    \includetikz{#1}
    \caption{#2}
    \label{fig:#1}
    \end{figure}
}

\newcommand{\includefigure}[3]{%
    \setlength\figurewidth{#2\columnwidth}
    \begin{figure}[!h]
    \centering
    \refstepcounter{figure}
    \includegraphics[width=\figurewidth]{figure\thefigure_#1}%
    \addtocounter{figure}{-1}
    \caption{#3}
    \label{fig:#1}
    \end{figure}
}

\pgfplotsset{select coords between index/.style 2 args={
    x filter/.code={
        \ifnum\coordindex<#1\fi
        \ifnum\coordindex>#2\fi
    }
}}

\tikzset{cross/.style={cross out, draw=black, minimum size=2*(#1-\pgflinewidth), inner sep=0pt, outer sep=0pt},
cross/.default={1pt}}

\definecolor{col1}{rgb}{0,0.4470,0.7410}
\definecolor{col2}{rgb}{0.8500,0.3250,0.0980}
\definecolor{col3}{rgb}{0.9290,0.6940,0.1250}
\definecolor{col4}{rgb}{0.4940,0.1840,0.5560}
\definecolor{col5}{rgb}{0.4660,0.6740,0.1880}
\definecolor{col6}{rgb}{0.3010,0.7450,0.9330}
\definecolor{col7}{rgb}{0.6350,0.0780,0.1840}

\usepackage [siunitx,european,cuteinductors]{circuitikz}

\usepackage{booktabs} 
\usepackage{multirow}

\usepackage{siunitx} 

\usepackage{cite}

\newif\ifPDF \ifx\pdfoutput\undefined\PDFfalse \else\ifnum\pdfoutput > 0\PDFtrue \else\PDFfalse \fi \fi
\ifPDF 
\usepackage[pdftex, plainpages = false, colorlinks=true, linkcolor=black, citecolor = green!50!blue, urlcolor = blue, filecolor=black, pagebackref=false, hypertexnames=false,  pdfpagelabels ]{hyperref}
\fi

\usepackage[capitalise,noabbrev]{cleveref} 

\usepackage{textcomp}
\newcommand\copyrighttext{%
	\fontsize{10}{8}\selectfont \textcopyright 2021 IEEE. Personal use of this material is permitted.  Permission from IEEE must be obtained for all other uses, in any current or future media, including reprinting/republishing this material for advertising or promotional purposes, creating new collective works, for resale or redistribution to servers or lists, or reuse of any copyrighted component of this work in other works. 
	DOI: 10.1109/MCS.2021.3107560}
\newcommand\copyrightnotice{%
	\begin{tikzpicture}[remember picture,overlay]
		\node[anchor=south,yshift=10pt] at (current page.south) {\fbox{\parbox{\dimexpr\textwidth-\fboxsep-\fboxrule\relax}{\copyrighttext}}};
	\end{tikzpicture}%
}

\title{Neuromorphic Control\\
\Large Designing multiscale mixed-feedback systems}
\author{Luka Ribar and Rodolphe Sepulchre\\
	\today }

\begin{document}
\maketitle
\copyrightnotice
\CSMsetup

\thispagestyle{empty}

Neuromorphic electronic engineering takes inspiration from the biological organization of nervous systems to rethink the technology of computing, sensing, and actuating. It started three decades ago with the realization by Carver Mead, a pioneer of VLSI technology, that the operation of a conventional transistor in the analog regime closely resembles the biophysical operation of a neuron \cite{mead1990}. Mead envisioned a novel generation of electronic circuits that would operate far more efficiently than conventional VLSI technology and would allow for a new generation of biologically inspired sensing devices. Three decades later, active vision has become a technological reality \cite{gallego2020, mead2020} and neuromorphic computing has emerged as a promising avenue to reduce the energy requirements of digital computers \cite{indiveri2011,cauwenberghs2013,furber2016,boahen2017}. These two applications could just be the tip of the iceberg.

Neuromorphic circuit architectures call for new computing, signal processing and control paradigms. A most pressing challenge of neuromorphic circuit design is to cope with the \textit{transistor mismatch} \cite{mead1989,sarpeshkar1998}, that is, the uncertainty that comes with operating in a mixed analog-digital regime that utilizes large ensembles of tiny devices with significant variability. Methods to mitigate the effects of this uncertainty have been investigated in specific applications, e.g. \cite{boahen1992a}. Yet, the transistor mismatch has been a main hurdle to the development of the field over the last two decades \cite{poon2011}. For control engineers, the challenge is certainly reminiscent of the uncertainty  bottleneck faced by the engineers of Bell Laboratories in the early days of long distance signal transmission, eventually solved by Black's invention of the negative feedback amplifier.

The new uncertainty bottleneck of neuromorphic engineering calls for novel control principles. Neuromorphic control aims at taking inspiration from biology to address this challenge.  Nervous systems do cope with the uncertainty and variability of their components. They are able to maintain robust and flexible function over an impressive span of spatial and temporal scales. Understanding what are the control mechanisms that allow the regulation of the robust global behavior through the cumulative effect of control actions at smaller scales could turn the variability of transistors into a feature rather than a hurdle. More generally, the growing understanding of how biology exploits variability could provide new inspiration for the increasingly multiscale nature of engineering design \cite{sepulchre2019} (see "\nameref{sb:multiscale}").

A control principle at the heart of biological systems is the prevalence of concurrent positive and negative feedback pathways, acting in different temporal and spatial scales (see "\nameref{sb:positive_negative_feedback}"). This mixed-feedback organization allows neural systems to colocalize memory and processing capabilities, generating discrete events with memory while at the same time continuously regulating  their internal properties to allow for learning and adaption. \textit{Neuromodulation} provides one important mechanism to tune those mixed feedback gains. Neuromodulation is a general term that encompasses a realm of neurochemical processes that control  the electrical activity of neurons.  These regulatory processes allow sensory networks to dynamically adjust their information processing capabilities \cite{krahe2004}, while providing the flexibility and robustness of biological clocks underlying repeated actions such as movement or breathing \cite{harris-warrick2011,marder2012}. The capability of neuromodulatory actions to target and modify local parts of neural networks provides the basis of the biological control of nervous systems across scales \cite{sepulchre2019,sepulchre2019a}.

The aim of this article is to highlight the potential of integrating neuromodulation mechanisms in the design and control of electronic circuits. It presents a simple circuit component introduced in \cite{ribar2017,ribar2019} that acts as an elementary feedback element providing localized positive or negative feedback. The parallel interconnection of such elements is sufficient to capture the feedback structure present in biological neurons, and thus presents an avenue for studying the principles of neuromorphic control. This interconnection structure augments the circuit with a simple control methodology through shaping its input-output current-voltage (I-V) characteristic.

The article is primarily based on the PhD thesis \cite{ribar2019a} and the recent work on neuromorphic design and control presented in \cite{ribar2017,ribar2019}. It is grounded in recent work aimed at understanding neuronal behaviors as feedback control systems and neuromodulation as controller design. The interested reader is referred to several review articles \cite{sepulchre2018,sepulchre2019,sepulchre2019a}, and to the technical contributions \cite{drion2012,franci2012a,franci2013,franci2014a,franci2014,drion2015,franci2017}.

The rest of the article is organized as follows. "\nameref{sec:synthesis}" reviews the key modeling principles of biological neurons in the form of excitable electrical circuits. It presents the novel circuit architecture introduced in \cite{ribar2017,ribar2019} that is amenable to control design and hardware realization while retaining  the core mixed feedback loop modules of biological neuronal networks. The architecture uses the traditional neuromorphic circuits introduced by Carver Mead. They are designed by a graphical input-output methodology reminiscent of loop shaping: it is the shape of the I-V curve of the circuit in distinct timescales that determines the closed-loop behavior, and each element of the circuit can be regarded as shaping a particular I-V curve. "\nameref{sec:control}" then discusses how the I-V curve shaping technique provides a way of implementing and interpreting the mechanisms of neuromorphic neuromodulation. The section highlights the control of important input-output neuronal properties and their significance for the control of network behavior. As a main illustration, the section shows how modulating the excitability properties of the nodes of a network provides a versatile control principle to continuously reconfigure the \textit{functional} topology of the network without altering its connectivity. This basic principle is illustrated on a five neuron network inspired from the crustacean stomatogastric ganglion (STG) that has served as a central model of neuromodulation for the last forty years \cite{nassim2018}. Concluding remarks summarize the potential of these design principles for control.

\section{Synthesis of neuromorphic circuits}
\label{sec:synthesis}

The neuromorphic approach aims to understand and mimic the biophysical mechanisms underlying neuronal behavior in order to build electrical systems with similar capabilities. The initial interest in bioinspired systems has sparked a rapidly developing research field (see "\nameref{sb:neuromorphic_overview}"). Neuromorphic hardware provides an exciting venue for bioinspired control systems, see for instance the recent survey \cite{gallego2020}. In spite of some early examples of joint efforts between the two communities ("\nameref{sb:astrom_neuromorphic}"), it is fair to say that the potential for novel biologically motivated control methods still remains largely unexplored.

The aim of this article is to simplify the apparent complexity of biophysical models through the prism of the feedback loops essential to excitability. This section highlights the core feedback structure of biophysical excitable models and presents the neuromorphic design methodology of \cite{ribar2019} inspired by the conductance-based structure of neuronal models. It shows how the fundamental dynamical features of neurons can be recreated in a circuit structure which is simple to manufacture and control with existing neuromorphic components. At the core of the framework are elementary feedback current elements that provide local positive or negative feedback, in a similar vein to the role of individual ionic currents in physiological models of neurons. The model provides a way of synthesizing electrical circuits with the mixed feedback structure that can be analyzed by shaping its input-output characteristic in the form of I-V curves. 

\subsection{Excitability: a mixed feedback principle}
The fundamental dynamical property of neurons is \textit{excitability}. Excitability of a system can be defined as an input-output property by considering the response of the system to pulse inputs of varying magnitude and duration. In the case of neurons, the external applied current is taken as the input variable, while the cell's membrane voltage is the output. A neuron is excitable because its response is all-or-none: for small, subthreshold current pulses, the system exhibits the stable response of a passive circuit with small variations in the membrane voltage. In contrast, a current above a certain threshold causes a \textit{spike}, or \textit{action potential} characterized by a large, well-defined excursion of the voltage (\cref{fig:hh_excitability_io}). Such specific nonlinear behavior comes from the appropriate balance of ionic currents that flow through the membrane with voltage-dependent dynamics. This mechanism is described in detail in the sidebar "\nameref{sb:hodgkin_huxley}", detailing the generation of the action potential in the squid giant axon.

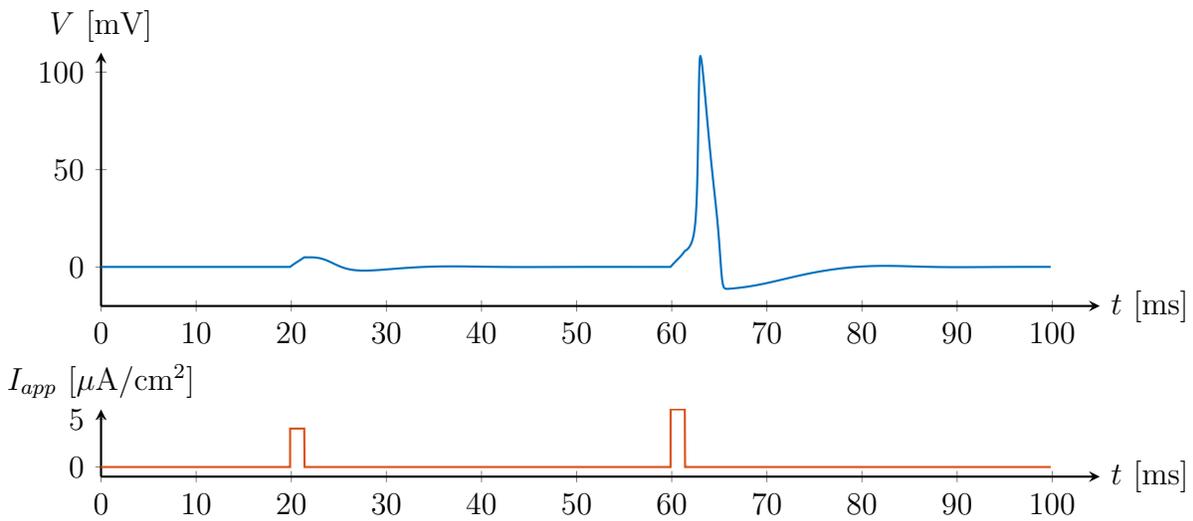
\includetikzfigure{hh_excitability_io}{Input-output characteristic of excitable systems. Subthreshold inputs generate subthreshold outputs, but suprathreshold inputs generate an all-or-none response in the form of one or several action potentials. Response shown for the Hodgkin-Huxley neural model.}

The underlying ionic mechanisms described in the seminal Hodgkin-Huxley paper generalize to all excitable neurons, but differ in their structure and complexity. At the core of the modeling of excitability is the representation of a neuron as a nonlinear circuit, known as a conductance-based model.

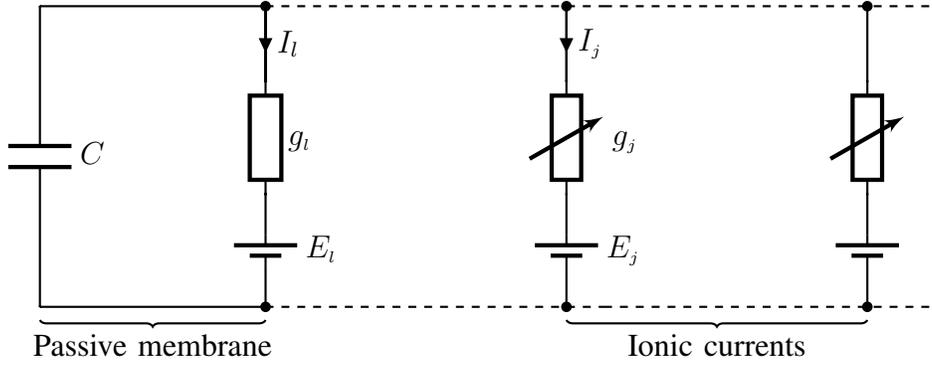
\includetikzfigure{conductance_circuit}{A general conductance-based circuit. The passive membrane consisting of the membrane capacitor and the leak current is interconnected with possibly many ionic currents.}

Every neuron is modeled as an electrical circuit describing the dynamical activity of its cellular membrane. This conductance-based structure means that the dynamics of each ionic current can be modeled as a time-varying conductance in series with a battery. The neural circuit is then modeled as a parallel interconnection of the membrane capacitor, a passive leak current, and the ionic currents (\cref{fig:conductance_circuit}). Kirchhoff's current law gives the membrane equation
\begin{equation}
\label{eq:conductance_model}
\begin{split}
C \dot{V} &= -I_l - \sum_j I_j \\
&= -\underbrace{g_l(V - E_l)\vphantom{\sum_j}}_{\text{Leak current}} - \underbrace{\sum_j g_j (V - E_j)}_{\text{Ionic currents}},
\end{split}
\end{equation}
where $E_j$ is the equilibrium potential for each ionic current $j$. While the leak conductance $g_l$ is constant, each conductance $g_j$ is voltage and time-dependent (see "\nameref{sb:hodgkin_huxley}" for an example), reflecting the dynamic opening and closing of individual ion channels. The activation and inactivation processes act in specific timescales, with inactivation usually acting in a significantly slower timescale, thus succeeding the activation.

Neurons utilize a large bank of ionic currents. The variety of ionic currents provides a variety of control parameters to achieve both flexible homeostatic regulation and robustness through redundancy. This apparent complexity may seem overwhelming.  However, all excitable systems share a remarkably simple and ubiquitous feedback structure. The core question thus boils down to understanding how the simple feedback structure can be extracted from complex neuronal models, and how the effect of each dynamic current maps to the distinct feedback loops.

Two main characteristics of neuron circuits that simplify the analysis are the parallel interconnection structure of the circuit, which means that ionic currents are additive, and a significant difference in the speed of activation and inactivation of different ionic currents, allowing for quasi steady-state analysis in distinct timescales. To this extent, standard linearization techniques can be utilized. They are described in "\nameref{sb:linearization}". The local behavior of the circuit around a given voltage is modeled by a parallel circuit interconnection of linear resistor, inductor and capacitor components, characterized by the total impedance of the circuit around the linearization point.

\subsection{Neuromorphic architecture}
\label{subsec:neuromorphic_architecture}

Mirroring the neuronal membrane organization where the ionic currents form a parallel interconnection structure, the neuromorphic architecture of \cite{ribar2019} also consists of a capacitor interconnected in parallel with a single resistive element, and a set of circuit elements emulating the active ionic currents. The general representation of the circuit is shown in \cref{fig:neuromorphic_circuit}.

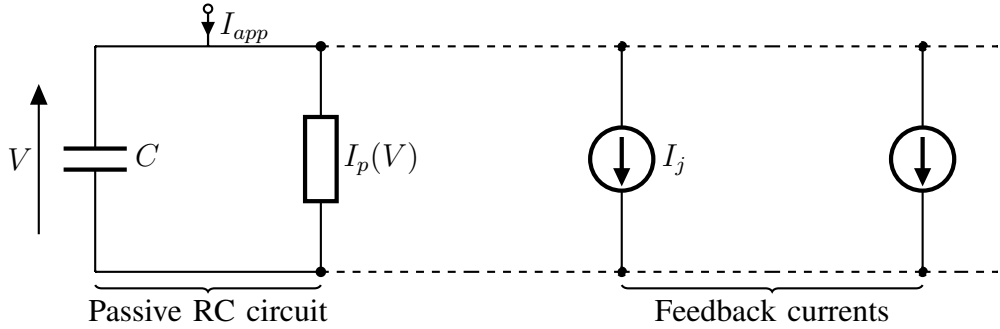
\includetikzfigure{neuromorphic_circuit}{A neuromorphic neuron circuit. A capacitor and a resistive element capturing the dissipative properties of the membrane are interconnected with elementary feedback current elements acting in different timescales.}

The resistive element has a monotonically increasing input-output relationship
\begin{equation}
	\frac{d I_p(V)}{dV} > 0, \quad \text{for all } V.
\end{equation}
Each element $I_j$ provides positive or negative conductance in a given voltage range and a given timescale. The elements are modelled as the cascade  of a linear filter and a nonlinear static function:
\begin{subequations}
	\label{eq:neuromorphic_current}
	\begin{align}
		I_j &= f_j (V_x), \\
		\tau_x \dot{V}_x &= V - V_x.
	\end{align}
\end{subequations}
A convenient way to model the input-output functions $f_j$ is to separate the elements into purely \textit{positive conductance} elements $I_j^+$ and purely \textit{negative conductance} elements $I_j^-$ that would therefore have the input-output characteristics
\begin{equation}
	\begin{aligned}
		\frac{df_j^+(V)}{dV} &> 0, \quad \text{for all } V, \\
		\frac{df_j^-(V)}{dV} &< 0, \quad \text{for all } V.
	\end{aligned}
\end{equation}
In addition, the elements can be grouped to act solely in a few specified timescales. This enables the analysis of the circuit through well-known techniques that utilize the timescale separation of the subsystems \cite{ermentrout2010}. By introducing the timescale of the membrane equation
\begin{equation}
	C / \tau_V = \frac{dI_p(V)}{dV} \bigg|_{V = V_{rest}},
\end{equation}
the elements can be grouped into three separate timescales: \textit{fast} ($\tau_f$), \textit{slow} ($\tau_s$) and \textit{ultra-slow} ($\tau_{us}$) satisfying the condition
\begin{equation}
	\label{eq:timescale_separation}
	\max (\tau_V, \tau_f) \ll \tau_s \ll \tau_{us}.
\end{equation}
The fast dynamics is usually taken to be of similar order as the dynamics of the voltage equation, and a commonly applied simplification is for the fast element dynamics to be instantaneous, i.e. $\tau_f = 0$. This assumption is used in the synthesis of excitable neuronal circuits in the following sections. In this setting every element can be indexed by the timescale in which it acts and the sign of its conductance, so that
\begin{equation}
	I_j = I_{x \in \{ f, s, us \}}^{\pm}.
\end{equation}

For a general neuromorphic model it is convenient to use the same form for all currents $I_j$, as well as consider an input-output mapping that is easily realizable in hardware, which is not necessarily the case for general polynomial forms. In this regard, a suitable choice for the functions $f_j^-$ and $f_j^+$ is a hyperbolic tangent function, so that
\begin{equation}
	\label{eq:current_source}
	\begin{aligned}
		f_j^+ &= \alpha_j \tanh (V_x - \delta_j), \\
		f_j^- &= - \alpha_j \tanh (V_x - \delta_j).
	\end{aligned}
\end{equation}
The function is defined by two parameters $\alpha_j$ and $\delta_j$ and is convenient for several reasons. Due to the saturation effect, the effect of each current is localized in the voltage range with the parameter $\delta_j$ controlling the position of the voltage window in which it acts, while the parameter $\alpha_j$ controls the strength of the current. The localization property is important as it captures the localized scope of the activation and inactivation functions. Secondly, hyperbolic tangent I-V characteristic is easily realizable in hardware, as presented in "\nameref{sb:subthreshold}" and "\nameref{sb:neuromorphic_building_blocks}".

\subsection{I-V curve characterization}
\label{subsec:i_v_curve_characterization}

The structure of the model \eqref{eq:neuromorphic_current} converts the analysis of the circuit into the shaping of its input-output characteristic or so-called \textit{I-V curves}. In addition to the static I-V curve which is measured by fixing the input voltage and measuring the corresponding static output current, transient properties can be characterized by defining the I-V curves in several distinct timescales. For this purpose, analysis can be conducted in the three timescales that are sufficient to capture spiking and bursting properties of neurons: fast, slow and ultra-slow. The corresponding fast, slow and ultra-slow I-V curves are then defined as
\begin{equation}
	\label{eq:i_v_curves}
	\mathcal{I}_x(V) = I_p(V) + \sum_{\tau_y \leq \tau_x} f^{\pm}_y(V),
\end{equation} 
so that $\mathcal{I}_x$ sums the input-output characteristics of all currents acting on the timescale $\tau_x$ or faster.

This set of input-output curves is sufficient for the analysis of the circuit due to the clear separation of the timescales in the model: for the quasi steady-state analysis of the model in timescale $\tau_x$ it is possible to freeze the effects of the currents with slower dynamics. This simplification enables the synthesis of excitable circuits to reduce to simple graphical conditions on its input-output I-V curves. In particular, generating excitable behavior involves introducing a region of negative conductance in a given timescale and restoring the positive conductance in a slower timescale, achieved by the appropriate interconnection of positive and negative conductance feedback elements. Controlling the circuit's behavior thus involves shaping the negative conductance region in different voltage ranges and timescales, enabling the control of multiscale excitable behaviors.
 
\subsection{Feedback structure of excitability}
\label{subsec:feedback_excitability}

The classical model of excitability, the Hodgkin-Huxley model, has three feedback currents: a resistive leak current, an inward sodium current, and an outward potassium current. Following the linearization procedure, it reduces to the linear circuit shown \cref{fig:hh_linearized_circuit}. The value of each component depends on the voltage around which the linearization is applied.

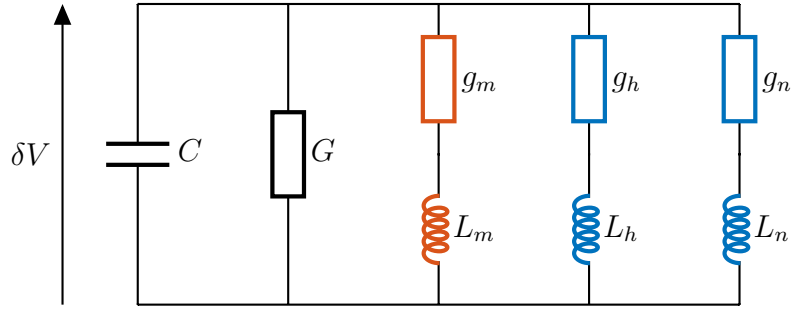
\includetikzfigure{hh_linearized_circuit}{Linearized Hodgkin-Huxley circuit. The impedance consists of the total resistive component $G$, in parallel with the first-order branches corresponding to the gating variables. The fast positive feedback appears as the fast negative conductance branch $g_m - L_m$ (red), while the slow negative feedback appears as the two positive conductance branches $g_h - L_h$ and $g_n - L_n$ (blue). All the values in the circuit are voltage-dependent, depending on the voltage point around which the linearization is considered.}

The interesting insight from this analysis comes by considering the steady state conductance functions $G(V)$, $g_m(V)$, $g_h(V)$ and $g_n(V)$ at every voltage in the physiological range. This is shown in \cref{fig:hh_linearized_conductances},  where the conductances are grouped based on the timescale in which they are acting:
\begin{itemize}
	\item The \textit{instantaneous} branch describes the passive dissipative property of the membrane.
	\item The \textit{fast} branch is due to the fast action of the sodium activation, which is several times faster than the sodium inactivation and potassium activation.
	\item The \textit{slow} branch is due to the slower action of the sodium inactivation and potassium activation.
\end{itemize}

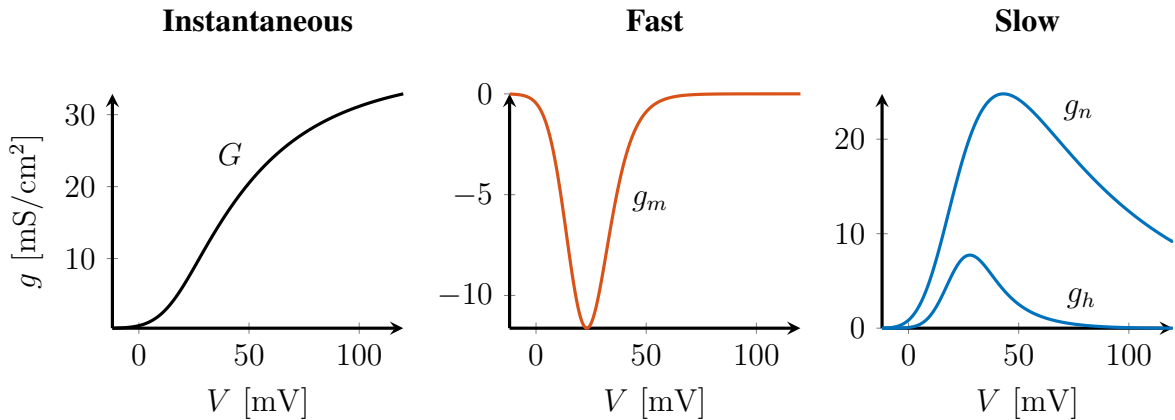
\includetikzfigure{hh_linearized_conductances}{Linearized conductances of the Hodgkin-Huxley model. The conductances are grouped in three parts: instantaneous (left), fast (middle) and slow (right). The excitability stems from the combination of the fast positive feedback (negative conductance) and the slow negative feedback (positive conductance).}

Linearization illuminates the dynamical signature of excitable systems in terms of the circuit structure. The conductance in the instantaneous branch is purely positive and models the passive properties of the membrane (\cref{fig:hh_linearized_conductances}, left). As the sodium current is inward, its activation provides positive feedback amplification modeled by the negative conductance of the circuit (\cref{fig:hh_linearized_conductances} middle), while its inactivation conversely provides positive conductance in the slower timescale (\cref{fig:hh_linearized_conductances}, right). Similarly, potassium is an outward current, so that its activation provides negative feedback amplification, modelled again as a slow positive conductance (\cref{fig:hh_linearized_conductances}, right). Excitability comes from the interconnection of the passive circuit with a \textit{fast negative} and a \textit{slow positive} conductance.

The second important point captured by the linearization is the local action of the dynamic currents. Both the fast negative conductance due to $g_m$ and the slow positive conductance due to $g_h$ and $g_n$ act in a voltage window defined by the their respective gating variables (i.e. where the derivatives of $m_\infty$, $h_\infty$ and $n_\infty$ are non-zero).

Analyzing the linearized model therefore highlights the following properties of the currents of an excitable neuron:
\begin{itemize}
	\item The feedback currents are \textit{localized} in \textit{amplitude}: they act in a limited voltage window.
	\item The feedback currents are \textit{localized} in \textit{time}: they have a well defined timescale in which they act.
\end{itemize}
These two properties allow the currents to shape the circuit's impedance in a local voltage and time range.

The basic structure of excitability has been captured in minimal dynamic models such as the famous FitzHugh-Nagumo model \cite{fitzhugh1961,nagumo1962} (see "\nameref{sb:min_models}"), and both Hodgkin-Huxley and FitzHugh-Nagumo equations have been successfully implemented in neuromorphic hardware \cite{mahowald1991,drakakis2000,linares-barranco1991}.

\subsection{Excitable circuit}

In the previous section, the core feedback components of an excitable neuronal membrane were highlighted to be an ionic current acting as a source of fast positive feedback, and a slower ionic current providing negative feedback. The minimal neuromorphic circuit is shown in \cref{fig:excitable_circuit}.

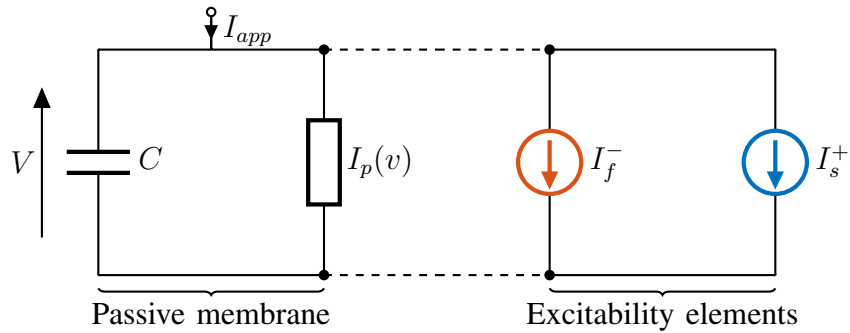
\includetikzfigure{excitable_circuit}{Neuromorphic excitable circuit. The circuit consists of a parallel interconnection of the passive membrane components with a fast negative conductance element ($I_f^-$) and a slow positive conductance element ($I_s^+$).}

The fast element $I_f^-$ generates a region of negative conductance in the fast I-V curve defining a region of bistability in the fast timescale (\cref{fig:excitable_bistability}). The slow element $I_s^+$ restores the monotonic I-V characteristic in the slow timescale by balancing the negative conductance of the fast I-V curve with positive conductance characteristic. The sufficient conditions for generating excitable behavior are thus
\begin{align}
	\frac{d\mathcal{I}_f}{dV} &< 0, \quad V \in (V_1^f, V_2^f), \\
	\frac{d\mathcal{I}_s}{dV} &> 0, \quad \text{for all } V,
\end{align}
which are shown in \cref{fig:excitable_iv_time}.

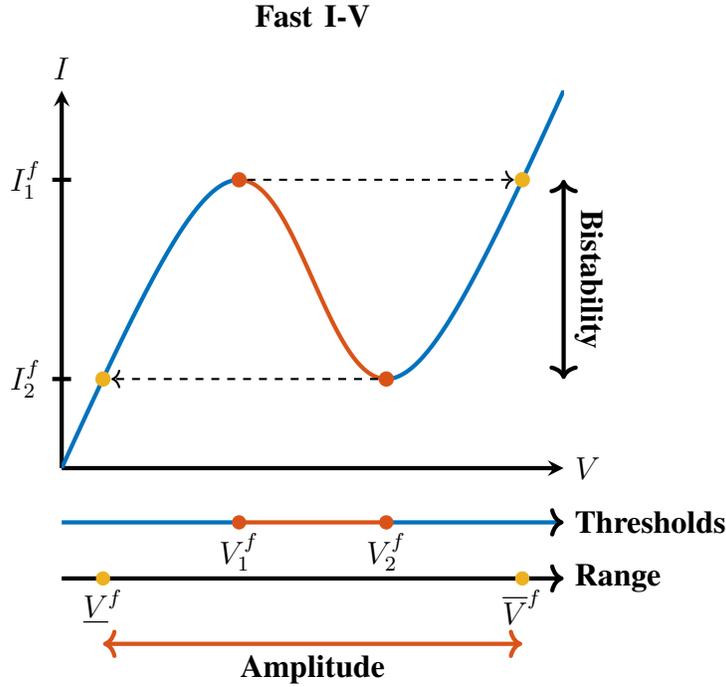
\includetikzfigure{excitable_bistability}{The N-shaped fast I-V curve. The curve represents a bistable fast system as for applied currents in the range $I_{app} \in (I_1^f, I_2^f)$ there are two stable equilibrium voltages separated by a middle unstable equilibrium. The range is determined by the \textit{threshold} voltages $V_1^f$ and $V_2^f$ representing the points where the slope is zero. Increasing the current above $I_1^f$ or decreasing it below $I_2^f$ leads to a 'jump' to the opposite branch of the curve, determined by the \textit{range} voltages $\overline{V}^f$ and $\underline{V}^f$ respectively. The amplitude of the spikes of the full system is determined by the voltage range $(\underline{V}^f,\overline{V}^f)$.}

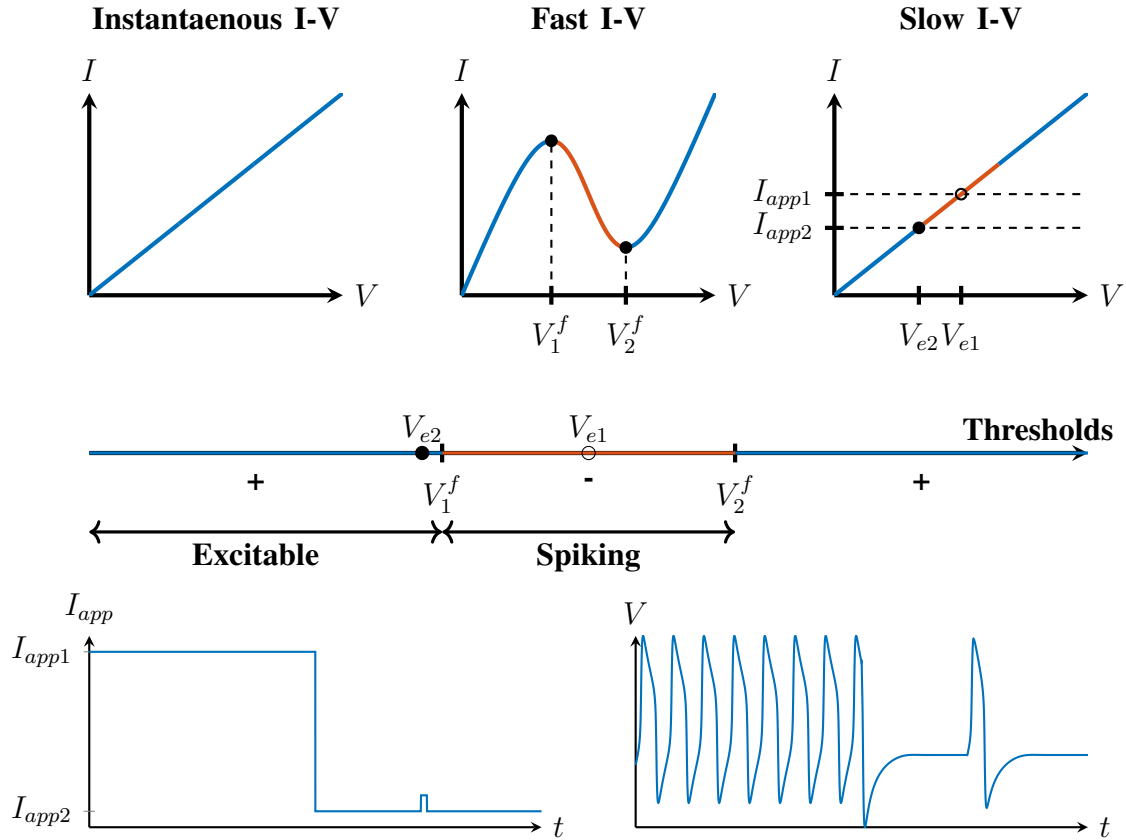
\includetikzfigure{excitable_iv_time}{Properties of the excitable circuit. Top: Excitability is characterized by the instantaneous passive I-V curve, N-shaped fast I-V curve and a monotonic slow I-V curve. The slow I-V curve is the steady-state characteristic of the system, so that the intersection of the line $I = I_{app}$ and the curve determines the equilibrium of the system $V_e$. Middle: If the equilibrium voltage lies within the negative-conductance region of the fast I-V curve, the equilibrium is unstable and the system exhibits constant spiking ($V_{e1}$). If the equilibrium voltage is outside this region, the equilibrium is stable and the system is excitable $V_{e2}$. Bottom: Changing the applied current switches the system between the two regimes.}

Given that the separation of timescales is satisfied \eqref{eq:timescale_separation}, the resulting behavior has the following characteristics:

\begin{itemize}
	\item The steady-state characteristic is given by the monotonic slow I-V curve, so that for each constant input $I_{app}$ there is a unique equilibrium point $(I_{app}, V_e)$. The equilibrium is stable except for a finite range of voltages within the interval $(V_1^f, V_2^f)$.
	\item The circuit has a stable spiking behavior for a finite range of constant applied input $I_{app}$. The spiking behavior is characterized by stable oscillations of the relaxation type, with the amplitude of the oscillation determined by the bistable range of the fast I-V curve (\cref{fig:excitable_bistability}).
	\item For stable equilibria near the unstable region contained in $(V_1^f, V_2^f)$, the system is in the \textit{excitable} state: short input pulses can trigger individual output \textit{spikes}, i.e. the transient manifestations of the spiking behavior.
\end{itemize}

The circuit in \cref{fig:excitable_circuit} shares the same essential structure as the FitzHugh-Nagumo model where the oscillatory spiking behavior appears through a subcritical Hopf bifurcation near the turning points of the fast I-V curve $V_1^f$ and $V_2^f$. The unstable range converges to the range $(V_1^f, V_2^f)$ as the timescales are more sharply separated, i.e. as $\max(\tau_V,\tau_f) / \tau_s \to 0$. The monotonicity of the slow I-V curve in conjunction with the N-shaped fast I-V curve and the sufficient separation in timescales guarantees the onset of spiking to appear through a Hopf bifurcation.

The specific bifurcation mechanism by which the rest-spike transition happens has important information processing consequences for the neuronal behavior \cite{drion2015b}. When the transition happens through a Hopf bifurcation, the oscillations emerge with a non-zero frequency. Instead, some neurons have a continuous frequency curve as the input is varied. This property is attributed to oscillations appearing through a saddle-node bifurcation \cite{izhikevich2007}, requiring non-monotonicity of the slow I-V curve. By ensuring that the turning point of the fast I-V curve coincides with the turning point of the slow I-V curve, the transition between rest and spiking instead appears through a saddle-node on an invariant circle bifurcation (SNIC). \Cref{fig:type1_type2} shows a comparison of the frequency curves for the two mechanisms.

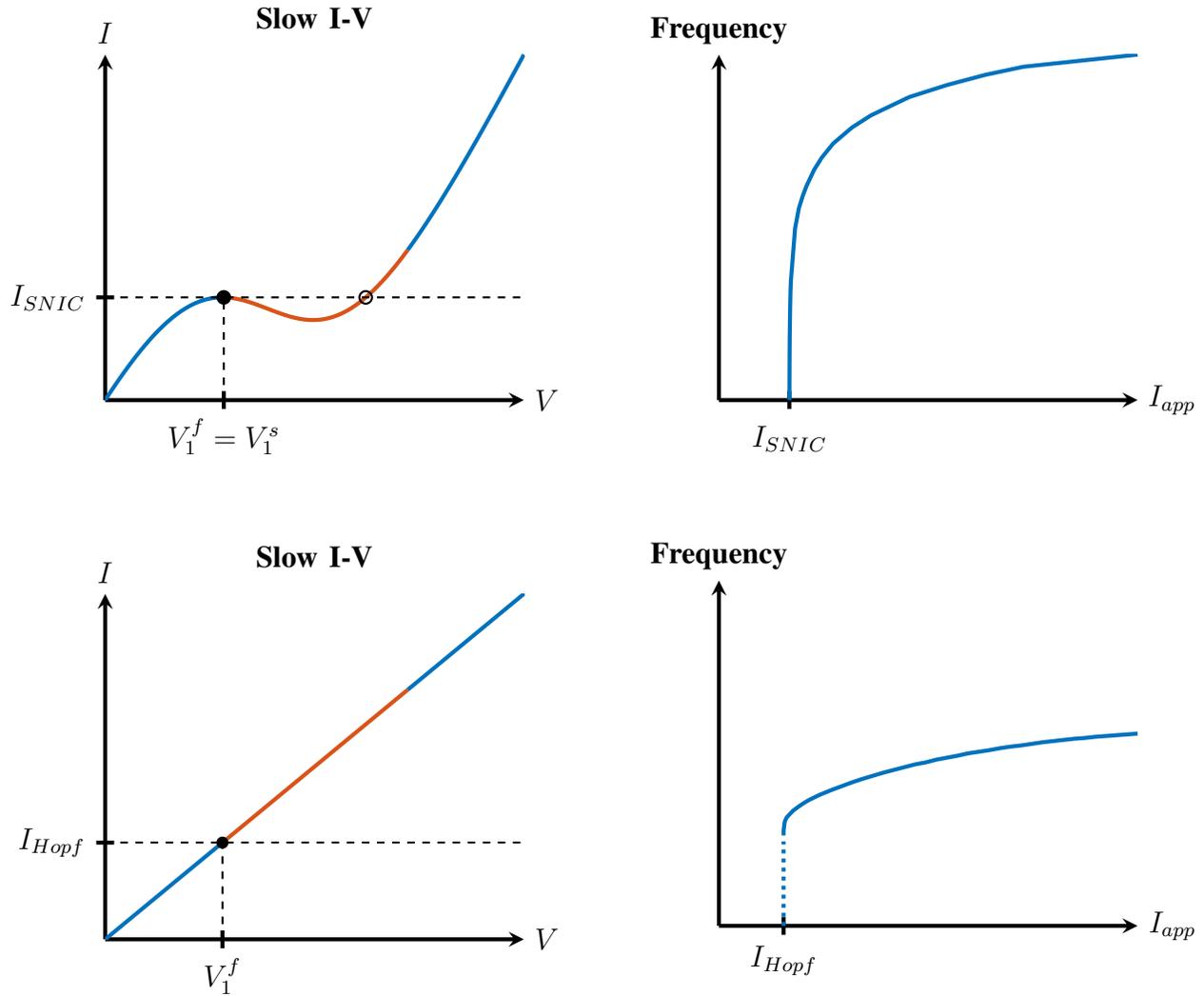
\includetikzfigure{type1_type2}{Different mechanisms of rest-spiking transition. The properties of the slow I-V curve determine the bifurcation mechanism of the transition between resting and spiking states, giving different input-output characteristics for the spiking frequency. Top: The transition point coinciding with the turning point of the slow I-V curve generates a SNIC bifurcation, characterized by a continuous frequency curve. Bottom: Monotonically increasing slow I-V curve generates a Hopf bifurcation at the transition point, characterized by a discontinuous frequency curve.}

The I-V curve interpretation is directly related to the standard phase portrait analysis as discussed in "\nameref{sb:phase_portraits}".

\subsection{Multiscale excitability: burst excitability}
\label{subsec:bursting}

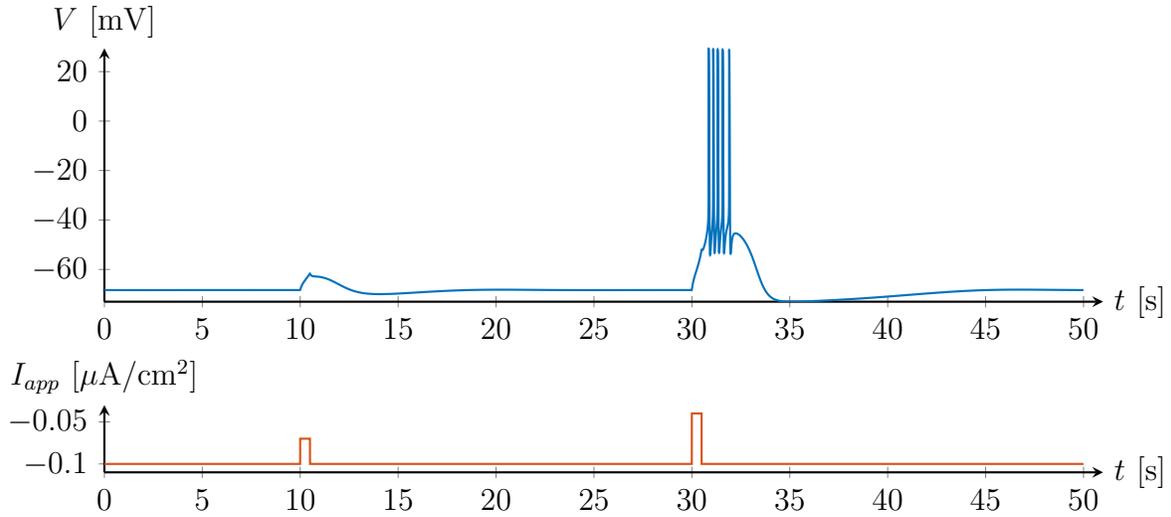
\includetikzfigure{burst_excitability_io}{Input-output characteristic of bursting systems. Similarly to the excitability property shown in \cref{fig:hh_excitability_io}, the subthreshold inputs generate small, subthreshold outputs. However, suprathreshold inputs generate a \textit{burst} of spikes. Response shown for the Aplysia R-15 model.}

The fundamental excitability motif of fast positive, slow negative feedback may appear repeated in a slower timescale, endowing the neuron with \textit{slow excitability} in addition to the excitable properties discussed previously. Slow excitability was discovered experimentally soon after the work of Hodgkin and Huxley \cite{moore1959}. It plays an important role in the regulation of excitability. This multiscale excitability motif leads to the specific form of excitable behavior called \textit{bursting}. Bursting is a prevalent signaling mechanism in neurons \cite{mccormick1992,beurrier1999,sherman2001,kepecs2002,krahe2004,marsat2006} and has important consequences on their synchronization properties as well as their information processing capabilities.

From the input-output perspective, burst excitability manifests as an all-or-none response in the form of a burst of spikes. This characteristic behavior is shown in \cref{fig:burst_excitability_io}. Like in the case of single-scale excitability, there is no unique physiological model of a bursting cell, and different neurons can have vastly different combinations of ionic currents that generate their activity. Still, an elementary feedback structure is common to all of them. A burst output in response to an input pulse consists of a slow spike that in turn activates fast spiking activity during the upstroke phase of the slow response. Thus, the underlying dynamical structure of bursting consists of a parallel interconnection of fast excitable and slow excitable subsystems, orchestrated by at least four distinct ionic currents. Linearization of conductance-based models of bursting neurons illuminates this dynamic structure as shown in the following section.

\subsection{Feedback structure of burst excitability}
\label{subsec:feedback_bursting}

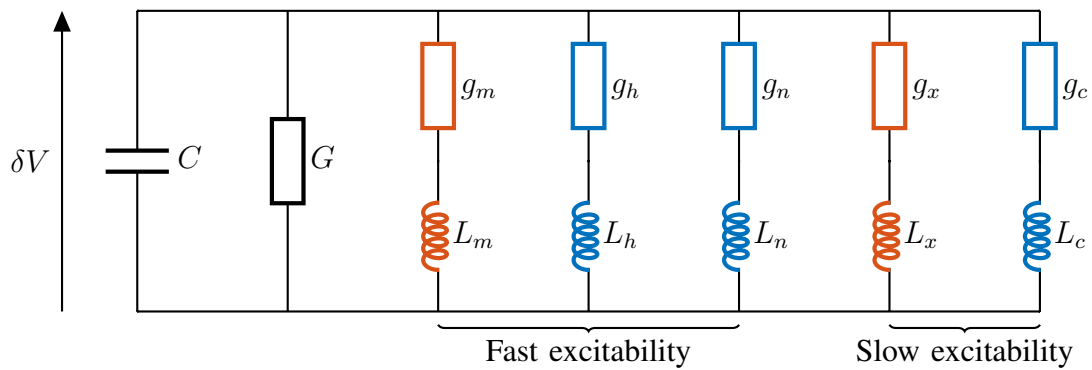
\includetikzfigure{rinzel_linearized_circuit}{Linearized bursting circuit of Aplysia R-15. The fast excitability component of the circuit impedance consists of the same elements as the Hodgkin-Huxley linearized circuit. The additional slow excitability elements provide slower first-order elements that make the system burst excitable.}

The feedback structure of bursting neurons is illustrated on the neuron R-15 of Aplysia \cite{plant1981,rinzel1987a}. The model has been one of the most extensively studied bursting neurons in the literature. The neuron has the same ionic currents as the Hodgkin-Huxley model, i.e. the sodium and potassium channels responsible for generating individual spikes, but in addition it has slower channels that are responsible for generating the slow variations of the membrane voltage. These channels are responsible for two additional ionic currents: an inward calcium current, and an outward calcium-activated potassium current. The model is described by
\begin{equation}
C\frac{dV}{dt} = - I_l - I_{Na} - I_K - I_{Ca} - I_{K-Ca} + I_{app},
\end{equation} 
where $I_l$, $I_{Na}$, and $I_K$ are as in \eqref{eq:hh_conductances} with the standard simplification that sodium activation is instantaneous (i.e. $m = m_{\infty}(V)$). The additional currents are modeled as
\begin{align}
\label{eq:calcium}
g_{Ca} &= \bar{g}_{Ca} \, x,\\
\label{eq:calcium_activated_potassium}
g_{K-Ca} &= \bar{g}_{K-Ca} \, \frac{c}{0.5 + c}.
\end{align}
The calcium activation variable $x$ and the calcium concentration $c$ can be written in the standard gating variable form (see gating variable equations in "\nameref{sb:hodgkin_huxley}"). Note that the calcium-activated potassium current does not have standard activation-inactivation form but it can be analyzed in the same way through linearization.

The time constants of the additional currents in the model $\tau_x$ and $\tau_{c}$ are significantly larger than the dynamics of the Hodgkin-Huxley currents. The calcium current $I_{Ca}$ therefore activates slowly compared to the potassium current $I_K$, while in turn the calcium-activated potassium current $I_{K-Ca}$ is the slowest in the model.

The two additional currents of the bursting model are modeled differently than in the Hodgkin-Huxley model but share the exact same mixed feedback structure to generate \textit{slow excitability}. The calcium current is is a slow analog of the sodium current: it is inward and its activation leads to a positive feedback loop. Likewise, the calcium-activated potassium current is a slow analog of the potassium current: it is outward and its activation provides negative feedback in the slowest timescale of the model. Bursting can therefore be effectively regarded as duplicating the feedback structure of Hodgkin-Huxley model: fast spiking results from the combination of fast positive and slow negative feedback, while bursting requires the same mixed feedback structure, but in a slower timescale. This structure is reminiscent of outer and inner control loops encountered in many feedback control systems.

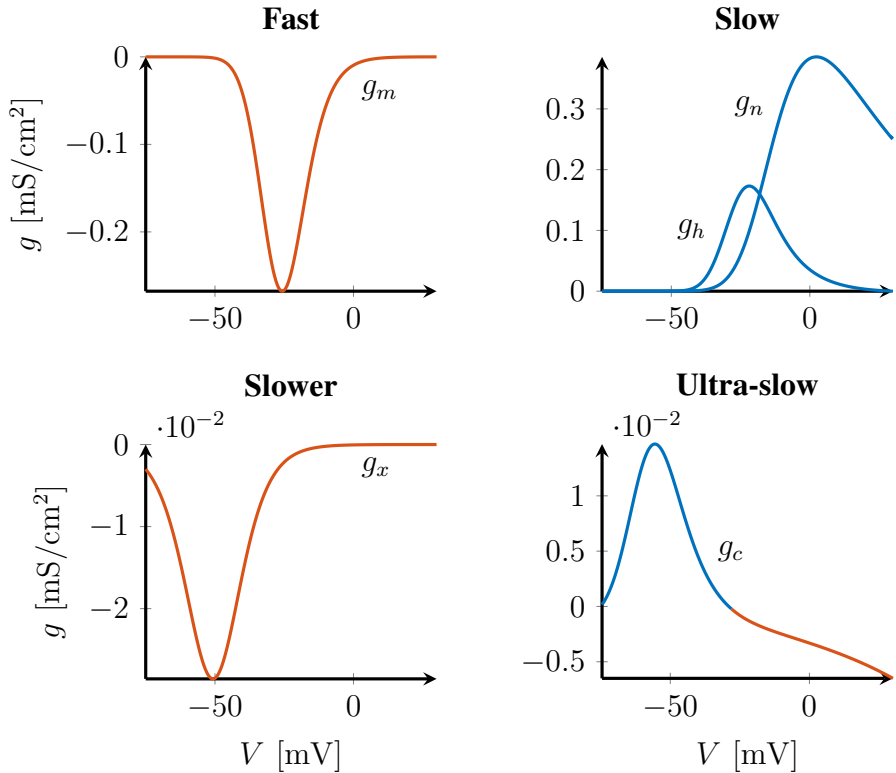
\includetikzfigure{rinzel_linearized_conductances}{Linearized conductances of the Aplysia R-15 bursting model. Building from the picture shown in \cref{fig:hh_linearized_conductances}, in addition to the fast negative and slow positive conductance necessary for spike generation (top), bursting requires additional slower negative and ultra-slow positive conductances (bottom). The two slower conductances activate at lower voltages than the two faster ones, so that the slow wave is generated in the lower voltage range. Note that at high voltage values the ultra-slow $g_c$ becomes negative: the excursions in this range are too fast for the ultra-slow variable to have an effect.}

The structure of the linearized circuit is illustrated in \cref{fig:rinzel_linearized_circuit}. Adding to the picture shown in \cref{fig:hh_linearized_circuit}, the two currents introduce two resistor-inductor branches to the linearized equation and their voltage dependence can be plotted as previously (\cref{fig:rinzel_linearized_conductances}).

Again, linearization illuminates the basic circuit structure of a bursting neuron. The \textit{fast negative}, \textit{slow positive} conductance of excitability is effectively repeated twice, so that the faster combination of the conductances generates the individual spikes within the burst, while the slower conductance combination generates the slow wave that switches the spiking on and off.

The nested structure is also reflected in the different localization in the voltage range: the fast conductances are localized in a higher voltage range than the slower conductances, which enables the distinct activation of the slow and fast thresholds.

A well-known minimal dynamical model that exploits this structure is the Hindmarsh-Rose model \cite{hindmarsh1984} (see "\nameref{sb:min_models}"), and several neuromorphic implementations of bursting neuronal models have been proposed since with differing levels of abstraction \cite{simoni2004,arthur2011,yu2011,wijekoon2008}.

\subsection{Bursting circuit}

Mirroring the structure of bursting neurons outlined in the previous section, a minimal neuromorphic bursting circuit is constructed by the interconnection of the excitable circuit discussed previuosly (\cref{fig:excitable_circuit}) with the additional feedback elements generating slow excitability. This structure is shown in \cref{fig:bursting_circuit}.

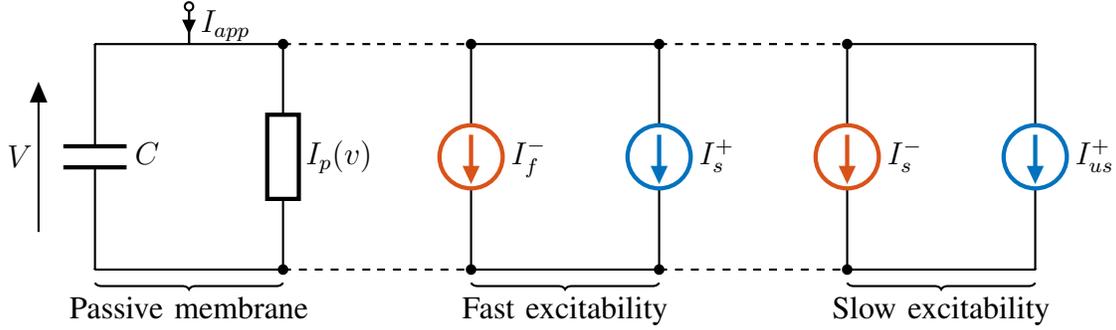
\includetikzfigure{bursting_circuit}{Neuromorphic bursting circuit. The circuit consists of the interconnection of a passive membrane with the feedback components generating both fast excitability ($I_f^-$ and $I_s^+$), as well as components generating slow excitability ($I_s^-$ and $I_{us}^+$).}

Slow excitability is generated with the same excitability interconnection structure of a positive feedback loop balanced by a slower negative feedback loop. Unlike fast excitability, slow excitability consists of the positive feedback acting on the slow timescale instead of being instantaneous, and the negative feedback is ultra-slow.

Similarly to how $I_f^-$ is responsible for generating bistability between a "low" and a "high" voltage state in the fast timescale through the N-shaped fast I-V curve, $I_s^-$ introduces a region of negative conductance in the slow I-V curve that gives rise to the bistability between a "low" voltage state and the "high" \textit{spiking} state, i.e. rest-spike bistability. This is shown in \cref{fig:slow_bistability_iv}. Both curves have regions of negative conductance, and importantly, the slow threshold is below the fast threshold, i.e
\begin{equation}
	V_1^s < V_1^f.
\end{equation}

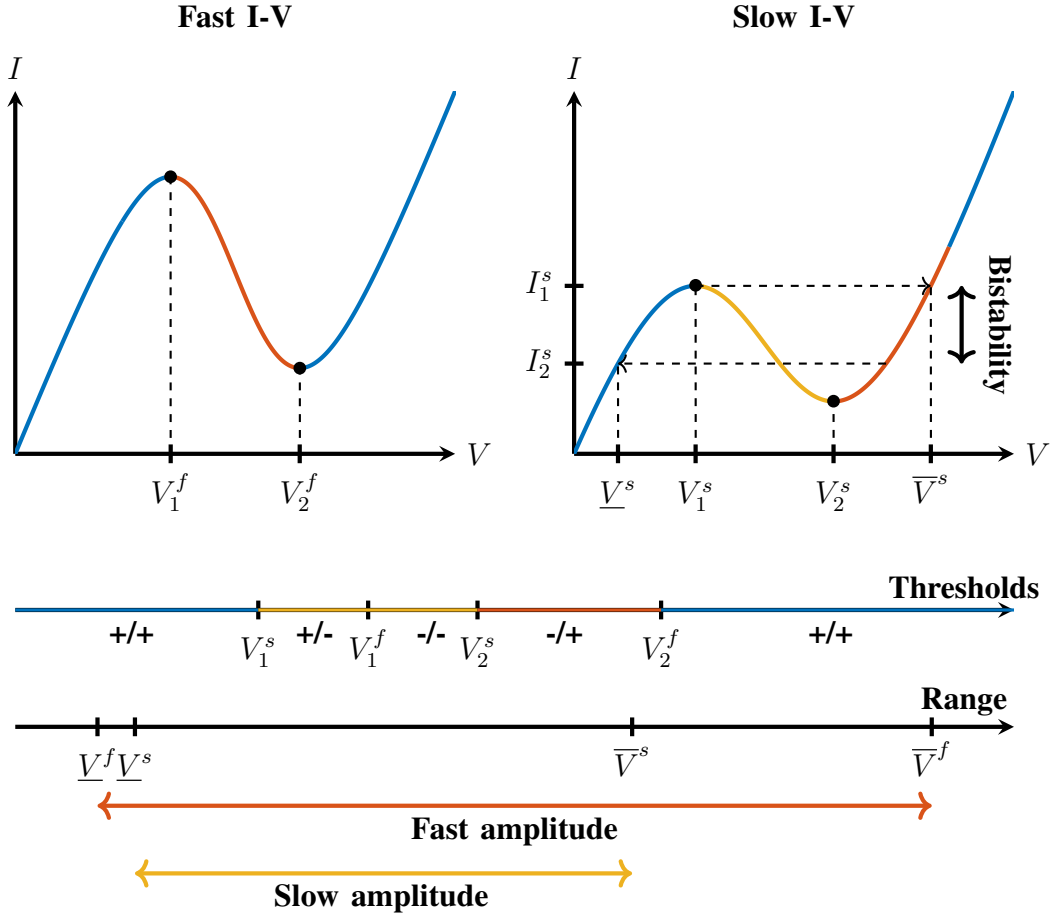
\includetikzfigure{slow_bistability_iv}{Slow bistability between stable resting and spiking states. Both fast and slow I-V curves of a slow bistable system are N-shaped signifying bistability. The "up" state of the slow I-V curve corresponds to the unstable region of the fast I-V curve, so that for currents $I_{app} \in (I_2^s, I_1^s)$ the system has coexisting stable equilibrium and a stable limit cycle. The fast and slow thresholds define different regions of positive/negative conductance in the fast (given by the first sign) and slow (given by the second sign) timescales respectively. Fast and slow bistability ranges then define the amplitudes of the fast and the slow spiking respectively of the full bursting system.}

Burst excitability is then achieved through the addition of the ultra-slow positive conductance element, and similarly to the case of an excitable system from previous section, a sufficient condition is for it to restore monotonicity in the ultra-slow I-V curve. The characteristics of a neuromorphic bursting circuit are shown in \cref{fig:bursting_iv_time}.

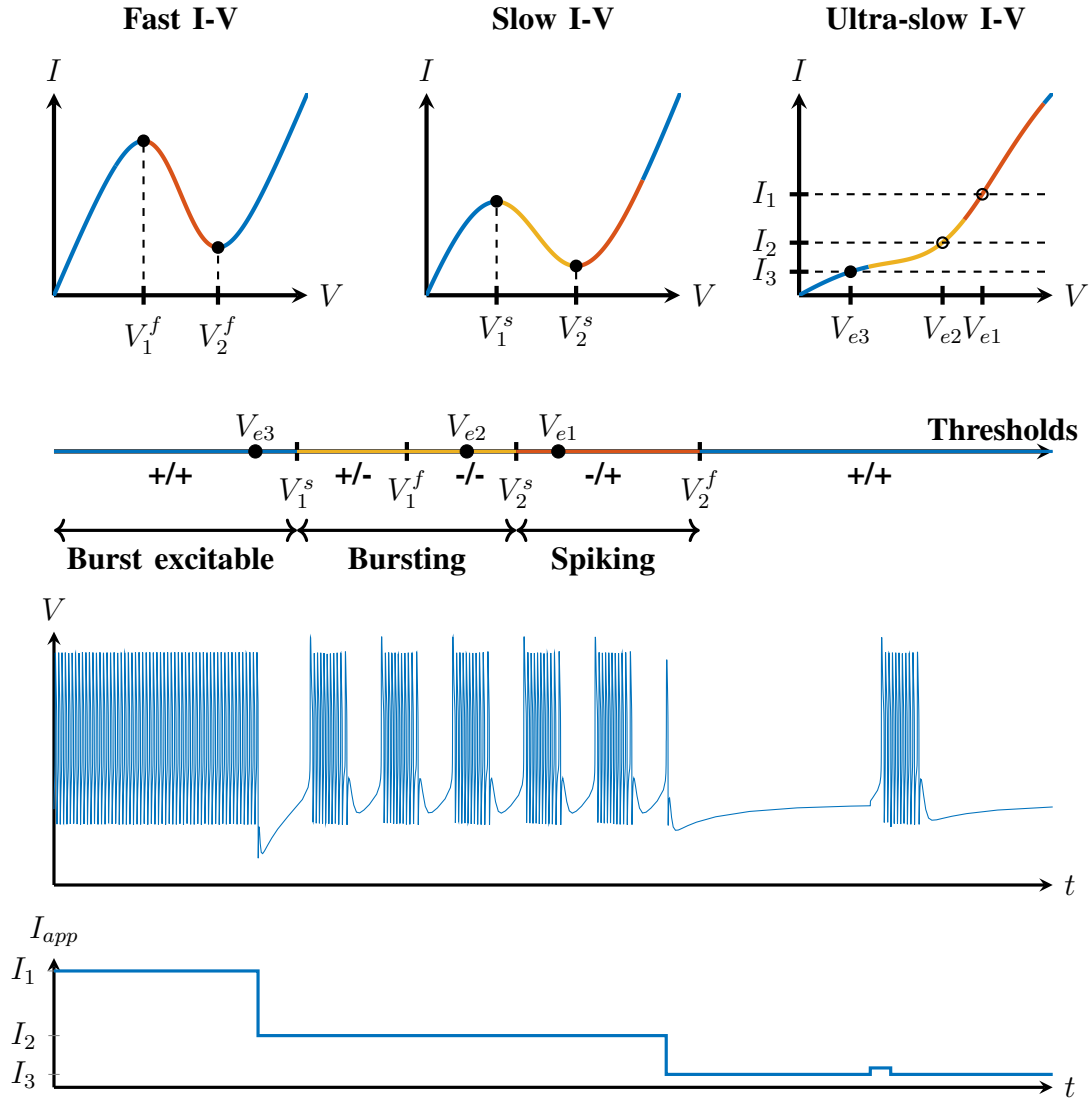
\includetikzfigure{bursting_iv_time}{Properties of the bursting circuit. Top: Burst excitability is characterized by the instantaneous passive I-V curve (not shown), N-shaped fast and slow I-V curves, and a monotonic ultra-slow I-V curve. The ultra-slow I-V curve is the steady state characteristic of the system, so that the intersection of the line $I=I_{app}$ and the curve determines the equilibrium of the system $V_e$. Middle: If the equilibrium voltage lies outside the negative conductance regions of fast and slow I-V curves, the system is burst excitable, if it lies in the negative conductance region of the slow I-V curve the system is bursting, and if it lies in the negative conductance region of the fast I-V curve above $V_2^s$ the system is spiking. Bottom: The system is switched between the three states through step changes in the input.}

Given that the separation of timescales is again satisfied \eqref{eq:timescale_separation}, the resulting behavior has the following characteristics:

\begin{itemize}
	\item The steady-state characteristic is given by the monotonic ultra-slow I-V curve, so that for each constant input $I_{app}$ there is a unique equilibrium point. The equilibrium is stable except for a finite range of voltages within the interval $(V_1^s, V_2^f)$.
	\item The circuit has a stable limit cycle characterized by the alternation between quiescence and spiking (bursting) for equilibrium voltages within the region $(V_1^s,V_2^s)$, and a stable spiking limit cycle within the range $(V_2^s,V_2^f)$.
	\item For stable equilibria near the turning point $V_1^s$ the system is in the \textit{burst excitable} state: short input pulses can trigger a burst of spikes. i.e. the transient manifestation of the bursting behavior.
\end{itemize}

The dynamical properties of the model are grounded in the analysis of a similar state-space model presented in \cite{drion2012,franci2012a,franci2014a}. The connection with the phase portrait analysis of this model is discussed in "\nameref{sb:phase_portraits}".

\subsection{Neural interconnections}

Even though single neurons exhibit complex behavior on their own, they never function in isolation. Instead, they are a part of richly interconnected neural networks where the behavior is critically dependent on the dynamical properties of the connecting elements.

Similarly to the feedback elements on the single-neuron level, neural interconnections consist of passive interconnections as well as active excitatory and inhibitory interconnections. Both types of connections are prevalent in biology with distinct effects on the net activity: the resistive connections are dissipative and tend to average out the activity across neurons, while synaptic connections can lead to diverse heterogeneous behaviors prevalent in neural activity. 

Passive interconnections are simply modeled as resistive connections between the voltage nodes of neurons, so that the currents flowing into neuron $j$ ($I_{ij}^p$) and neuron $i$ ($I_{ji}^p$) due to a passive connection between them are:
\begin{equation}
	I_{ij}^p = - I_{ji}^p = g_{ij}^p (V_i - V_j),
\end{equation}
where $g_{ij}^p$ is the conductance of the passive connection. They are the network analog of the leaky current of isolated neurons.

The active connections model the effects of chemical synapses which act as directed connections where the activity of the \textit{presynaptic} neuron affects the behavior of the \textit{postsynaptic} neuron. They are the network analog of the active current sources of an isolated neuron. Such components can be modeled in the same framework of \eqref{eq:current_source} as current source elements in parallel with the membrane elements of the postsynaptic neuron whose output depends on the voltage of the presynaptic neuron. The synaptic current $I_{ij}^{syn}$ in neuron $j$ thus depends on the activity of neuron $i$ as
\begin{subequations}
	\label{eq:synapse}
	\begin{align}
		I_{ij}^{syn} &= \alpha_{ij}^{syn} S(\beta (V_i^x - \delta^{syn})), \\
		\tau_x \dot{V_i}^x &= V_i - V_i^x,
	\end{align}
\end{subequations}
where $S(x)$ is the sigmoid function, $\alpha_{ij}^{syn}$ is the synaptic weight, $\beta$ is the steepness factor, $\delta^{syn}$ is the voltage offset of the synaptic current. This form is used to model both inhibitory and excitatory synaptic connections with the only difference lying in the sign of the synaptic weight $\alpha_{ij}$. In the case of an excitatory synapse, the spike in the presynaptic neuron induces an inward current $I_{ij}^{exc}$ in the postsynaptic neuron so that
\begin{equation}
	\alpha_{ij} > 0,
\end{equation}
and in the case of an inhibitory synapse, the spike in the presynaptic neuron induces an outward current $I_{ij}^{inh}$in the postsynaptic neuron, so that
\begin{equation}
	\alpha_{ij} < 0.
\end{equation}

The reader will note that the only difference between intrinsic and extrinsic current sources is that the current depends on the intrinsic voltage in the former case, while it depends on the extrinsic voltage in the latter case. The emphasis in the design of artificial neural networks is usually on controlling the extrinsic (or synaptic) conductances. The aim of the present article is to highlight the equally important role of intrinsic conductances for the control of networks. Biophysical neural networks rely as much on the modulation of intrinsic nodal feedback loops as on the modulation of the network synaptic connectivity \cite{nadim2014}.

\section{Control of neuromorphic circuits}
\label{sec:control}

The circuit synthesis framework presented in "\nameref{sec:synthesis}" amends neuromorphic circuits with a simplified methodology for controlling the excitability properties of neurons, mirroring the biophysical control mechanisms of neuromodulation. The I-V curves offer a set of simple input-output conditions that determine the characteristics of the circuit, thus connecting the simple interpretability of classical phase portrait models with the neuromodulatory principles of biological circuits. The first part of this section highlights the main mechanisms for controlling the excitability properties of a neuromorphic neuron.

The second part of this section discusses the importance of capturing the single neuron control properties for the behavior of networks. Controlling the excitability of a neuron drastically alters its response to external stimuli \cite{sherman2001,krahe2004,kepecs2002,marsat2006}. This not only changes the input-output processing characteristics at the single-neuron level, but changes the synchronization susceptibility of the individual nodes in the network setting. This \textit{excitability switch} offers a fundamental mechanism by which individual nodes can locally control the global network behavior through the appropriate mixed feedback balance at the cellular level. These principles are first shown in a two-neuron inhibitory circuit, an elementary oscillatory module that is a ubiquitous building block of more complex network structures. Finally, the scaling of these principles to interactions of these modules is demonstrated on a case-study of the well-known STG central pattern generating network.

\subsection{Neuromodulation as a control problem}

Neuromodulation is a general term describing the effects of various neurochemicals such as neurotransmitters and neuromodulators on the excitability of neurons. Neuromodulators have profound effects on the behavior of neural networks \cite{mccormick1992,harris-warrick2011,marder2012,nadim2014}, and understanding the effects of neuromodulation from a control perspective is essential in order to characterize the coexisting robustness and sensitivity of neuronal behaviors.

The detailed molecular mechanisms of neuromodulation are complex, but their aggregate effect at the cellular scale is that they modulate both the expression of ion channels as well as the sensitivity of their receptors to various neurotransmitters. In the electrical circuit model of a neuron, the effect of neuromodulation is modeled by varying the maximal conductance parameters of the currents. Understanding how neuromodulators can drive the behavior in the desired way therefore boils down to understanding which regions in the parameter space of the maximal conductances correspond to which behavior. This is a difficult task due to the highly nonlinear nature of the currents, but recent work \cite{drion2015,drion2015b} has proposed a way of viewing the modulation of maximal conductances as a form of \textit{loop-shaping} of the basic feedback loops responsible for generating the behavior.

The sensitivity of most parameters to a specific neuromodulator can be superimposed through the linearization technique into the aggregate conductances in representative timescales, mainly: fast, slow and ultra-slow. Spiking then requires the appearance of fast negative conductance, and slow positive conductance, while bursting requires the addition of the slow negative conductance and ultra-slow positive conductance, appropriately localized in the voltage range.

Modulating the maximal conductances thus shapes the conductance in each timescale, changing the properties such as frequency, duty cycle, as well as the qualitative properties of the waveform. The critical properties of both fast and slow excitability are primarily controlled by balancing the total positive and negative conductances in the fast and the slow timescale, which provides a highly redundant tuning mechanism, robust to large uncertainties in the individual elements.

For any proposed neural model, its modulation can be viewed as shaping the balance of positive and negative conductances in the appropriate timescale. This viewpoint can be directly applied to the neuromorphic architecture presented previously by considering the corresponding I-V curves of the model.

\subsection{Control of neuronal properties}

Each circuit element of the model \eqref{eq:neuromorphic_current} has a simple interpretation: it provides either positive or negative feedback in a specific voltage range and a specific timescale. The natural control parameters of this architecture are the maximal conductance parameters. \Cref{fig:bursting_amplitude,fig:bursting_frequency} illustrate that modulating those few parameters indeed shapes the spiking and bursting properties of the circuit. The available applet \cite{ribar2020} allows the interested reader to explore the ease of tuning the circuit properties by modulating the maximal conductance parameters. This tuning methodology is neuromorphic in that the biophysical mechanism of neuromodulation precisely modulates the maximal conductances of specific ionic currents.

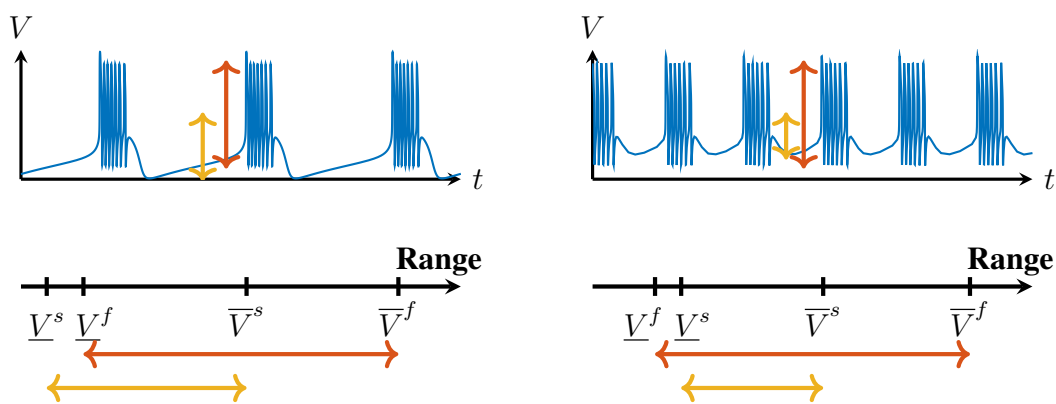
\includetikzfigure{bursting_amplitude}{Controlling the bursting waveform. The bursting oscillation can be designed by concurrently shaping the fast and slow I-V curves, thus changing the amplitudes of the fast and slow spiking. This example shows the difference between plateau (left) and non-plateau (right) oscillations by moving the negative conductance regions in the fast and slow timescales with respect to each other.}
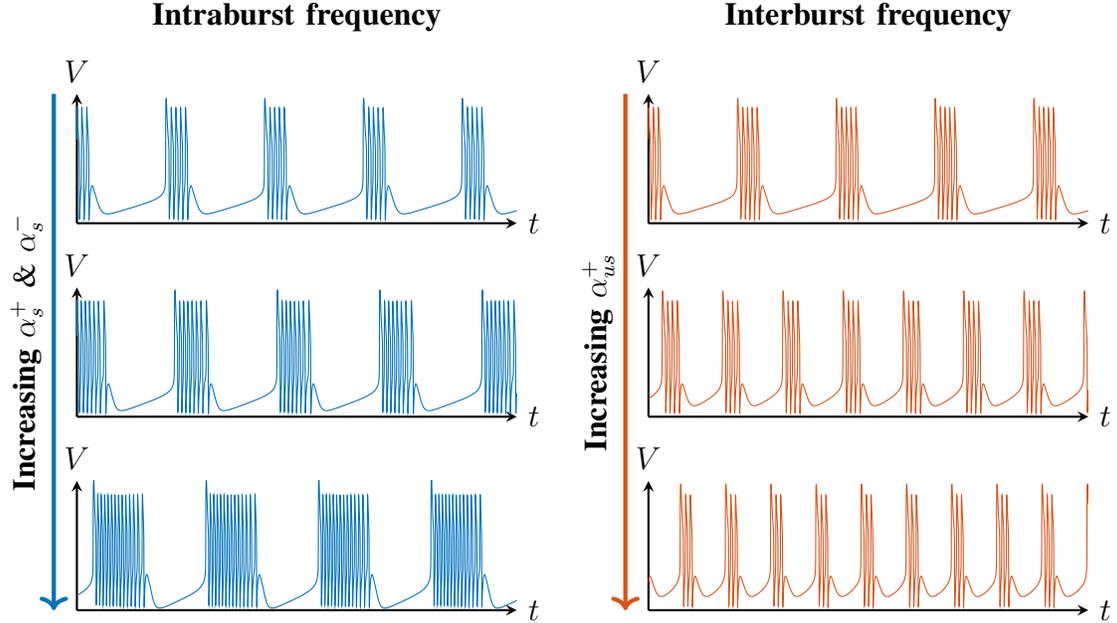
\includetikzfigure{bursting_frequency}{Controlling the frequency of the oscillation. Modulating the gains of the elements providing positive conductance determines the frequency: increasing the gain of the slow positive conductance element increases the intraburst frequency (left), while increasing the gain of the ultra-slow positive conductance increases the interburst frequency (right). In the first case, the slow negative conductance gain is increased proportionally as well in order to keep the interburst frequency constant.}

A parameter of particular importance is the balance between positive and negative conductance in the slow timescale. It is indeed the range of slow negative conductance which controls the transition between \textit{slow spiking} (characterized by the absence of negative conductance in the slow I-V curve) and \textit{bursting} (characterized by the presence of negative conductance in the slow I-V curve).

The transition between slow spiking and bursting will be shown to be of particular significance in the following sections. Modeling the transition between spiking and bursting has been a bottleneck of many mathematical models of bursting. It is for instance presented as an open question on the Scholarpedia webpage on bursting \cite{izhikevich2020}. The reason is that many mathematical models of bursting lack a source of negative conductance in the slow timescale. The reader will note that there is no possibility of tuning slow negative conductance without the current source $I_s^-$. In contrast, the current source $I_s^-$ directly controls the transition.

The point of transition between the two states can be studied mathematically by considering the local analysis of the slow I-V curve around its critical points \cite{ribar2019}. This interpretation can be directly derived from the singularity theory analysis of the phase plane model in \cite{franci2014a}. By aligning the threshold points of the fast and the slow I-V curves
\begin{equation}
V_1^f = V_1^s,
\end{equation}
the transition between the two regimes reduces to a condition on the concavity of the slow I-V curve around this point: in the bursting regime, the curve is locally concave, and in the spiking regime, the curve is locally convex. The two are separated by the condition
\begin{equation}
\frac{d^2 \mathcal{I}_s}{dV^2} = 0,
\end{equation}
which corresponds to the pitchfork bifurcation in the analysis of the phase plane model \cite{franci2014a}. This is illustrated in \cref{fig:pitch} which shows the control of the neuronal behavior through the sole parameter $\alpha_s^-$ controlling the gain of the slow negative conductance element.

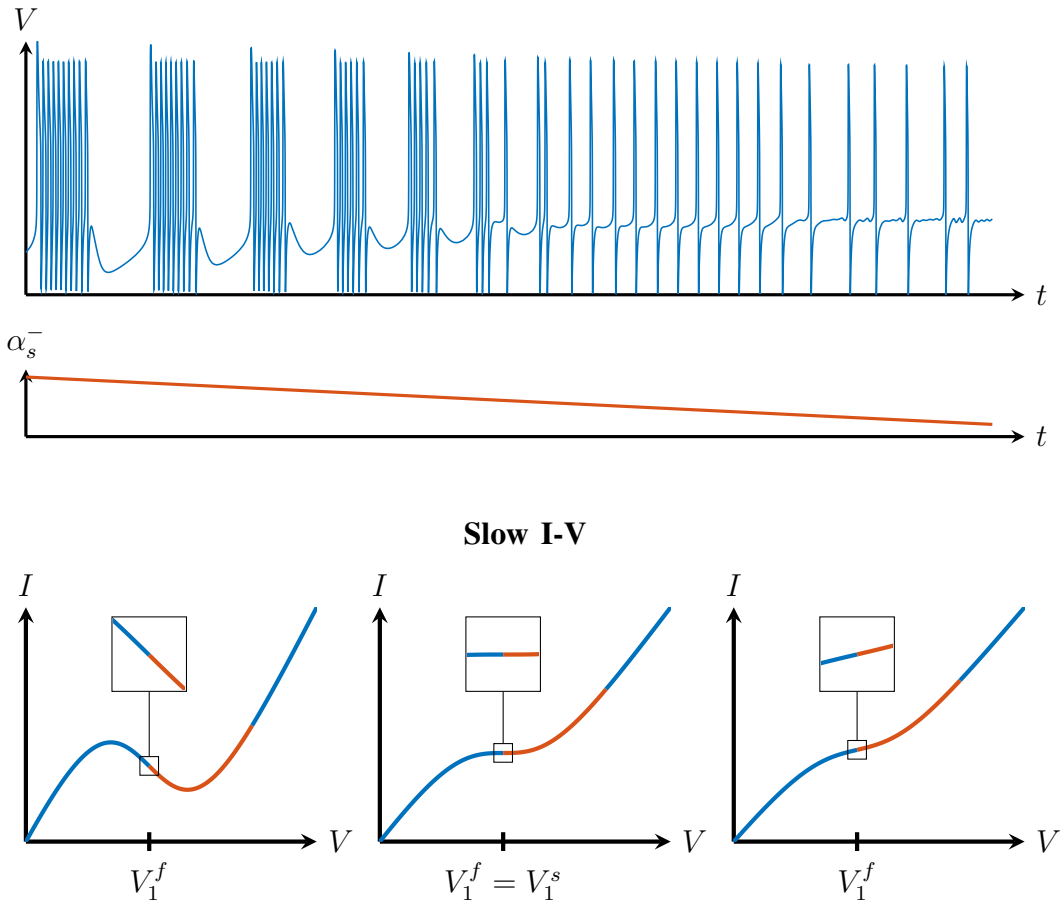
\includetikzfigure{pitch}{Controlling the neuronal oscillation mode. The transition between bursting and slow spiking modes is determined by the magnitude of the negative conductance region of the slow I-V curve (bottom). Tracing it locally around the point $V = V_1^f = V_1^s$ (middle), increasing the slow negative conductance gain (left) generates bursting, while decreasing it moves the system into slow spiking (right). The transition is shown at the top for a ramp input in $\alpha_s^-$, where the decrease in the slow bistable region gradually decreases the length of individual bursts, eventually leading to spiking with the disappearance of the slow bistability.}

The circuit structure of \cref{fig:bursting_circuit} can be easily realized in hardware using the basic principles of neuromorphic engineering, and the sidebar "\nameref{sb:circuit_implementation}" shows how standard circuit components can be utilized to implement a circuit implementation amendable to the neuromodulation principles discussed in this section.

\subsection{Controlling the relay properties of a neuron: a nodal mode and a network mode}

The robustness of neural communication lies in the excitable nature of neurons. The threshold property of firing neurons allows neural connections to reliably send information about significant events across long distances in spite of the signal degradation and noise characteristic of analog systems. Controlling the excitability of a neuron modulates the properties of this relay transmission.

The combination of the fast and slow excitability feedback elements endows the neuron with four possible discrete dynamical states, depending if the slow and fast excitability are present. Of particular importance is the control of the slow excitability of the node while the cell remains fast excitable. This mechanism manifests as the transition between bursting and slow spiking behaviors presented in \cref{fig:pitch}. In the first approximation, the role of fast excitability can be interpreted as to relay action potentials between neurons, whereas the role of slow excitability is to modulate the gain of the relay transmission.

\Cref{fig:neuron_io} illustrates the significance of this transition for the relay properties of a neuron. In the slow spiking state, the system acts as a relay to the individual spiking events: input spikes directly correlate with the spikes at the output. In contrast, in the bursting mode, individual spikes are effectively low-pass filtered and do not appear at the output. 

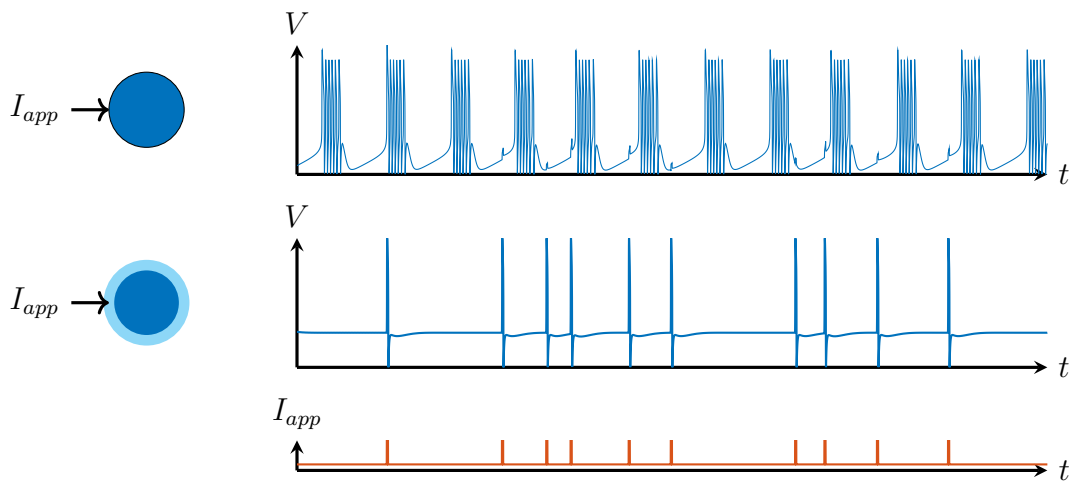
\includetikzfigure{neuron_io}{Relay properties in the bursting and slow spiking regimes. Subjecting the neuron to a train of spikes, when the system is in the bursting mode, the fast input spikes are rejected and the oscillating bursting behavior is uninterrupted. When the system is in the slow spiking state, each individual spike is relayed at the output. The switch is controlled through a single parameter corresponding to the gain of the slow negative conductance. The light blue border around the node represents the spiking state of the neuron.}

This excitability switch has important consequences for the single cell behavior in the network setting. In the bursting mode, a neuron tends to synchronize with the slow rhythm of the network and effectively blocks the transmission of local spiking inputs. In contrast, in the spiking mode, a neuron can switch off from the network rhythm and participate in the local relay of spiking inputs. This elementary mechanism suggests a key role for the slow negative conductance of a neuron: it controls the transition between a \textit{nodal} mode, where the neuron relays local information at the cellular scale, and a \textit{network} mode, where the individual neuron becomes an internal unit of a larger ensemble.

\subsection{Controlling the rebound properties of a neuron}

In its nodal mode, the input-output response of a neuron is primarily characterized by \cref{fig:hh_excitability_io}, that is, by how it responds to a short depolarizing input current. In contrast, how a neuron responds to a longer hyperpolarizing pulse turns out to be critical for its network mode. This property is known as the \textit{post-inhibitory rebound}. A robust generation of the post-inhibitory rebound response is essential for the generation of oscillations in networks where the individual neurons do not oscillate in isolation (see "\nameref{sb:cpg}").

In neurophysiology this mechanism has been linked to two specific inward currents, known as the 'hyperpolarization-activated cation current' $I_h$ and the 'low-threshold T-type calcium current' $I_{Ca,T}$ \cite{bucher2015}. The low threshold inward calcium current only activates under hyperpolarization. Its activation provides the slow negative conductance necessary for bursting. The  $I_h$  current adds to the total inward current to regulate the response to hyperpolarization.

The rebound mechanism is first illustrated in \cref{fig:pir_fast_bistable} on a spiking neuron. Starting from the bistable circuit of \cref{fig:excitable_bistability}, the rebound spike is obtained by adding a slow negative feedback current with low activation range. In the same manner, the rest-spike bistable system of \cref{fig:slow_bistability_iv} can provide a transient bursting response with an additional ultra-slow negative feedback current \cref{fig:pir_slow_bistable}.

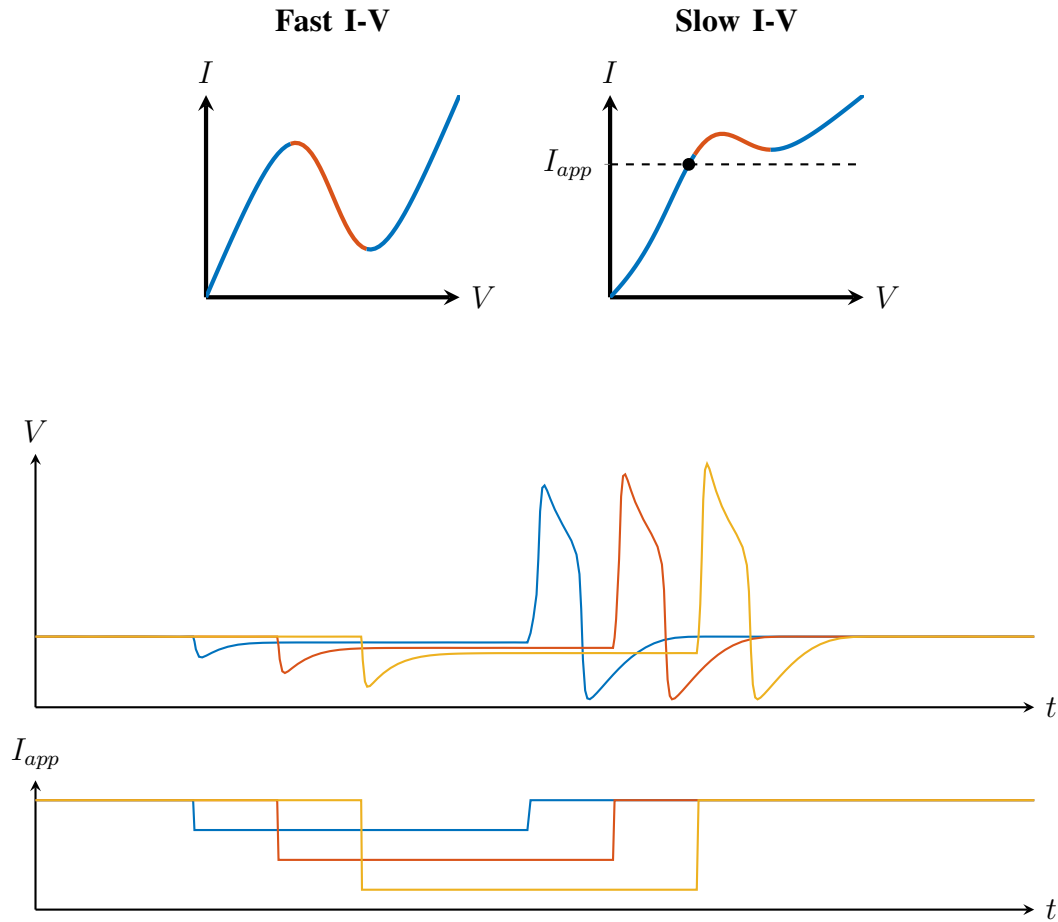
\includetikzfigure{pir_fast_bistable}{Post inhibitory rebound spike in a fast bistable system. By introducing a slow negative feedback current with a low activation to the bistable system of \cref{fig:excitable_bistability}, the system exhibits a robust transient spiking response to a prolonged negative input pulse in the absence of endogenous spiking oscillation.}

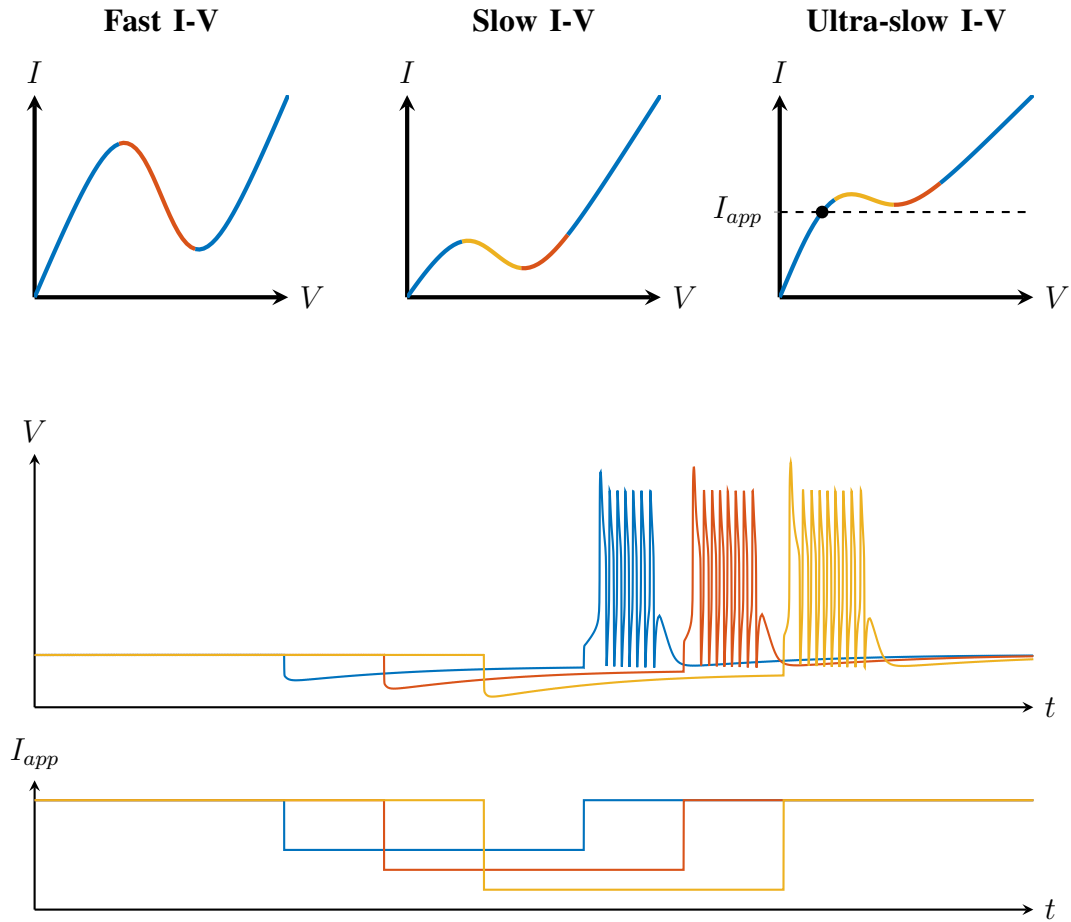
\includetikzfigure{pir_slow_bistable}{Post inhibitory rebound burst in a slow rest-spike bistable system. Mirroring the construction in \cref{fig:pir_fast_bistable}, by introducing an ultra-slow negative feedback current with a low activation to the bistable system of \cref{fig:slow_bistability_iv}, the system exhibits a robust transient burst response to a prolonged negative input pulse in the absence of endogenous bursting oscillation. This is the essential mechanism that allows networks of neurons to generate robust rhythm in the absence of endogenous oscillatory behavior in the individual nodes.}

\subsection{Turning a rhythmic circuit on and off}

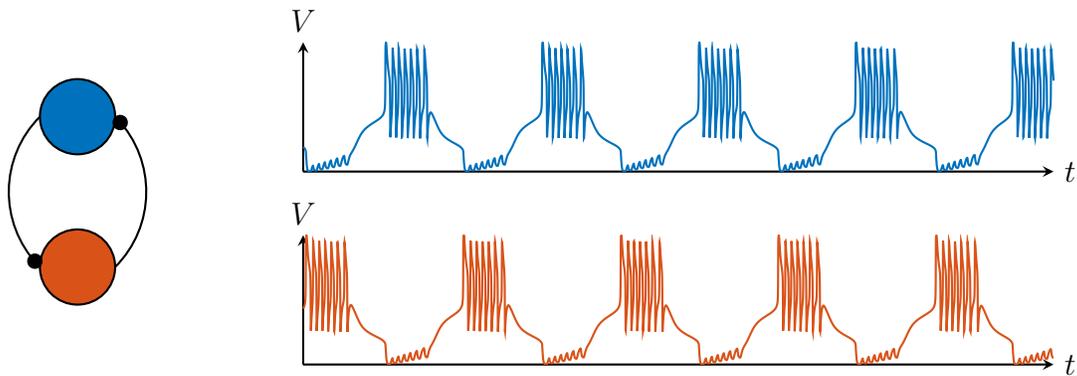
\includetikzfigure{hco_pir}{Half-center oscillator. Using the cellular mechanism described in \cref{fig:pir_slow_bistable}, anti-phase oscillation in a two-neuron network is achieved by interconnecting the neurons with inhibitory synapses. When one neuron fires a burst of spikes it introduces a transient negative input current into the other neuron. Once this input terminates, the other neuron fires a burst of spikes in response, inhibiting the first neuron in alternation. If the synaptic connections are strong enough, this alternative spiking produces a robust anti-phase oscillation of the network.}

The post-inhibitory rebound input-output response of a neuron was recognized more than a century ago as central to rhythm generation in neuronal circuits (see "\nameref{sb:cpg}"). The simplest illustration is provided by the circuit made of two neurons that inhibit each other. The anti-phase rhythm illustrated in \cref{fig:hco_pir} is a direct consequence of the rebound properties of each neuron. Even if the two neurons do not burst in isolation, their mutual inhibition generates an autonomous rhythm in the circuit. Each neuron of this network exhibits its network mode, that is, becomes an internal component of a rhythm at the circuit scale.

The two inhibitory connections define the phase relationship between the individual oscillations so that the neurons synchronize in anti-phase, even when the neurons oscillate endogenously. The excitability switch shown in \cref{fig:neuron_io} effectively switches the neuron from an endogenous source of oscillation, robust to external disturbance, to an exogenous relay mode, sensitive to external inputs.

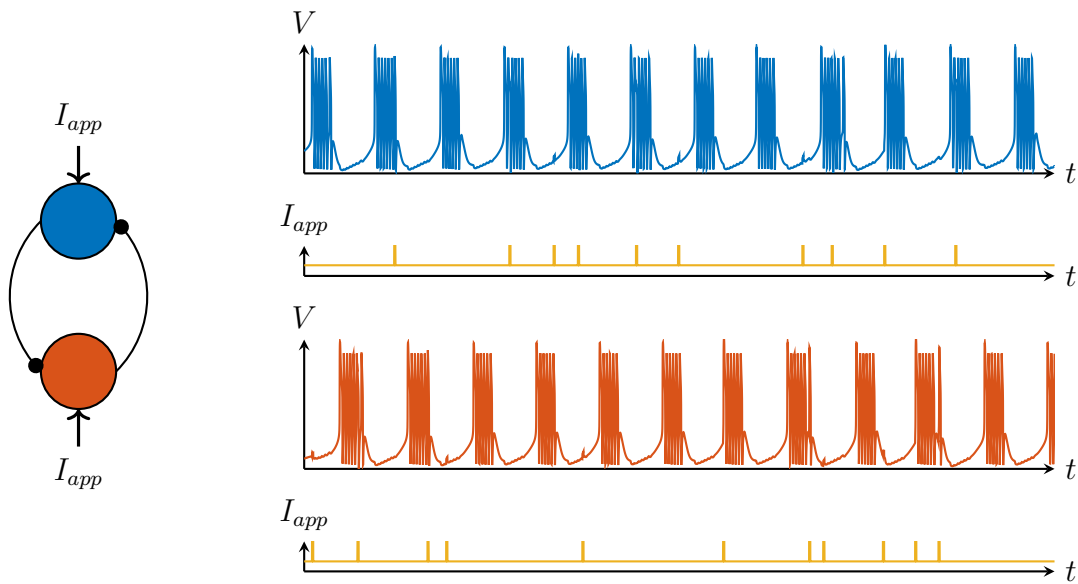
\includetikzfigure{hco_switch_bursting}{Two-neuron inhibitory network switched ON. When the individual neurons are in the bursting regime, the neurons oscillate in anti-phase defined by the inhibitory connections. The rhythm is robust to external disturbance, so that spike trains to individual neurons do not significantly alter the behavior.}
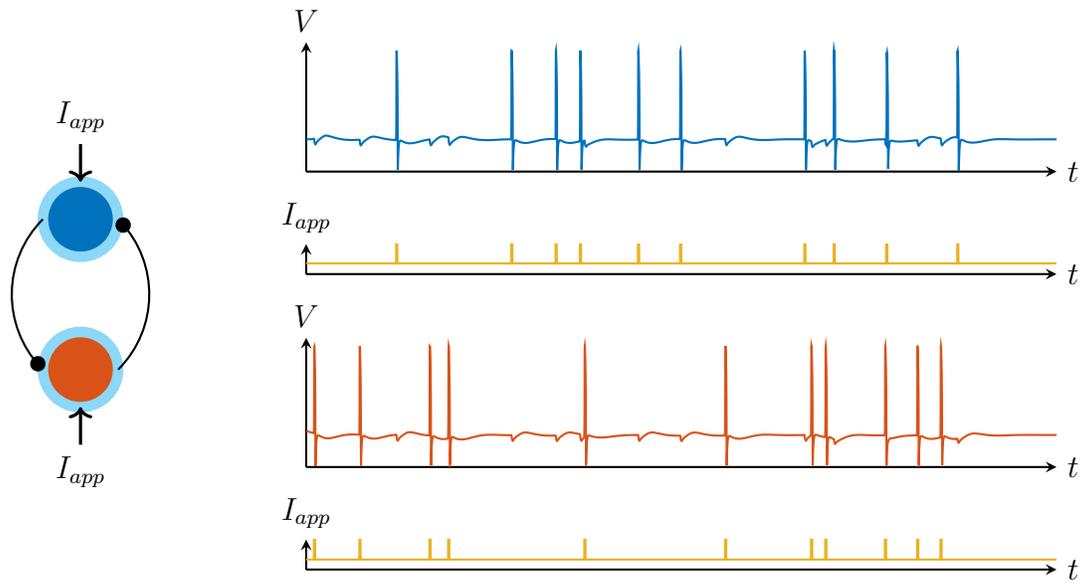
\includetikzfigure{hco_switch_spiking}{Two-neuron inhibitory network switched OFF. Decreasing a single parameter corresponding to the slow negative conductance gain of each neuron, neurons are switched from the bursting to the spiking regime. This effectively switches the rhythm off and neurons are able to relay individual incoming spikes, similarly to \cref{fig:neuron_io}.}

In \cref{fig:hco_switch_bursting}, the gain of the slow negative conductance of the individual neurons is high and they generate robust synchronized bursting oscillations. When subjected to external disturbance in the form of train of pulses in the applied currents, the behavior is largely independent of the changes at the inputs. Lowering the gain of the slow negative conductance of the two neurons (\cref{fig:hco_switch_spiking}), the endogenous bursting oscillation is turned off, and the neurons act as independent relaying systems. The effect of the input train of pulses is drastically different in this mode, as individual spikes are all transmitted.

The nodal excitability switch provides a way of controlling the network rhythm through a control of a single nodal parameter. By scaling up the same principles, more complex networks that consist of interconnections of such rhythm-generating modules can switch between different rhythms through the control of excitability properties of the individual nodes.

The reader will note that the switch between the nodal and network mode of a neuron is itself a control mechanism that can be intrinsic or extrinsic to the neuron: it is intrinsic when controlled via the modulation of the slow negative conductance of the neuron, and is extrinsic when controlled via hyperpolarization of the neuron through the applied current. The versatile control of this neuronal switch by a diversity of different mechanisms indeed suggests its key significance in controlling the properties of a neural network.

\subsection{Nodal control of a network}
\label{subsec:nodal_control}

This final section aims at illustrating that the basic control principle described in the previous section is eminently scalable and modular. A network can be dynamically reconfigured into a variety of different networks by simply modulating which nodes are "in" at a given time and a given location. Here, a five-neuron network is considered, inspired by the well-known crustacean stomatogastric ganglion (STG) structure \cite{marder2007,gutierrez2013,gutierrez2014}. The STG has been a central biological model over the last forty years in the discovery of the significance of neuromodulation. The interested reader is refered to the inspiring essay \cite{nassim2018}, a fascinating account of the scientific journey of Eve Marder into the world of neuromodulation as a key control principle of biological networks. The network below is a crude simplification of the biological STG network, but retains its core structure: it consists of an interconnection of two central pattern generating circuits that oscillate at different frequencies. The central pattern generating networks are modeled by two two-neuron inhibitory modules that oscillate in anti-phase. The two CPGs interact with each other through resistive connections with a middle "hub" neuron, and thus the network rhythm is given by a combination of the two individual rhythms.

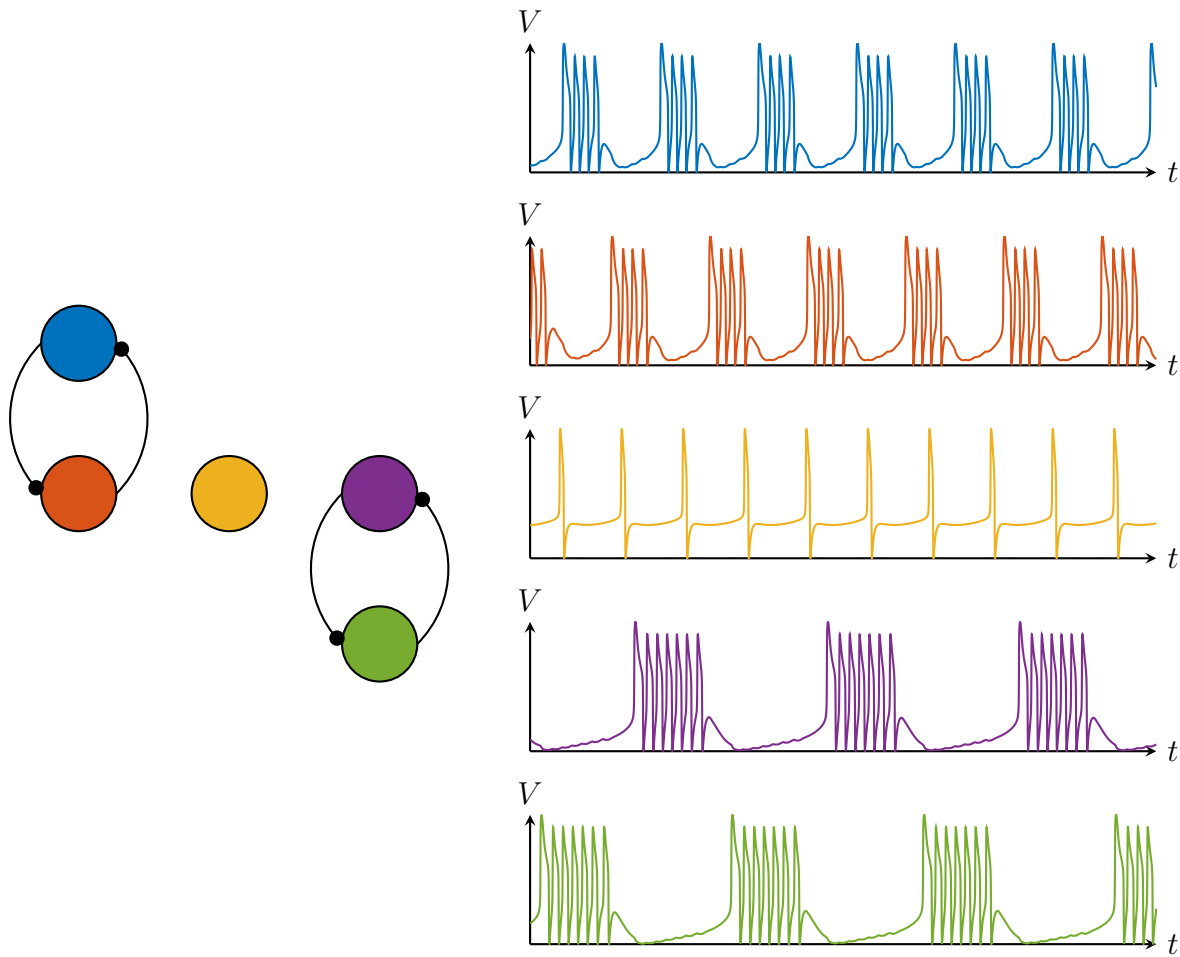
\includetikzfigure{cpg_disconnected}{Rhythm-generating network disconnected. The network consists of a pair of two-neuron oscillators and a middle hub neuron. The parameters are chosen so that the left oscillator is fast and the right oscillator is slow, while the hub neuron is in the slow spiking mode.}

The network is recreated with the neuromorphic model in \cref{fig:cpg_disconnected}. The parameters of the neurons are chosen so that the two inhibitory circuits generate a fast and a slow rhythm, respectively. The difference between the two subnetworks involves no change in the time constants of the individual currents, but only a different value of the ultra-slow feedback gain, which was shown to control the interburst frequency in the single neuron analysis. The middle hub neuron is set to its slow spiking mode, so that it acts as a relay between the two independent rhythms.

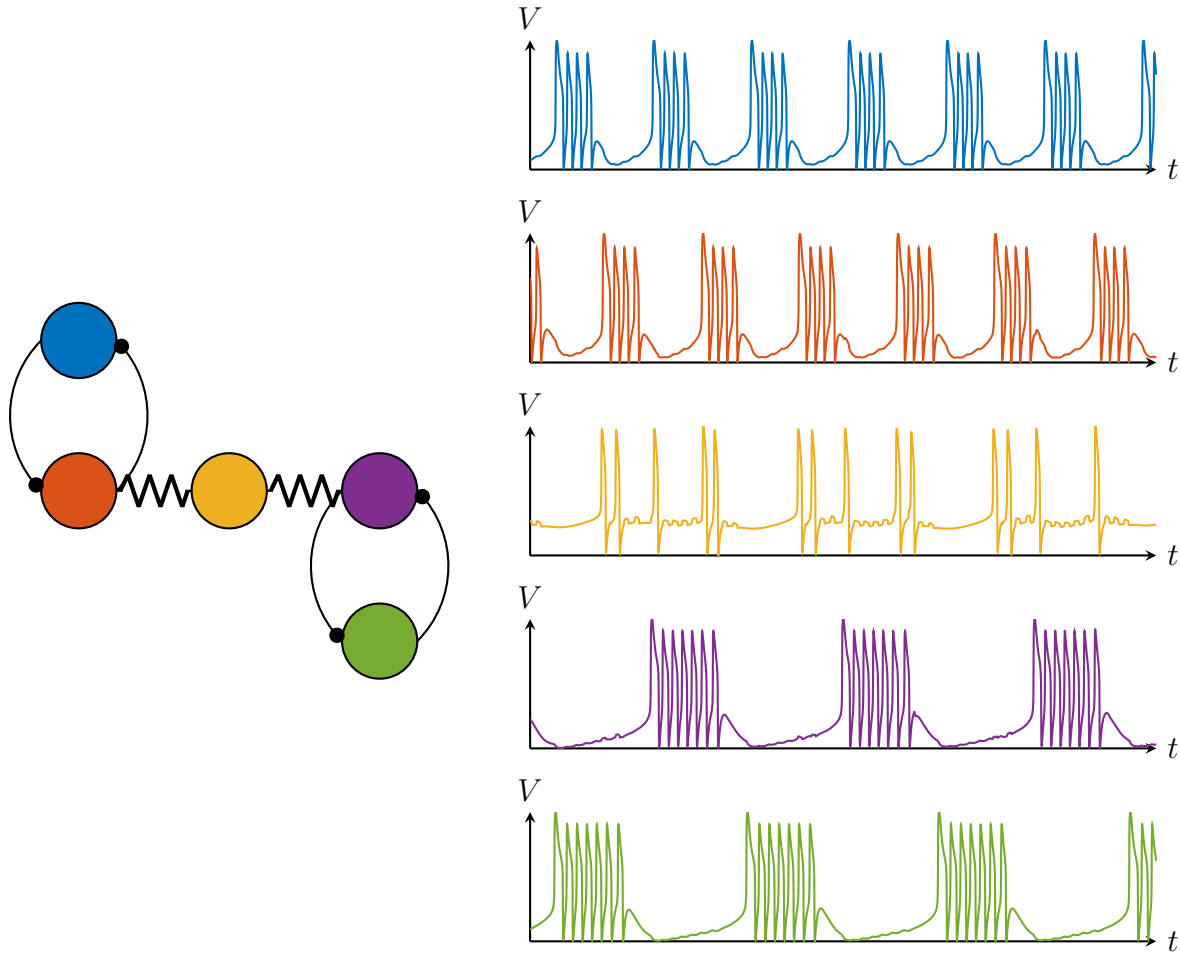
\includetikzfigure{cpg_connected}{Rhythm-generating network connected. Hub neuron is connected to individual neurons of the two-neuron oscillators through electrical connections. The resistive connections enable the interaction between the two rhythms. Due to the presence of slow negative conductance in the rhythm generating neurons, the bursting rhythms are robust to the input disturbances, so that the fast and slow rhythms coexist, while the hub neuron experiences a mix of the two oscillations.}

\Cref{fig:cpg_connected} shows the two rhythms interacting under the addition of resistive connections between the hub neuron and the individual CPGs. When the resistive connections are weak, the individual rhythms are mostly unaffected by the connection, and the fast and the slow rhythm within the network coexist. The hub neuron exhibits a mix of the two frequencies.

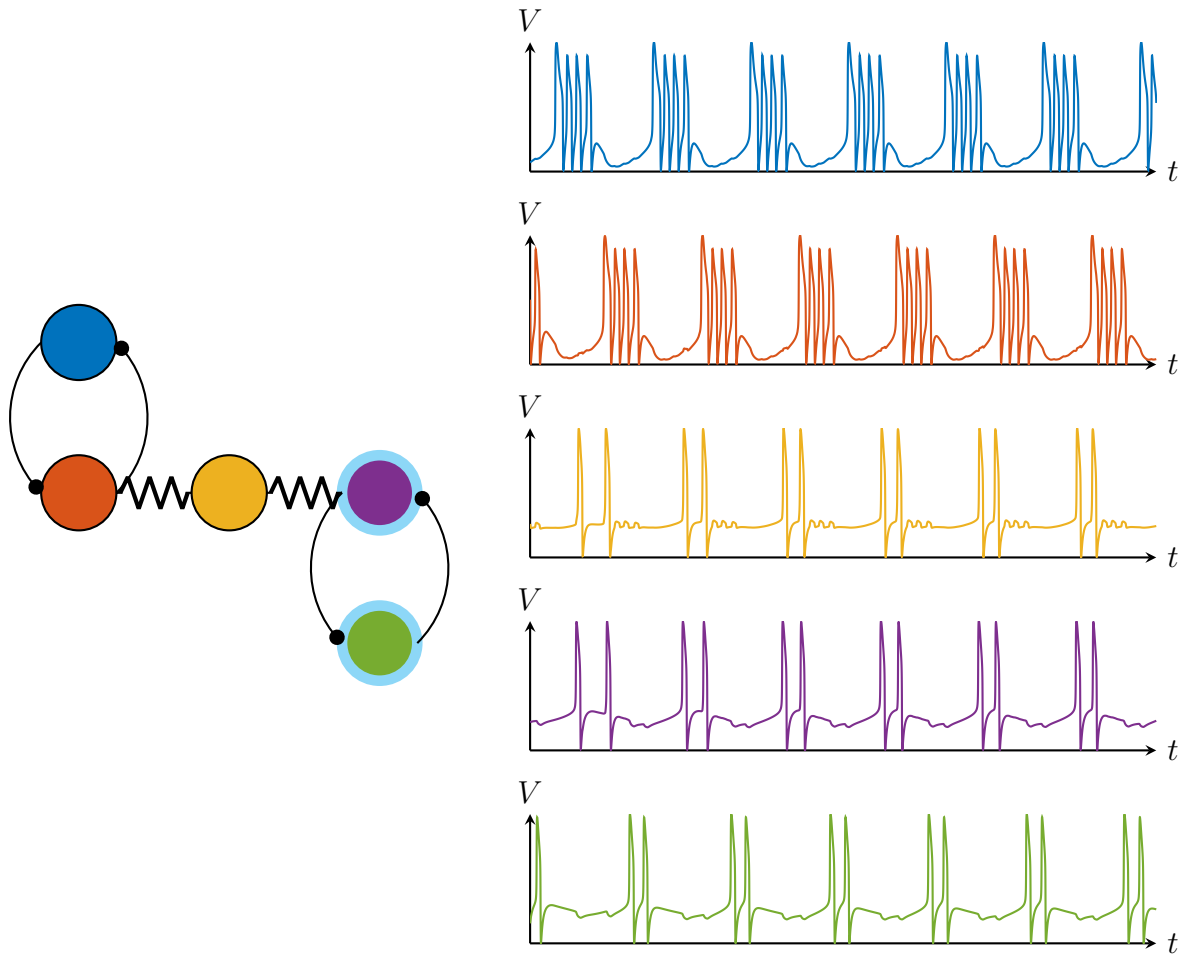
\includetikzfigure{cpg_modulated}{Disconnecting a rhythm from the network. By modulating the slow two-neuron oscillator through the decrease in the slow negative conductance gain, the slow cells are switched to the slow spiking relay mode. The fast bursting rhythm effectively propagates through the hub neuron, and the network is globally oscillating with the fast rhythm.}

By using the mechanism shown previously in \cref{fig:hco_switch_spiking}, the individual rhythms can be effectively disconnected from the network by modulating the two-neuron oscillators between the bursting and slow spiking modes. This is shown in \cref{fig:cpg_modulated} where the slow negative conductance gain is decreased for the neurons within the slow oscillator. The slow oscillator neurons then act as followers, and the fast rhythm propagating through the hub neuron now has sufficient strength to synchronize these neurons. In this case the whole network oscillates with the fast rhythm.

A more extensive investigation of the various configurations of the circuit is provided in \cite{drion2018a}. The key message is that the control of each node in its nodal versus network modes enables dynamic transitions between networks that share the same \textit{hardware} network topology but exhibit a different \textit{functional} topology controlled by which node is "in" (network mode) and which node is "out" (nodal mode). This suggests that the control of excitability at the nodal level provides a rich network control principle.

\section{Conclusion}

The article presented a framework for designing and controlling mixed-feedback systems inspired by biological neuronal networks. This neuromorphic approach aims at mimicking the feedback structure, as well as the control mechanisms that make neurons robust, adaptable and energy efficient.

The dynamical structure of excitable neurons was presented through the fundamental feedback organization consisting of interlocked positive and negative feedback loops acting in distinct timescales. By using well-known low-power analog circuits, these elementary feedback loops can be efficiently implemented in hardware while retaining the interconnection structure of biological neurons. At the same time, their unique dynamical structure makes neuromorphic neurons amendable to an input-output analysis that reframes the control problem as static loop shaping of the input-output characteristic in distinct timescales.

The article further discussed how this mixed-feedback organization of the neuronal nodes endows them with a unique dynamical switch in the network setting. Biological neural networks constantly adapt to the changing conditions and requirements through appropriate modulation of neurons and their interconnections. By controlling the excitability of the nodes, the article showed how individual circuits can turn local rhythms within the network on and off, offering a robust mechanism for controlling the global network behavior through the appropriate mixed feedback balance at the individual nodes.

The main obstacle in the development of neuromorphic circuits has proven to be their increased fragility to noise and fabrication deviations that is an inevitable consequence of their analog mode of operation. However, biological neurons effectively cope with the same limitations through their unique multiscale mixed-feedback architecture. By combining the expertise of both control and electronics experts, neuromorphic control shows great promise in helping to solve future large-scale challenges in energy efficient ways using direct inspiration from biology.

\section{Acknowledgements}

This work was supported by the European Research Council under the Advanced ERC Grant Agreement Switchlet n.670645.

\clearpage
\newpage

\bibliographystyle{ieeetr}
\bibliography{references}

\begin{thebibliography}{10}

\bibitem{mead1990}
C.~Mead, ``Neuromorphic electronic systems,'' {\em Proceedings of the IEEE},
  vol.~78, no.~10, pp.~1629--1636, 1990.

\bibitem{gallego2020}
G.~Gallego, T.~Delbruck, G.~Orchard, C.~Bartolozzi, B.~Taba, A.~Censi,
  S.~Leutenegger, A.~Davison, J.~Conradt, K.~Daniilidis, and D.~Scaramuzza,
  ``Event-based {{Vision}}: {{A Survey}},'' {\em IEEE Transactions on Pattern
  Analysis and Machine Intelligence}, pp.~1--1, 2020.

\bibitem{mead2020}
C.~Mead, ``How we created neuromorphic engineering,'' {\em Nature Electronics},
  vol.~3, pp.~434--435, July 2020.

\bibitem{indiveri2011}
G.~Indiveri, B.~{Linares-Barranco}, T.~J. Hamilton, A.~van Schaik,
  R.~{Etienne-Cummings}, T.~Delbruck, S.-C. Liu, P.~Dudek, P.~H{\"a}fliger,
  S.~Renaud, J.~Schemmel, G.~Cauwenberghs, J.~Arthur, K.~Hynna, F.~Folowosele,
  S.~Saighi, T.~{Serrano-Gotarredona}, J.~Wijekoon, Y.~Wang, and K.~Boahen,
  ``Neuromorphic {{Silicon Neuron Circuits}},'' {\em Frontiers in
  Neuroscience}, vol.~5, 2011.

\bibitem{cauwenberghs2013}
G.~Cauwenberghs, ``Reverse engineering the cognitive brain,'' {\em Proceedings
  of the National Academy of Sciences}, vol.~110, pp.~15512--15513, Sept. 2013.

\bibitem{furber2016}
S.~Furber, ``Large-scale neuromorphic computing systems,'' {\em Journal of
  Neural Engineering}, vol.~13, p.~051001, Oct. 2016.

\bibitem{boahen2017}
K.~Boahen, ``A {{Neuromorph}}'s {{Prospectus}},'' {\em Computing in Science \&
  Engineering}, vol.~19, no.~2, pp.~14--28, 2017.

\bibitem{mead1989}
C.~Mead, {\em Analog {{VLSI}} and Neural Systems}, vol.~1.
\newblock {Reading: Addison-Wesley}, 1989.

\bibitem{sarpeshkar1998}
R.~Sarpeshkar, ``Analog {{Versus Digital}}: {{Extrapolating}} from
  {{Electronics}} to {{Neurobiology}},'' {\em Neural Computation}, vol.~10,
  pp.~1601--1638, Oct. 1998.

\bibitem{boahen1992a}
K.~A. Boahen and A.~G. Andreou, ``A {{Contrast Sensitive Silicon Retina}} with
  {{Reciprocal Synapses}},'' in {\em Advances in {{Neural Information
  Processing Systems}} 4 ({{NIPS}} 1991)} (J.~E. Moody, S.~J. Hanson, and R.~P.
  Lippmann, eds.), no.~4, pp.~764--770, {San Mateo, CA}: {Morgan Kaufmann},
  1992.

\bibitem{poon2011}
C.-S. Poon and K.~Zhou, ``Neuromorphic {{Silicon Neurons}} and
  {{Large}}-{{Scale Neural Networks}}: {{Challenges}} and {{Opportunities}},''
  {\em Frontiers in Neuroscience}, vol.~5, 2011.

\bibitem{sepulchre2019}
R.~Sepulchre, G.~Drion, and A.~Franci, ``Control {{Across Scales}} by
  {{Positive}} and {{Negative Feedback}},'' {\em Annual Review of Control,
  Robotics, and Autonomous Systems}, vol.~2, pp.~89--113, May 2019.

\bibitem{krahe2004}
R.~Krahe and F.~Gabbiani, ``Burst firing in sensory systems,'' {\em Nature
  Reviews Neuroscience}, vol.~5, pp.~13--23, Jan. 2004.

\bibitem{harris-warrick2011}
R.~M. {Harris-Warrick}, ``Neuromodulation and flexibility in {{Central Pattern
  Generator}} networks,'' {\em Current Opinion in Neurobiology}, vol.~21,
  pp.~685--692, Oct. 2011.

\bibitem{marder2012}
E.~Marder, ``Neuromodulation of {{Neuronal Circuits}}: {{Back}} to the
  {{Future}},'' {\em Neuron}, vol.~76, pp.~1--11, Oct. 2012.

\bibitem{sepulchre2019a}
R.~Sepulchre, T.~O'Leary, G.~Drion, and A.~Franci, ``Control by
  neuromodulation: {{A}} tutorial,'' in {\em 2019 18th {{European Control
  Conference}} ({{ECC}})}, pp.~483--497, June 2019.

\bibitem{ribar2017}
L.~Ribar and R.~Sepulchre, ``Bursting through interconnection of excitable
  circuits,'' in {\em 2017 {{IEEE Biomedical Circuits}} and {{Systems
  Conference}} ({{BioCAS}})}, pp.~1--4, Oct. 2017.

\bibitem{ribar2019}
L.~Ribar and R.~Sepulchre, ``Neuromodulation of {{Neuromorphic Circuits}},''
  {\em IEEE Transactions on Circuits and Systems I: Regular Papers}, pp.~1--13,
  2019.

\bibitem{ribar2019a}
L.~Ribar, {\em Synthesis of Neuromorphic Circuits with Neuromodulatory
  Properties}.
\newblock Thesis, University of Cambridge, 2019.

\bibitem{sepulchre2018}
R.~Sepulchre, G.~Drion, and A.~Franci, ``Excitable {{Behaviors}},'' in {\em
  Emerging {{Applications}} of {{Control}} and {{Systems Theory}}}, Lecture
  {{Notes}} in {{Control}} and {{Information Sciences}} - {{Proceedings}},
  pp.~269--280, {Springer, Cham}, 2018.

\bibitem{drion2012}
G.~Drion, A.~Franci, V.~Seutin, and R.~Sepulchre, ``A novel phase portrait for
  neuronal excitability,'' {\em PloS one}, vol.~7, no.~8, p.~e41806, 2012.

\bibitem{franci2012a}
A.~Franci, G.~Drion, and R.~Sepulchre, ``An {{Organizing Center}} in a {{Planar
  Model}} of {{Neuronal Excitability}},'' {\em SIAM Journal on Applied
  Dynamical Systems}, vol.~11, pp.~1698--1722, Jan. 2012.

\bibitem{franci2013}
A.~Franci, G.~Drion, V.~Seutin, and R.~Sepulchre, ``A {{Balance Equation
  Determines}} a {{Switch}} in {{Neuronal Excitability}},'' {\em PLoS
  Computational Biology}, vol.~9, p.~e1003040, May 2013.

\bibitem{franci2014a}
A.~Franci, G.~Drion, and R.~Sepulchre, ``Modeling the {{Modulation}} of
  {{Neuronal Bursting}}: {{A Singularity Theory Approach}},'' {\em SIAM Journal
  on Applied Dynamical Systems}, vol.~13, pp.~798--829, Jan. 2014.

\bibitem{franci2014}
A.~Franci and R.~Sepulchre, ``Realization of nonlinear behaviors from
  organizing centers,'' in {\em 53rd {{IEEE Conference}} on {{Decision}} and
  {{Control}}}, pp.~56--61, Dec. 2014.

\bibitem{drion2015}
G.~Drion, A.~Franci, J.~Dethier, and R.~Sepulchre, ``Dynamic {{Input
  Conductances Shape Neuronal Spiking}},'' {\em eNeuro}, vol.~2, June 2015.

\bibitem{franci2017}
A.~Franci, G.~Drion, and R.~Sepulchre, ``Robust and tunable bursting requires
  slow positive feedback,'' {\em Journal of Neurophysiology}, vol.~119,
  pp.~1222--1234, Dec. 2017.

\bibitem{nassim2018}
C.~Nassim, {\em Lessons from the {{Lobster}}: {{Eve Marder}}'s {{Work}} in
  {{Neuroscience}}}.
\newblock {MIT Press}, June 2018.

\bibitem{ermentrout2010}
G.~B. Ermentrout and D.~H. Terman, {\em Mathematical {{Foundations}} of
  {{Neuroscience}}}.
\newblock {Springer Science \& Business Media}, July 2010.

\bibitem{fitzhugh1961}
R.~FitzHugh, ``Impulses and physiological states in theoretical models of nerve
  membrane,'' {\em Biophysical journal}, vol.~1, no.~6, p.~445, 1961.

\bibitem{nagumo1962}
J.~Nagumo, S.~Arimoto, and S.~Yoshizawa, ``An active pulse transmission line
  simulating nerve axon,'' {\em Proceedings of the IRE}, vol.~50, no.~10,
  pp.~2061--2070, 1962.

\bibitem{mahowald1991}
M.~Mahowald and R.~Douglas, ``A silicon neuron,'' {\em Nature}, vol.~354,
  no.~6354, pp.~515--518, 1991.

\bibitem{drakakis2000}
E.~M. Drakakis and A.~J. Payne, ``A {{Bernoulli Cell}}-{{Based Investigation}}
  of the {{Non}}-{{Linear Dynamics}} in {{Log}}-{{Domain Structures}},'' in
  {\em Research {{Perspectives}} on {{Dynamic Translinear}} and
  {{Log}}-{{Domain Circuits}}} (W.~A. Serdijn and J.~Mulder, eds.), pp.~21--40,
  {Boston, MA}: {Springer US}, 2000.

\bibitem{linares-barranco1991}
B.~{Linares-Barranco}, E.~{S{\'a}nchez-Sinencio},
  A.~{Rodr{\'i}guez-V{\'a}zquez}, and J.~L. Huertas, ``A {{CMOS}}
  implementation of {{FitzHugh}}-{{Nagumo}} neuron model,'' {\em IEEE Journal
  of Solid-State Circuits}, vol.~26, no.~7, pp.~956--965, 1991.

\bibitem{drion2015b}
G.~Drion, T.~O'Leary, and E.~Marder, ``Ion channel degeneracy enables robust
  and tunable neuronal firing rates,'' {\em Proceedings of the National Academy
  of Sciences}, vol.~112, pp.~E5361--E5370, Sept. 2015.

\bibitem{izhikevich2007}
E.~M. Izhikevich, {\em Dynamical Systems in Neuroscience}.
\newblock {MIT press}, 2007.

\bibitem{moore1959}
J.~W. Moore, ``Excitation of the {{Squid Axon Membrane}} in {{Isosmotic
  Potassium Chloride}},'' {\em Nature}, vol.~183, pp.~265--266, Jan. 1959.

\bibitem{mccormick1992}
D.~A. McCormick, ``Neurotransmitter actions in the thalamus and cerebral cortex
  and their role in neuromodulation of thalamocortical activity,'' {\em
  Progress in Neurobiology}, vol.~39, pp.~337--388, Oct. 1992.

\bibitem{beurrier1999}
C.~Beurrier, P.~Congar, B.~Bioulac, and C.~Hammond, ``Subthalamic nucleus
  neurons switch from single-spike activity to burst-firing mode,'' {\em The
  Journal of neuroscience}, vol.~19, no.~2, pp.~599--609, 1999.

\bibitem{sherman2001}
S.~M. Sherman, ``Tonic and burst firing: Dual modes of thalamocortical relay,''
  {\em Trends in neurosciences}, vol.~24, no.~2, pp.~122--126, 2001.

\bibitem{kepecs2002}
A.~Kepecs, X.-J. Wang, and J.~Lisman, ``Bursting {{Neurons Signal Input
  Slope}},'' {\em Journal of Neuroscience}, vol.~22, pp.~9053--9062, Oct. 2002.

\bibitem{marsat2006}
G.~Marsat and G.~S. Pollack, ``A {{Behavioral Role}} for {{Feature Detection}}
  by {{Sensory Bursts}},'' {\em Journal of Neuroscience}, vol.~26,
  pp.~10542--10547, Oct. 2006.

\bibitem{plant1981}
R.~E. Plant, ``Bifurcation and resonance in a model for bursting nerve cells,''
  {\em Journal of Mathematical Biology}, vol.~11, pp.~15--32, Jan. 1981.

\bibitem{rinzel1987a}
J.~Rinzel and Y.~S. Lee, ``Dissection of a model for neuronal parabolic
  bursting,'' {\em Journal of mathematical biology}, vol.~25, no.~6,
  pp.~653--675, 1987.

\bibitem{hindmarsh1984}
J.~L. Hindmarsh and R.~M. Rose, ``A model of neuronal bursting using three
  coupled first order differential equations,'' {\em Proceedings of the Royal
  Society of London B: Biological Sciences}, vol.~221, no.~1222, pp.~87--102,
  1984.

\bibitem{simoni2004}
M.~Simoni, G.~Cymbalyuk, M.~Sorensen, R.~Calabrese, and S.~DeWeerth, ``A
  {{Multiconductance Silicon Neuron With Biologically Matched Dynamics}},''
  {\em IEEE Transactions on Biomedical Engineering}, vol.~51, pp.~342--354,
  Feb. 2004.

\bibitem{arthur2011}
J.~V. Arthur and K.~A. Boahen, ``Silicon-{{Neuron Design}}: {{A Dynamical
  Systems Approach}},'' {\em IEEE Transactions on Circuits and Systems I:
  Regular Papers}, vol.~58, pp.~1034--1043, May 2011.

\bibitem{yu2011}
T.~Yu, T.~J. Sejnowski, and G.~Cauwenberghs, ``Biophysical {{Neural Spiking}},
  {{Bursting}}, and {{Excitability Dynamics}} in {{Reconfigurable Analog
  VLSI}},'' {\em IEEE Transactions on Biomedical Circuits and Systems}, vol.~5,
  pp.~420--429, Oct. 2011.

\bibitem{wijekoon2008}
J.~H. Wijekoon and P.~Dudek, ``Compact silicon neuron circuit with spiking and
  bursting behaviour,'' {\em Neural Networks}, vol.~21, pp.~524--534, Mar.
  2008.

\bibitem{nadim2014}
F.~Nadim and D.~Bucher, ``Neuromodulation of neurons and synapses,'' {\em
  Current Opinion in Neurobiology}, vol.~29, pp.~48--56, Dec. 2014.

\bibitem{ribar2020}
L.~Ribar, ``Python environment for neuromorphic neuromodulation.'' Available
  online at https://github.com/lukaribar/Circuit-Neuromodulation.
\newblock (last accessed June, 2021).

\bibitem{izhikevich2020}
E.~M. Izhikevich, ``Bursting.'' Available online at
  http://www.scholarpedia.org/article/Bursting.
\newblock (last accessed June, 2021).

\bibitem{bucher2015}
D.~Bucher, G.~Haspel, J.~Golowasch, and F.~Nadim, ``Central {{Pattern
  Generators}},'' in {\em {{eLS}}}, pp.~1--12, {American Cancer Society}, 2015.

\bibitem{marder2007}
E.~Marder and D.~Bucher, ``Understanding {{Circuit Dynamics Using}} the
  {{Stomatogastric Nervous System}} of {{Lobsters}} and {{Crabs}},'' {\em
  Annual Review of Physiology}, vol.~69, no.~1, pp.~291--316, 2007.

\bibitem{gutierrez2013}
G.~J. Gutierrez, T.~O'Leary, and E.~Marder, ``Multiple {{Mechanisms Switch}} an
  {{Electrically Coupled}}, {{Synaptically Inhibited Neuron}} between
  {{Competing Rhythmic Oscillators}},'' {\em Neuron}, vol.~77, pp.~845--858,
  Mar. 2013.

\bibitem{gutierrez2014}
G.~J. Gutierrez and E.~Marder, ``Modulation of a {{Single Neuron Has
  State}}-{{Dependent Actions}} on {{Circuit Dynamics}},'' {\em eNeuro},
  vol.~1, Nov. 2014.

\bibitem{drion2018a}
G.~Drion, A.~Franci, and R.~Sepulchre, ``Cellular switches orchestrate rhythmic
  circuits,'' {\em Biological Cybernetics}, Sept. 2018.

\end{thebibliography}


\begin{thebibliography}{99}
	\bibitem[S1]{sb_sejnowski2020}
	T.~J. Sejnowski, ``The unreasonable effectiveness of deep learning in
	artificial intelligence,'' {\em Proceedings of the National Academy of
		Sciences}, p.~201907373, Jan. 2020.
\end{thebibliography}

\begin{thebibliography}{99}
	\bibitem[S2]{sb_maxwell1868}
	J.~C. Maxwell, ``I. {{On}} governors,'' {\em Proceedings of the Royal Society
		of London}, vol.~16, pp.~270--283, Jan. 1868.
	\bibitem[S3]{sb_tucker1972}
	D.~Tucker, ``The history of positive feedback: {{The}} oscillating audion, the
	regenerative receiver, and other applications up to around 1923,'' {\em Radio
		and Electronic Engineer}, vol.~42, pp.~69--80, Feb. 1972.
\end{thebibliography}

\begin{thebibliography}{99}
	\bibitem[S4]{sb_maher1989}
	M.~Maher, S.~Deweerth, M.~Mahowald, and C.~Mead, ``Implementing neural
	architectures using analog {{VLSI}} circuits,'' {\em IEEE Transactions on
		Circuits and Systems}, vol.~36, pp.~643--652, May 1989.
	
	\bibitem[S5]{sb_lyon1988}
	R.~Lyon and C.~Mead, ``An analog electronic cochlea,'' {\em IEEE Transactions
		on Acoustics, Speech, and Signal Processing}, vol.~36, pp.~1119--1134, July
	1988.
	
	\bibitem[S6]{sb_watts1992}
	L.~Watts, D.~A. Kerns, R.~F. Lyon, and C.~A. Mead, ``Improved implementation of
	the silicon cochlea,'' {\em IEEE Journal of Solid-State Circuits}, vol.~27,
	pp.~692--700, May 1992.
	
	\bibitem[S7]{sb_mead1988}
	C.~A. Mead and M.~A. Mahowald, ``A silicon model of early visual processing,''
	{\em Neural Networks}, vol.~1, no.~1, pp.~91--97, 1988.
	
	\bibitem[S8]{sb_serrano-gotarredona1999}
	T.~{Serrano-Gotarredona}, B.~{Linares-Barranco}, and A.~G. Andreou, ``A general
	translinear principle for subthreshold {{MOS}} transistors,'' {\em IEEE
		transactions on circuits and systems I: fundamental theory and applications},
	vol.~46, no.~5, pp.~607--616, 1999.
	
	\bibitem[S9]{sb_andreou1996}
	A.~G. Andreou and K.~A. Boahen, ``Translinear circuits in subthreshold
	{{MOS}},'' {\em Analog Integrated Circuits and Signal Processing}, vol.~9,
	pp.~141--166, Mar. 1996.
	
	\bibitem[S10]{sb_douglas1995a}
	R.~Douglas, M.~Mahowald, and C.~Mead, ``Neuromorphic {{Analogue VLSI}},'' {\em
		Annual Review of Neuroscience}, vol.~18, no.~1, pp.~255--281, 1995.
	
	\bibitem[S11]{sb_boahen2000}
	K.~A. Boahen, ``Point-to-point connectivity between neuromorphic chips using
	address events,'' {\em IEEE Transactions on Circuits and Systems II: Analog
		and Digital Signal Processing}, vol.~47, no.~5, pp.~416--434, 2000.
	
	\bibitem[S12]{sb_qiao2015}
	N.~Qiao, H.~Mostafa, F.~Corradi, M.~Osswald, F.~Stefanini, D.~Sumislawska, and
	G.~Indiveri, ``A reconfigurable on-line learning spiking neuromorphic
	processor comprising 256 neurons and {{128K}} synapses,'' {\em Frontiers in
		Neuroscience}, vol.~9, 2015.
	
	\bibitem[S13]{sb_yu2012}
	T.~Yu, J.~Park, S.~Joshi, C.~Maier, and G.~Cauwenberghs, ``65k-neuron
	integrate-and-fire array transceiver with address-event reconfigurable
	synaptic routing,'' in {\em 2012 {{IEEE Biomedical Circuits}} and {{Systems
				Conference}} ({{BioCAS}})}, pp.~21--24, Nov. 2012.
			
	\bibitem[S14]{sb_merolla2014}
	P.~A. Merolla, J.~V. Arthur, R.~{Alvarez-Icaza}, A.~S. Cassidy, J.~Sawada,
	F.~Akopyan, B.~L. Jackson, N.~Imam, C.~Guo, Y.~Nakamura, B.~Brezzo, I.~Vo,
	S.~K. Esser, R.~Appuswamy, B.~Taba, A.~Amir, M.~D. Flickner, W.~P. Risk,
	R.~Manohar, and D.~S. Modha, ``A million spiking-neuron integrated circuit
	with a scalable communication network and interface,'' {\em Science},
	vol.~345, pp.~668--673, Aug. 2014.
	
	\bibitem[S15]{sb_benjamin2014}
	B.~V. Benjamin, P.~Gao, E.~McQuinn, S.~Choudhary, A.~R. Chandrasekaran, J.-M.
	Bussat, R.~{Alvarez-Icaza}, J.~V. Arthur, P.~A. Merolla, and K.~Boahen,
	``Neurogrid: {{A}} mixed-analog-digital multichip system for large-scale
	neural simulations,'' {\em Proceedings of the IEEE}, vol.~102, no.~5,
	pp.~699--716, 2014.
	
	\bibitem[S16]{sb_schemmel2010}
	J.~Schemmel, D.~Briiderle, A.~Griibl, M.~Hock, K.~Meier, and S.~Millner, ``A
	wafer-scale neuromorphic hardware system for large-scale neural modeling,''
	in {\em Proceedings of 2010 {{IEEE International Symposium}} on {{Circuits}}
		and {{Systems}}}, ({Paris, France}), pp.~1947--1950, {IEEE}, May 2010.
	
	\bibitem[S17]{sb_boahen2005}
	K.~Boahen, ``Neuromorphic {{Microchips}},'' {\em Scientific American},
	vol.~292, no.~5, pp.~56--63, 2005.
	
	\bibitem[S18]{sb_liu2010}
	S.-C. Liu and T.~Delbruck, ``Neuromorphic sensory systems,'' {\em Current
		Opinion in Neurobiology}, vol.~20, pp.~288--295, June 2010.
	
	\bibitem[S19]{sb_cassidy2013}
	A.~S. Cassidy, J.~Georgiou, and A.~G. Andreou, ``Design of silicon brains in
	the nano-{{CMOS}} era: {{Spiking}} neurons, learning synapses and neural
	architecture optimization,'' {\em Neural Networks}, vol.~45, pp.~4--26, 2013.
	
	\bibitem[S20]{sb_sanni2019}
	K.~A. Sanni and A.~G. Andreou, ``A {{Historical Perspective}} on {{Hardware AI
			Inference}}, {{Charge}}-{{Based Computational Circuits}} and an 8 bit
	{{Charge}}-{{Based Multiply}}-{{Add Core}} in 16 nm {{FinFET CMOS}},'' {\em
		IEEE Journal on Emerging and Selected Topics in Circuits and Systems},
	vol.~9, no.~3, pp.~532--543, 2019.
	
	\bibitem[S21]{sb_jo2010}
	S.~H. Jo, T.~Chang, I.~Ebong, B.~B. Bhadviya, P.~Mazumder, and W.~Lu,
	``Nanoscale {{Memristor Device}} as {{Synapse}} in {{Neuromorphic
			Systems}},'' {\em Nano Letters}, vol.~10, pp.~1297--1301, Apr. 2010.
	
	\bibitem[S22]{sb_indiveri2013}
	G.~Indiveri, B.~{Linares-Barranco}, R.~Legenstein, G.~Deligeorgis, and
	T.~Prodromakis, ``Integration of nanoscale memristor synapses in neuromorphic
	computing architectures,'' {\em Nanotechnology}, vol.~24, p.~384010, Sept.
	2013.
	
	\bibitem[S23]{sb_carver_mead_image}
	``National Science and Technology Medals Foundation''. Available online at https://nationalmedals.org/laureate/carver-a-mead/ (last accessed June, 2021).
	
\end{thebibliography}

\begin{thebibliography}{99}
	\bibitem[S24]{sb_deweerth1991}
	S.~P. DeWeerth, L.~Nielsen, C.~A. Mead, and K.~J. Astrom, ``A simple neuron
	servo,'' {\em IEEE Transactions on Neural Networks}, vol.~2, no.~2,
	pp.~248--251, 1991.
\end{thebibliography}

\begin{thebibliography}{99}
	\bibitem[S25]{sb_hodgkin1952}
	A.~L. Hodgkin and A.~F. Huxley, ``A quantitative description of membrane
	current and its application to conduction and excitation in nerve,'' {\em The
		Journal of physiology}, vol.~117, no.~4, p.~500, 1952.
	
	\bibitem[S26]{sb_hodgkin_huxley_image}
	R. M. Simmons, ``Sir Andrew Fielding Huxley OM. 22 November 1917—30 May 2012,'' {\em Biographical Memoirs of Fellows of the Royal Society}, vol.~65, pp.~179–215, 2018. Available online at https://royalsocietypublishing.org/doi/10.1098/rsbm.2018.0012 (last accessed June 2021).
\end{thebibliography}

\begin{thebibliography}{99}
	\bibitem[S27]{sb_koch2004}
	C.~Koch, {\em Biophysics of Computation: Information Processing in Single
		Neurons}.
	\newblock {Oxford university press}, 2004.
\end{thebibliography}

\begin{thebibliography}{99}
	\bibitem[S28]{sb_morris1981}
	C.~Morris and H.~Lecar, ``Voltage oscillations in the barnacle giant muscle
	fiber.,'' {\em Biophysical journal}, vol.~35, no.~1, p.~193, 1981.
	
	\bibitem[S29]{sb_izhikevich2000}
	E.~M. Izhikevich, ``Neural excitability, spiking and bursting,'' {\em
		International Journal of Bifurcation and Chaos}, vol.~10, no.~06,
	pp.~1171--1266, 2000.
\end{thebibliography}

\begin{thebibliography}{99}
	\bibitem[S30]{sb_rinzel1985a}
	J.~Rinzel, ``Excitation dynamics: Insights from simplified membrane models,''
	in {\em Fed. {{Proc}}}, vol.~44, pp.~2944--2946, 1985.
	
	\bibitem[S31]{sb_kepler1992}
	T.~B. Kepler, L.~F. Abbott, and E.~Marder, ``Reduction of conductance-based
	neuron models,'' {\em Biological Cybernetics}, vol.~66, no.~5, pp.~381--387,
	1992.
	
	\bibitem[S32]{sb_rinzel1987}
	J.~Rinzel, ``A formal classification of bursting mechanisms in excitable
	systems,'' in {\em Mathematical Topics in Population Biology, Morphogenesis
		and Neurosciences}, pp.~267--281, {Springer}, 1987.
	
	\bibitem[S33]{sb_rinzel1989}
	J.~Rinzel and G.~B. Ermentrout, ``Analysis of neural excitability and
	oscillations,'' in {\em Methods in Neuronal Modeling}, pp.~135--169, {MIT
		press}, 1989.
	
	\bibitem[S34]{sb_cirillo2020}
	G.~I. Cirillo and R.~Sepulchre, ``The geometry of rest\textendash spike
	bistability,'' {\em The Journal of Mathematical Neuroscience}, vol.~10,
	p.~13, Dec. 2020.
\end{thebibliography}

\begin{thebibliography}{99}
	\bibitem[S35]{sb_marder1996}
	E.~Marder and R.~L. Calabrese, ``Principles of rhythmic motor pattern
	generation,'' {\em Physiological Reviews}, vol.~76, pp.~687--717, July 1996.
	
	\bibitem[S36]{sb_brown1911}
	T.~G. Brown and C.~S. Sherrington, ``The intrinsic factors in the act of
	progression in the mammal,'' {\em Proceedings of the Royal Society of London.
		Series B, Containing Papers of a Biological Character}, vol.~84,
	pp.~308--319, Dec. 1911.
\end{thebibliography}


\sidebars 

\sbsection{Summary}{summary}

Neuromorphic engineering is a rapidly developing field that aims to take inspiration from the biological organization of neural systems to develop novel technology for computing, sensing, and actuating. The unique properties of such systems call for new signal processing and control paradigms. 

The article introduces the mixed feedback organization of excitable neuronal systems, consisting of interlocked positive and negative feedback loops acting in distinct timescales. The principles of biological neuromodulation suggest a methodology for designing and controlling mixed-feedback systems \textit{neuromorphically}. The proposed design consists of a parallel interconnection of elementary circuit elements that mirrors the organization of biological neurons and utilizes the hardware components of neuromorphic electronic circuits. The interconnection structure endows the neuromorphic systems with a simple control methodology which we showcase on elementary network examples that suggest the scalability of the mixed-feedback principles.

One of the main obstacles in the development of analog neuromorphic hardware has proven to be their increased fragility to noise and fabrication mismatch. Biological systems however cope with similar limitations with amazing effectiveness and efficiency through their unique multi-scale mixed-feedback organizational principles. By combining the expertise of control and electronics experts, neuromorphic control approaches can provide efficient solutions to emerging large-scale challenges.

\sbend

\sbsection{Multiscale in biology and engineering}{multiscale}

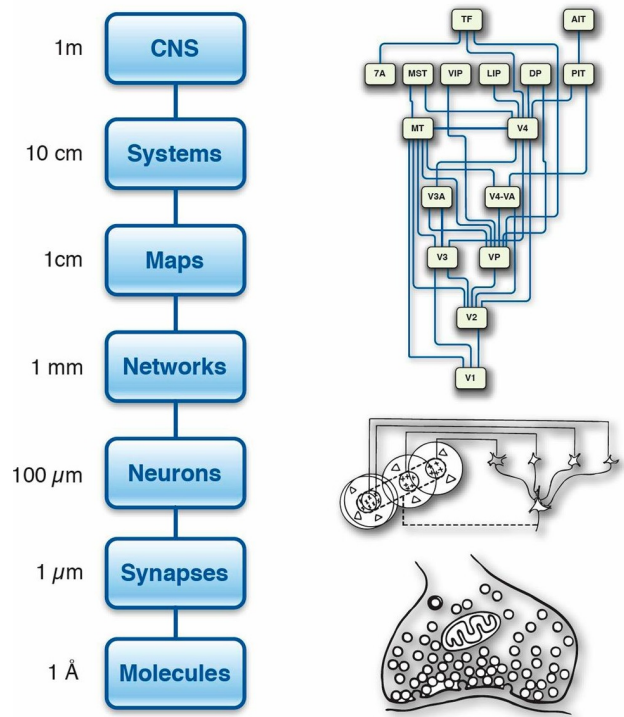
\includefigure{sejnowski_multiscale}{0.5}{The multiscale organization of the brain. The picture shows the hierarchy of neural organization, starting from the smallest molecular level all the way to the whole central nervous system (CNS) that generates the behavior. (Adapted from \cite{sb_sejnowski2020} with permission.)}

Designing and controlling systems across scales is a rich and ongoing area of interest within the control community. Control of vehicle platoons for automated highway systems, microeletromechanical systems (MEMS), segmented large telescope mirrors, as well as bionspired applications such as soft robotics are all examples of distributed systems where the desired system behavior is achieved through control and sensing at the local level. Due to the abundance of cheap actuators and sensors, there is a need for a theoretical and design methodology for building robust complex systems with imprecise individual components. The ability of constructing robust and adaptable systems using noisy and uncertain building blocks is precisely the defining characteristic of biological systems. Understanding the organizational principles that govern biological behaviors could thus inform the design of systems with such remarkable capabilities.

Control across scales is at the heart of the organization of neural networks (see \cref{fig:sejnowski_multiscale}). At the nanoscale, special proteins called ion channels open and close in a stochastic manner and thus generate inward and outward currents that control the potential difference across a cell's membrane. At the cellular scale, the flow of ionic currents is controlled through mass opening and closing of these channels, enabling the control of cell excitability. At the network scale, the control of individual excitability properties of the cells, together with the constant modulation of the connectivity strength between the cells, defines the pattern of the network behavior. These networks can function as robust and adaptable clocks that generate rhythms for repeatable actions such as breathing and walking, while others may filter, process and relay incoming motor and sensory signals. Finally, the interaction of many such networks enables the higher cognitive functions such as awareness, task selection and learning. The brain thus encompasses a vast range of spatial and temporal scales, from microseconds and nanometers of ion transport to hours and meter scales of complex brain functions.

One of the main challenges associated with distributed sensing and actuating is the inherent local action and measurement involved: the errors at the small scales are aggregated at the large scale, making the global control through the cumulative action of individual agents a difficult task. This does not seem to be a challenge for biological systems which retain robust global function while consisting of a large interconnection of noisy, imprecise, but remarkably energy efficient components. The core of this discrepancy appears to be the difference in the feedback structure of the systems: engineered systems mostly involve a clear separation between positive and negative feedback pathways, while neural systems involve interlocked mixed feedback at all scales. This enables biological networks to operate in a continuum between the purely analog and the purely digital world, a key concept behind neuromorphic design and control.

\sbend

\sbsection{Positive and negative feedback}{positive_negative_feedback}

Both positive and negative feedback have a long history of application in engineering sciences, even though studying positive feedback is much less prominent nowadays in control system design. From Maxwell's pioneering work on stabilization of governors \cite{sb_maxwell1868}, negative feedback has been recognized as the basis of robust control system design for regulation or homeostasis. This realization came somewhat later to the world of electronics, where initially, positive feedback was utilized in order to increase the sensitivity of amplifiers, in turn bringing the unwanted destabilizing effects \cite{sb_tucker1972}. Negative feedback became key to designing robust and stable amplifying circuits using unreliable components, and positive feedback was mainly utilized as a design principle for designing memory storage units in digital electronics or autonomous oscillators. This development has lead to a clear distinction between the applications of the two feedback loops. Positive feedback is utilized to generate bistable switches and clocks, while negative feedback enables robust regulation. Biological systems do not make a rigid separation between the negative and positive feedback pathways, but utilize both in a combined manner. Every neuronal system contains both memory and processing capabilities mixed within the same structure, and this unique organization gives the robust, adaptable and efficient characteristics of neural systems. Such mixed feedback principles could provide a promising new direction in engineering and control, enabling versatile control principles for multiscale designs \cite{sepulchre2019}.

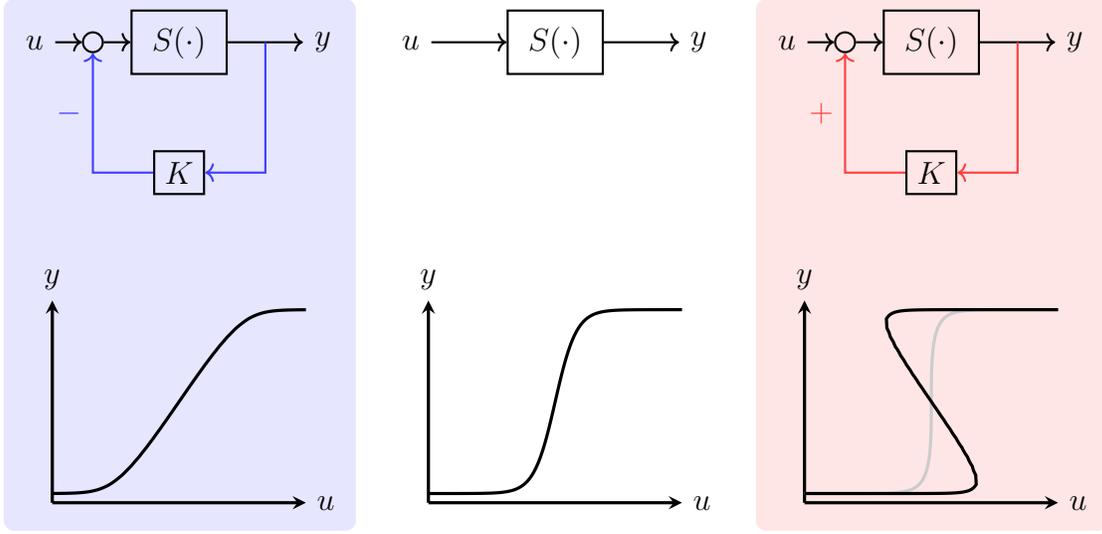
\includetikzfigure{feedback}{Feedback effect on static input-output functions. A system with a sigmoidal input-output mapping $S(u) = \tanh(u)$ is connected in a negative feedback (left) and a positive feedback (right) configuration, with the strength of the feedback connections controlled by a proportional gain $K$. In the negative feedback configuration, feedback linearizes the nonlinear mapping, stretching the linear region of the amplifier, while at the same time decreasing the sensitivity. On the other hand, positive feedback increases the sensitivity, and at the singular point, the slope becomes infinite (grey line). Increasing the magnitude of the feedback beyond this point brings the system into the bistable regime, where the hysteretic region defines the bivalued range where the system can be either in the low or the high state.}

In order to illuminate the complementarity of positive and negative feedback, \cref{fig:feedback} presents a simple toy example that nonetheless captures the fundamentals of the two feedback configurations on an arbitrary input-output system. The simplified model of the \textit{plant} is a sigmoidal mapping which provides localized amplification and saturates for small and large inputs. It models the natural saturation that occurs in any physical or biological process. When negative feedback is applied to the static plant, it linearizes the input-output characteristic as observed by the extended linear region of the amplifier (left). The sensitivity of the feedback system to an input change is thus decreased. On the other hand, positive feedback configuration has the opposite effect on the input-output characteristic by increasing the sensitivity, thus reducing the linear range of the amplifier. If the positive feedback is sufficiently large, the amplifying region shrinks to zero, while the sensitivity becomes arbitrarily large around $0$. This point of \textit{ultra-sensitivity} divides the input-output behavior between two distinct regimes: a continuous amplifying regime and a bistable discrete regime. Interestingly, excitable systems essentially operate on a continuum between these two extremes by having coexisting positive and negative feedback loops with differing dynamical properties.

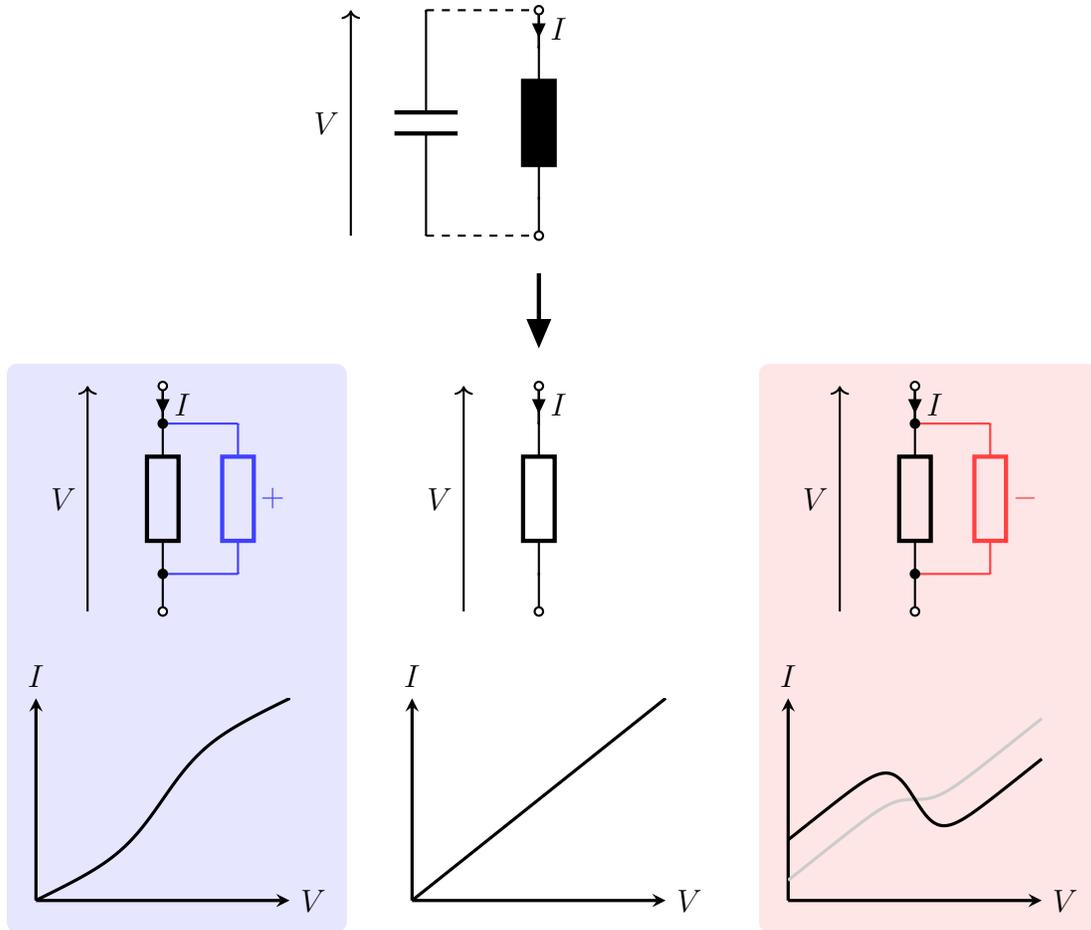
\includetikzfigure{feedback_circuit}{Feedback as circuit interconnection. Top represents the basic feedback configuration where the voltage across the capacitor is controlled by means of a single circuit element. The middle area represents the input-output characteristic of the initial passive element. Applying the input voltage $V$ across its terminals and measuring the passing current $I$, the obtained characteristic is a linear I-V curve for a simple resistor. Interconnecting a resistor with localized positive conductance (blue) and repeating the procedure produces a nonlinear, but again, purely monotonically increasing I-V curve characterizing again a passive interconnection. On the other hand interconnecting a resistor with localized negative conductance (red) locally decreases the slope of the measured I-V curve. At the singular point, there is a point of zero conductance (light gray). Increasing the negative resistance of the element further generates a nonlinear N-shaped I-V curve which is a signature of bistability.}

In physical devices, positive and negative feedback results from suitable component interconnections. An elementary circuit implementation is the parallel interconnection of a capacitor with a positive or negative resistance element (\cref{fig:feedback_circuit}). Here, the effect of a single feedback element is considered, so that the voltage across the capacitor is given by
\begin{equation}
C \dot{V} = - F(V),
\end{equation}
where $F(V)$ is the current-voltage (I-V) characteristics of the nonlinear resistor. The conductance of the resistor is defined as the derivative of its I-V characteristic around some voltage $F'(V)$. The linearized circuit can be regarded as the negative feedback interconnection of two input-output systems: a capacitor with a transfer function $1 / Cs$ and a nonlinear resistor with local gain $F'(V)$. This relationship directly relates the sign of the conductance to the sign of the feedback: \textit{positive} conductance elements provide \textit{negative} feedback, and \textit{negative} conductance elements provide \textit{positive} feedback.

Whether in biophysical or engineered circuits, the static characteristic $F(V)$ is only approximate as it neglects the dynamics of the device. Timescale separation nevertheless allows for a quasi-static analysis of feedback loops in distinct timescales. In each temporal scale, the static characteristic then provides an estimation of the local feedback gain via the linearized conductance $F'(V)$. Thus, this simplified circuit interpretation is a useful aid for both understanding the effect of different ionic currents in the generation of the neural behavior, as well as for developing a synthesis methodology for the design of excitable circuits.

\sbend

\sbsection{Three decades of neuromorphic engineering}{neuromorphic_overview}

The foundations for the neuromorphic approach to designing electrical circuits were laid out in the pioneering work of Carver Mead (\cref{fig:carver_mead}) and colleagues in the late 1980s \cite{sb_maher1989,mead1989,mead1990}. By making an analogy between the behavior of MOS transistors at low operating voltages and the channel dynamics of neurons, they provided a novel methodology for synthesizing bioinspired systems by using the physics of the devices as a computational resource. This analog way of computing is in contrast with the established digital technology where transistors are purely viewed as on/off switches, and computation is achieved by abstracting their behavior and utilizing the principles of Boolean algebra.

\includefigure{carver_mead}{0.5}{Carver Mead. (Source \cite{sb_carver_mead_image}, unknown photographer.)}

One of the main advantages of computing in this low-voltage analog regime is the energy efficiency. By making the currents orders of magnitude smaller than in conventional digital electronics, neuromorphic circuits are able to naturally compute continuous operations such as exponentiation, multiplications and summations using little power. This discovery has highlighted the potential of low-power analog circuits in emulating the efficiency and efficacy of biological systems in silicon chips.

Initial studies into neuromorphic architectures led to the first developments of electrical circuits emulating the structure and operation of neurons, sensory organs and the fundamental organizational principles of neural networks. These included the first developments on implementing the conductance-based structure of neurons, which led to the first silicon Hodgkin-Huxley based neuron \cite{mahowald1991}, as well as replicating the auditory and vision sensory systems through the silicon cochlea \cite{sb_lyon1988,sb_watts1992} and the silicon retina \cite{sb_mead1988}. All of these devices aimed to mimic the analog computation that is achieved by biological sensors, thus drastically reducing the redundancy in the sensory information collected. The developments in neuromorphic analog circuits cleverly exploited the nonlinear nature of transistor components and have since been successfully used in various implementations \cite{sb_serrano-gotarredona1999,sb_andreou1996,drakakis2000,arthur2011}. Apart from computational, novel communication methods were developed to circumvent the inability of implementing the massively interconnected neural structures found in biology. This was achieved by cleverly combining the analog processing of individual neurons with digital communication through a central communication hub in the address-event representation \cite{sb_douglas1995a,sb_boahen2000}.

The initial research into neuromorphic computing has since influenced many other approaches for developing neuroinspired hardware. Several neuromorphic chips have been developed \cite{furber2016}, such as \cite{sb_qiao2015,sb_yu2012}, as well as the larger-scale IBM True North \cite{sb_merolla2014}, Neurogrid \cite{sb_benjamin2014} and BrainScaleS \cite{sb_schemmel2010}. All of these use different levels of abstraction, as well as a different digital/analog mix for implementing the neural computations. A significant advance has been made in the area of neuromorphic vision, allowing for commercial products that use the retinal principles \cite{sb_boahen2005,sb_liu2010}. The rapid development of artificial neural networks has been followed by research into energy efficient neuromorphic implementations that would achieve economic learning and computation \cite{sb_cassidy2013,sb_sanni2019}. At the same time, novel devices such as memristors have been investigated as efficient implementations of synaptic neural interconnections that would self-adapt and learn \cite{sb_jo2010,sb_indiveri2013}. Future advancements in this direction may lead to intelligent systems that would have learning and adaptation capabilities present on the physical level, providing engineering solutions with possibly unparalleled efficiency.

\sbend

\sbsection{The first neuromorphic controller}{astrom_neuromorphic}

An early example illustrating the potential of neuromorphic architectures in tackling multiscale control can be traced to \cite{sb_deweerth1991}, where the authors presented the first neuromorphic design of a servo controller.

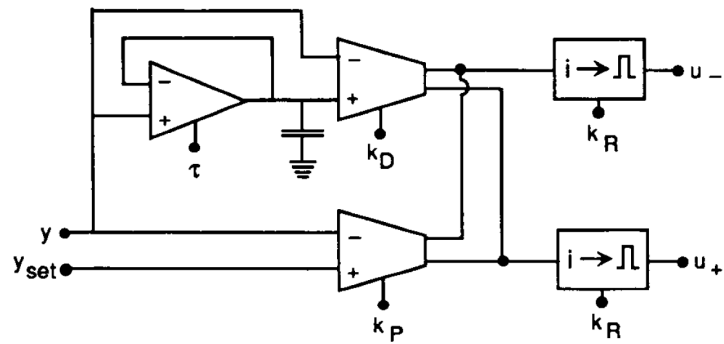
\includefigure{neuron_servo_circuit}{0.6}{Neuromorphic proportional-derivative (PD) motor controller. Differential output of a conventional PD circuit is passed through two neuron-like pulse generators. (Adapted from \cite{sb_deweerth1991} with permission.)}

The structure of the controller is shown in Fig. \ref{fig:neuron_servo_circuit}. The output of a conventional proportional-derivative (PD) action controller is put through a neuronal circuit which converts the output of the controller into a pulse train. The output is differential so that the positive and negative values of the output are sent through two independent pulse channels. The controller output is then fed into a conventional DC motor.

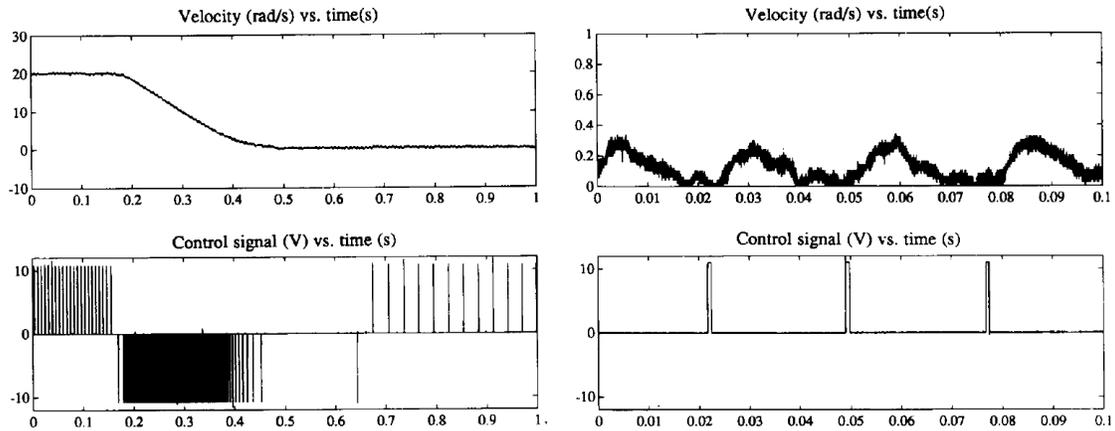
\includefigure{neuron_servo_time}{0.9}{Speed control of the neuron servo. The neural controller provides a pulse-width modulation control scheme at high speeds, and a stepper control scheme at lower speeds. The right figure provides the zoomed in time plot of the low speed regulation. (Adapted from \cite{sb_deweerth1991} with permission.)}

The speed regulation of the neuromorphic controller can be seen in Fig. \ref{fig:neuron_servo_time}. The remarkable aspect of the controller can be observed by comparing its operation at a high speed of rotation and at a low speed. At medium and high speeds both the conventional analog and the neuromorphic controller behave reliably: the neuromorphic controller effectively operates with a pulse-width modulation scheme, as the average number of pulses per second determines the DC value supplied to the motor. However, lowering the speed set point to a significantly smaller value, the conventional analog controller fails to rotate the motor due to its inability to overcome static friction. In sharp contrast, the neuromorphic controller naturally adapts its operation to a scheme resembling the operation of a stepper motor: each individual pulse has enough energy to overcome the friction, and therefore the controller provides repeating kicks to the motor that allow it to rotate at a small average frequency (Fig. \ref{fig:neuron_servo_time}, right).

This work came about through the collaboration of, among others, Karl Astrom and Carver Mead, pioneers in control and neuromorphic engineering respectively. It highlights the revolutionary potential of neuromorphic control. The mixed analog-digital operation of neural systems allows for novel adaptation schemes not possible in the purely analog or the purely digital world.

\sbend

\sbsection{The Hodgkin-Huxley model}{hodgkin_huxley}
The classical model of excitability comes from the pioneering experimental work of Hodgkin and Huxley (\cref{fig:hodgkin_and_huxley}) of the squid giant axon in 1952 \cite{sb_hodgkin1952}. Although the analysis concentrated on this particular neuron, the methodology was later applied to explain the behavior of other neural cells. The starting point of the analysis is the observation that the excitable membrane can be modeled as an electrical circuit (\cref{fig:hh_circuit}).

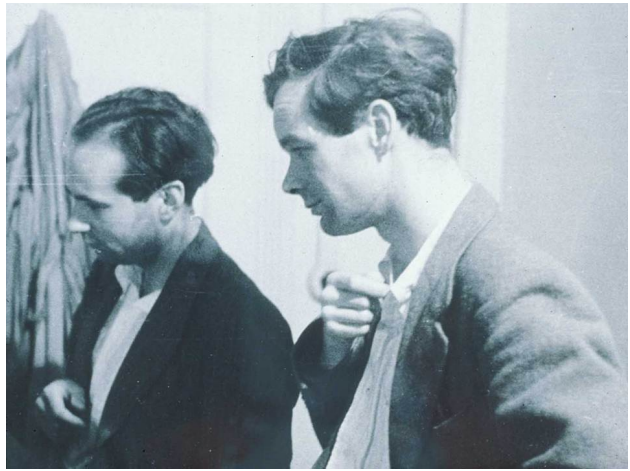
\includefigure{hodgkin_and_huxley}{0.5}{Andrew Huxley (left) and Alan Hodgkin (right). (Source \cite{sb_hodgkin_huxley_image}, unknown photographer.)}

Firstly, every cell consists of an impermeable membrane which is able to maintain an electrical potential difference between the intercellular and extracellular environments. This property is modeled with a capacitor which stores the charge between the two media. In addition to this, the membrane is equipped with special proteins called \textit{ion channels} which are selectively permeable to specific ions in the environment. Due to the different concentrations of ions inside and outside the cell, the cell dynamically controls its membrane voltage by opening and closing these channels and thus controlling the flow of ionic currents through the membrane. Hodgkin and Huxley identified two key players for the generation of electrical pulses in the squid axon: potassium ($K^{+}$) and sodium ($Na^{+}$) ionic currents. Remaining currents were lumped into a third, leak component. The selective permeability to each ion is captured by an individual conductance element in the circuit, while the equilibrium (Nernst) potential where diffusion exactly balances the electrical force is modeled by a battery. There is a higher concentration of sodium outside the cell than inside, and vice-versa for potassium. Within the circuit modeling framework, this translates to a high sodium Nernst potential ($\SI{40}{\milli\volt}$ in the Hodgkin-Huxley model) and a low potassium Nernst potential ($\SI{-70}{\milli\volt}$ in the Hodgkin-Huxley model). Thus, sodium current is always inward (negative by convention) and acts to increase the voltage, while potassium current is outward (positive by convention) and acts to decrease the voltage.

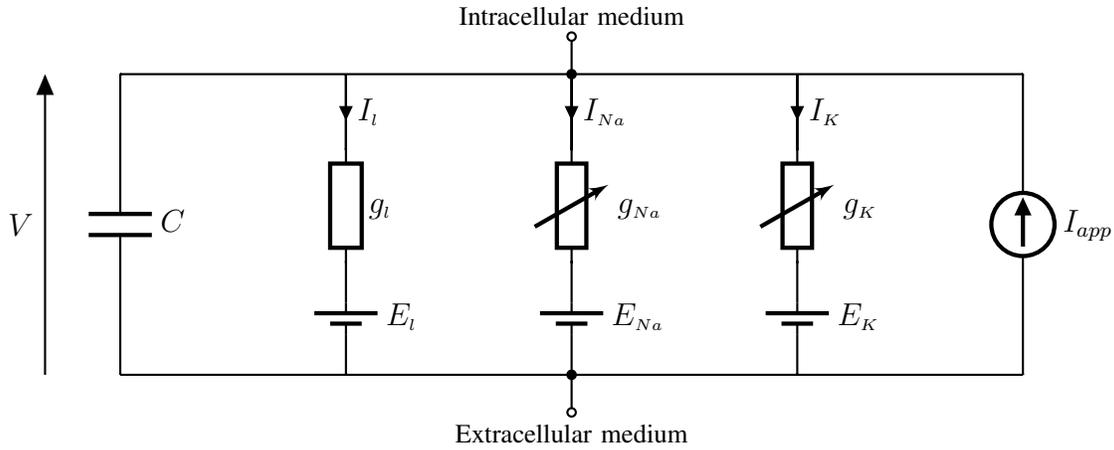
\includetikzfigure{hh_circuit}{The Hodgkin-Huxley circuit. The neural membrane, separating the intracellular and the extracellular media, is modeled as a parallel interconnection of the passive capacitor and leak current $I_{l}$, together with the active sodium $I_{Na}$ and potassium $I_K$ currents. External current applied to the cell is represented with the current source $I_{app}$.}

The dynamics of the circuit are governed by the membrane equation
\begin{equation}
\label{eq:hodgkin_huxley}
C\frac{dV}{dt} = -g_{l}(V-E_{l})- g_{Na}(V-E_{Na}) - g_{K}(V-E_{K})  + I_{app},
\end{equation}
where $C$ is the membrane capacitance, $V$ is the membrane voltage, $g_l$, $g_{Na}$, and $g_{K}$ are the conductances corresponding to leak, sodium, and potassium respectively, with $E_{l}$, $E_{Na}$ and $E_{K}$ being their corresponding equilibrium potentials, and $I_{app}$ is the externally injected current into the cell.

Sodium and potassium conductances are \textit{active} i.e. they are voltage and time-dependent. This reflects the continuous opening and closing of the their ion channels, in contrast to the leak conductance which is constant, and accounts for the passive properties of the membrane. By using the technique known as \textit{voltage clamping}, Hodgkin and Huxley were able to keep the membrane voltage fixed at different levels and measure different step responses from the resting state of the system. Fitting the data to the simplest form, they obtained the equations for the two conductances:
\begin{align}
\label{eq:hh_conductances}
g_{Na} &= \overline{g}_{Na} m^3 h, \\
g_{K} &= \overline{g}_{K} n^4,
\end{align}
where $\overline{g}_{Na}$ and $\overline{g}_{K}$ are the maximal conductances of sodium and potassium respectively, and $m$, $h$, and $n$ are the gating variables that follow first-order dynamics
\begin{equation}
\begin{aligned}
\label{eq:hh_first_order}
\tau_{m}(V)\dot{m} &= m_{\infty}(V) - m, \\
\tau_{h}(V)\dot{h} &= h_{\infty}(V) - h, \\
\tau_{n}(V)\dot{n} &= n_{\infty}(V) - n.
\end{aligned}
\end{equation}
Each gating variable has a value between 0 and 1 and thus represents the continuous tuning of the ion channels between being fully closed and fully open. The steady-state functions $m_{\infty}(V)$, $h_{\infty}(V)$, and $n_{\infty}(V)$ have a sigmoidal shape (\cref{fig:hh_inf_tau}, left), while the voltage-dependent time constants $\tau_{m}$, $\tau_{h}$, and $\tau_{n}$ have a Gaussian shape (\cref{fig:hh_inf_tau}, right).

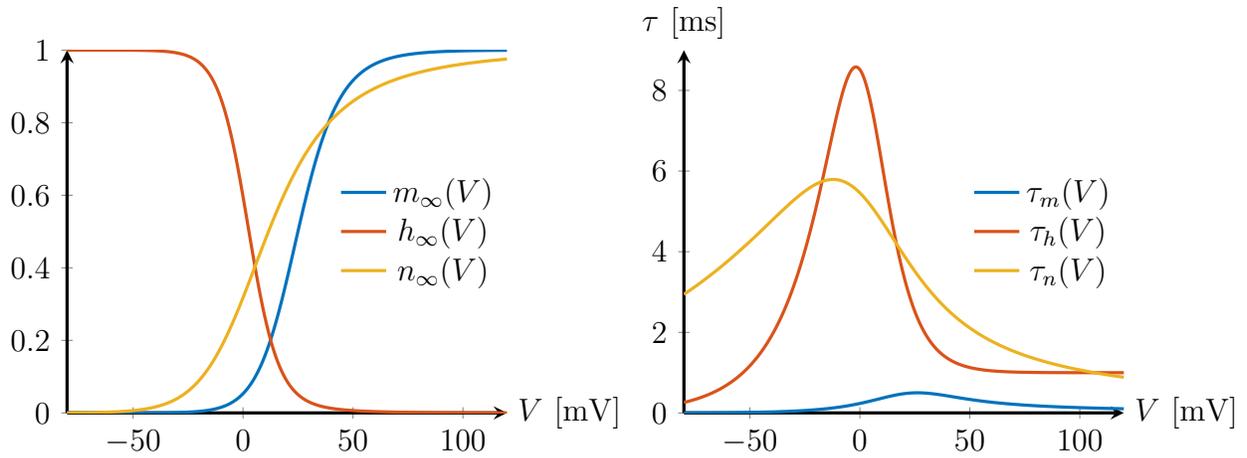
\includetikzfigure{hh_inf_tau}{Steady-state and time constant functions of the gating variables. The steady-state functions (left) have a sigmoidal shape and are monotonically increasing for activation variables, and monotonically decreasing for inactivation variables. The time constants (right) have a Gaussian shape.}

The dynamics of each gating variable is fully characterized by these two voltage-dependent functions. The steady-state functions are monotonic and the voltage range in which the slope of these functions is non-zero defines the window in which the currents are active, while the slope defines the sign of the feedback. The voltage-dependent time constants define the timescale window in which the currents operate.

\sbend

\sbsection{Linearizing conductance-based models}{linearization}

Conductance-based models can be analyzed by considering small perturbations around every equilibrium voltage \cite{sb_koch2004,drion2015}. This technique characterizes the circuit based on how its impedance function changes with the equilibrium voltage and frequency, illuminating important properties of possibly highly nonlinear and complex systems that are generally hard to analyze.

When a small change in the membrane voltage is applied, it induces a current change that has two components: an instant change in the current, due to the resistive properties of each ionic current, and a slower component, due to the active properties of the gating variables. In circuit terms, the former leads to a simple resistor that gives an instantaneous change in current for a change in voltage, and each gating variable leads to a resistor-inductor branch, capturing the slower, first-order dynamics that arise. The value of each component will depend on the voltage around which the linearization is applied. When considering an arbitrary ionic current with activation and inactivation gating variables
\begin{equation}
	I_j = m_j^p h_j^q (V - E_j),
\end{equation}
linearization gives the change in current
\begin{equation}
\delta I_j = \frac{\partial I_j}{\partial V}\delta V + \frac{\partial I_j}{\partial m_j}\delta m_j + \frac{\partial I_j}{\partial h_j}\delta h_j,
\end{equation}
where the first term represents the resistive component of the current, while the second and the third term are due to the activation and inactivation function. Taking the Laplace transform, the full linearized membrane model thus becomes:
\begin{equation}
s C \delta V = - \underbrace{\bigg[G(V) + \sum_j \Big( \frac{g_{m_j}(V)}{\tau_{m_j}(V)s + 1} + \frac{g_{h_j}(V)}{\tau_{h_j}(V)s + 1} \Big)\bigg]}_{\text{\normalsize $Z(s,V)$}} \delta V.
\end{equation}

The behavior of the linearized membrane model is fully characterized by the capacitance $C$ and the total impedance $Z(s,V)$ which consists of the total resistive part $G(V)$ accumulating all resistive components of the ionic currents
\begin{equation}
G(V) = g_l + \sum_j \frac{\partial I_j}{\partial V},
\end{equation}
and the first-order terms arising from the dynamics of the gating variables
\begin{align}
g_{m_j}(V) &= \frac{\partial I_j}{\partial m_j} \frac{d m_{j,\infty} (V)}{dV}, \\
g_{h_j}(V) &= \frac{\partial I_j}{\partial h_j} \frac{d h_{j,\infty} (V)}{dV}.
\end{align}

Linearization of ionic currents has a simple circuit interpretation shown in \cref{fig:current_linearization}, with the single resistor capturing the resistive term $G_j(V)$, and a series connection of a resistor and an inductor capturing the first order terms due to the activation and inactivation functions. This interpretation illuminates the synthesis problem, as it decouples the input-output properties of the circuit elements from their internal physical realization. Linearization is a indispensable tool for tractable analysis of excitable behaviors as it allows the effects of the dynamic variables to be separated, and contributions from many different ionic currents to be grouped together according to their temporal range of activation.

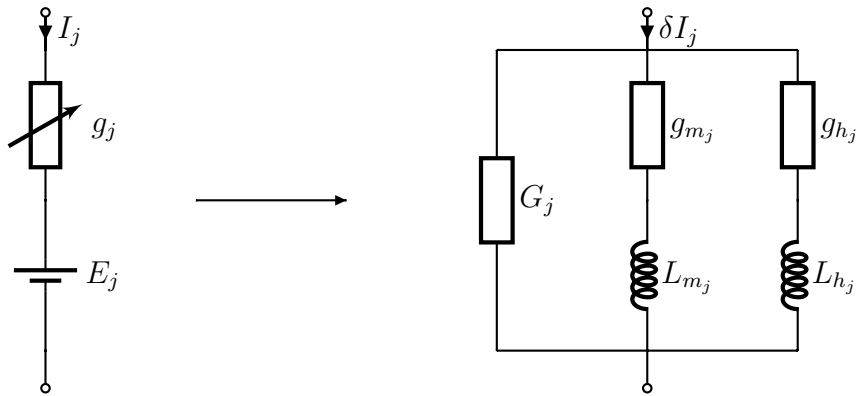
\includetikzfigure{current_linearization}{Linearization of an arbitrary ionic current with an activation and an inactivation variable. The resistive term is represented by a single resistor $G_j$, while each of the serial connections of a resistor and an inductor represents the terms stemming from the activation function $m_j$ and the inactivation function $h_j$.}

\sbend

\sbsection{Using physics to compute: subtreshold electronics}{subthreshold}

The fundamental idea behind the neuromorphic approach pioneered by Carver Mead and colleagues was using physics to compute, i.e. using the native relationships between currents and voltages in electronic devices in order to implement computational primitives. At the core of this approach is the use of standard metal-oxide-semiconductor (MOS) technology operating in the low-voltage \textit{subthreshold} regime.

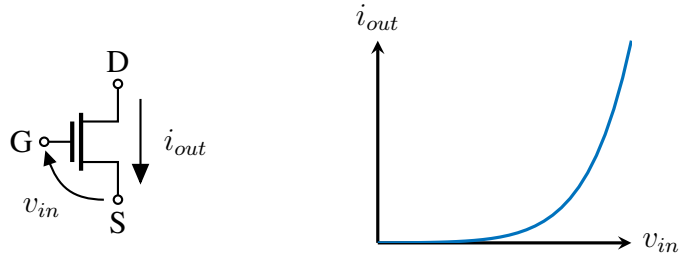
\includetikzfigure{nmos_symbol}{An n-channel MOSFET. The output current $i_{out}$ between its drain (D) and source (S) is controlled by the input voltage $v_{in}$ between its gate (G) and source, and is largely independent of the voltage between its output nodes. In the subthreshold regime, the input-output relationship takes an exponential form.}

A transistor is a three-terminal current source element (\cref{fig:nmos_symbol}): the current going between its drain and its source (output current) is controlled by the voltage applied between its gate and source (input voltage). In traditional digital electronics, MOSFET is only considered on if the input voltage is sufficiently high, disregarding the small output current when the input is below the threshold voltage. In this regime the output current is a quadratic function of the input voltage and the charge is mainly carried by drift. In turn, in the subthreshold regime the output current is orders of magnitude smaller and there is an exponential input-output relationship. In this region, the charge is mainly carried by diffusion and the transistor's input-output characteristic can be reduced to the form
\begin{equation}
	i_{out} = i_0 e^{\frac{v_{in}}{v_0}},
\end{equation}
where $i_0$ is the zero-bias current, an inherent property of the transistor, and $v_0$ is the voltage constant depending on the properties of the integrated design process and the operating temperature.

There are two main advantages of operating in this regime:
\begin{itemize}
	\item The voltages and the currents are low. This leads to circuits operating with very low power requirements in the order of \si{\micro\watt}s.
	\item The exponential input-output relationship is a powerful analog primitive for synthesizing sigmoidal activation functions.
\end{itemize}

Analog subthreshold circuits compute fundamental functions such as the exponential and logarithmic function, as well as standard multiplication, division, addition and subtraction in an exceptionally energy efficient way, and thus come close to emulating the way biological systems achieve analog computation. They offer an exciting avenue for designing novel neuromorphic control schemes for mixed analog-digital excitable circuits.
\sbend

\sbsection{Neuromorphic building blocks}{neuromorphic_building_blocks}

Two basic subthreshold circuit building blocks are sufficient for the design of neuromorphic circuits using the I-V curve synthesis methodology. These are the transconductance amplifier, a device that generates a sigmoidal input-output relationship, and the integrator-follower circuit that provides temporal integration.

The transconductance amplifier generates an output current that depends on the difference between the two input voltages. The schematic of the differential amplifier is shown in \cref{fig:transconductance_amplifier}.

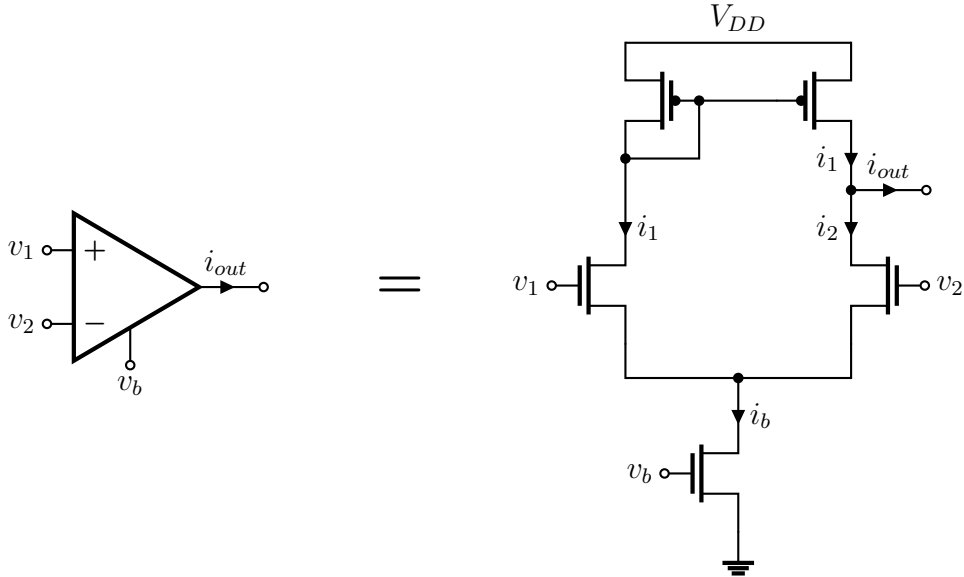
\includetikzfigure{transconductance_amplifier}{Transconductance amplifier. The circuit realizes a hyperbolic tangent mapping from the differential voltage input to the current output. The gain of the function is determined by the current flowing through the base transistor ($i_b$), controlled by its base voltage ($v_b$) that acts as the control input to the amplifier.}

In the subthreshold regime, the input transistors generate currents $i_1$ and $i_2$ that are are exponentially related to the two input voltages. The total sum of the two currents is controlled through the bottom bias transistor that provides the control current. These two currents are subtracted to form the output current through the current mirror formed by the two upper transistors which copies the current $i_1$ into the output node. This simple circuit architecture conveniently generates a hyperbolic tangent mapping
\begin{equation}
	\label{eq:amplifier_io}
	i_{out} = i_b \tanh \bigg(\frac{v_1 - v_2}{2 v_0} \bigg),
\end{equation}
where $v_0$ is the voltage constant depending on the integrated process parameters and the circuit temperature.

The circuit provides a versatile building block for the synthesis of I-V curves, as it provides localized conductance characteristic whose gain can be controlled externally through the input $v_b$ over several orders of magnitude. These two characteristics make it a sufficient building block for defining the local action of the positive and negative feedback elements of the neuromorphic architecture.

The same element can be used to generate temporal filtering. Since the time constants of neuromorphic systems generally require high resistor values in order to align with the frequencies at which biological systems operate, the circuit can be utilized as a resistive element with variable resistance. This is shown in \cref{fig:follower_integrator}.

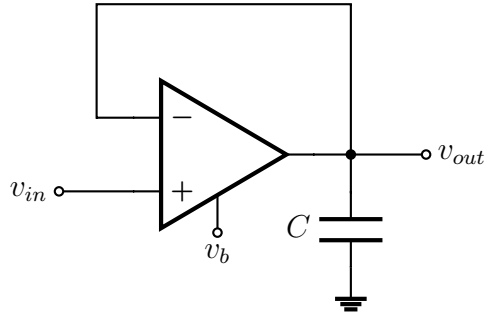
\includetikzfigure{follower_integrator}{Follower-integrator circuit. The circuit implements nonlinear first-order filtering of the input voltage $v_{in}$. The time constant of the filter can be controlled by changing the voltage $v_{b}$, which effectively controls the output resistance of the amplifier.}

The output voltage of the circuit is determined by
\begin{equation}
	C \frac{dv_{out}}{dt} = i_b \tanh \bigg( \frac{v_{in} - v_{out}}{2 v_0} \bigg).
\end{equation}
For small changes in the voltage the equation takes the form of a linear first-order filter, and the time constant of the circuit can be appropriately defined as
\begin{equation}
	T = \frac{2 v_0}{i_b} C.
\end{equation}
The circuit therefore provides an effective dynamical primitive for defining the time constants of the neuromorphic feedback elements.
\sbend

\sbsection{Minimal models of excitability}{min_models}

The classical minimal model of excitability comes in the form of the FitzHugh-Nagumo equations \cite{fitzhugh1961,nagumo1962}
\begin{equation}
\label{eq:fitzhugh}
\begin{aligned}
\dot{V} &= - \frac{V^3}{3} + V -n + I_{app}, \\
\tau \dot{n} &= V + a - bn.
\end{aligned}
\end{equation}
The two-dimensional model qualitatively captures the basic structure of excitable systems: the instantaneous dissipative properties are captured by the term $\frac{V^3}{3}$, the negative conductance by the instantaneous term $V$, and the slow positive conductance by the term $n$. The time constant $\tau \gg 1$ models the timescale separation between the fast variable $V$ and the slow variable $n$.

Model \eqref{eq:fitzhugh} provides the fundamental phase portrait of excitability. In addition, the simple structure of the model allowed for straightforward circuit realizations, starting with the contribution from Nagumo and colleagues \cite{nagumo1962}. 

Another important minimal model is the Morris-Lecar model \cite{sb_morris1981}. Unlike the FitzHugh-Nagumo model, the model has a conductance-based structure which is why it has commonly been used when biophysical interpretability is needed. The model is minimal as it is two-dimensional and has the necessary and sufficient components for excitability: a fast negative conductance component, represented by the instantaneously activated inward calcium current, and slow positive conductance component, represented by the slowly activating potassium current. The model is described by
\begin{equation}
\begin{aligned}
C \frac{dV}{dt} &= - \overline{g}_l (V - E_l) - \overline{g}_{Ca} m_{\infty}(V) (V - E_{Ca}) - \overline{g}_K n (V - E_{K}) + I_{app}, \\
\tau_n \dot{n} &= n_{\infty}(V) - n.
\end{aligned}
\end{equation}
From the dynamical systems perspective, the model is also important because for different parameters it can model different types of excitability in a minimal representation amenable to phase plane analysis.

There are various proposed mathematical models of bursting in literature, capturing the particular dynamical features of different neuronal bursting cells \cite{izhikevich2007}. From the mathematical viewpoint, the main classification between bursting types lies in the mechanism of the transition from resting to spiking and back. If the fast spiking system has a region of bistability between the resting and spiking states, then an additional slower variable is sufficient to drive the system around the hysteresis and generate a bursting attractor. However, if the fast system is monostable, then the slow system needs to be at least two-dimensional in order to generate semi-autonomous slow oscillations that push the system between spiking and rest. Analysis of these mechanisms has led to an extensive classification that focuses on the bifurcations that lead to the transition between resting and spiking and vice-versa \cite{sb_izhikevich2000}.

An early and key example of a minimal bursting model came from the analysis of Hindmarsh and Rose \cite{hindmarsh1984}. The model can be seen as a modification of the minimal spiking model of FitzHugh that establishes bistability between stable rest and spiking states by introducing a quadratic slow variable instead of a linear one. The variable thus captures both negative and positive feedback in a single variable. In addition to modifying the slow variable, the model introduces ultra-slow adaptation that provides negative feedback in the ultra-slow timescale and generates the oscillation between the rest and spiking states. The full model is
\begin{equation}
\label{eq:hindmarsh_rose}
\begin{aligned}
\dot{V} &= - a V^3 + b V^2 - n - z + I_{app}, \\
\dot{n} &= d V^2 - c - n, \\
\tau_z \dot{z} &= s(V - v_1) - z.
\end{aligned}
\end{equation}

Recent work has concentrated on modelling the biophysical transition between spiking and bursting modes in a minimal dynamical model. This has led to the proposed three-timescale model based on the transcritical singularity \cite{franci2014a}. The model is based on the reduction of the Hodgkin-Huxley model with the addition of a slowly-activated calcium current \cite{drion2012}, and it shares a similar structure as the FitzHugh-Nagumo model of spiking and the Hindmarsh-Rose model of bursting. The model is given by
\begin{equation}
\begin{aligned}
\label{eq:transcritical_bursting}
\dot{V} &= - \frac{V^3}{3} + V - (n + n_0)^2 - z + I_{app}, \\
\tau_n \dot{n} &= n_{\infty}(V) - n, \\
\tau_z \dot{z} &= z_{\infty}(V) - z.
\end{aligned}
\end{equation}

Both model \eqref{eq:hindmarsh_rose} and model \eqref{eq:transcritical_bursting} can be studied by treating the ultra-slow variable $z$ as a constant parameter and analyzing the family of phase portraits parameterized by $z$. Bursting is obtained from slow adaptation of $z$ over the range of parameters where the phase portraits exhibit bistability between spiking and resting.

The sidebar "\nameref{sb:phase_portraits}" provides further insight into the phase portrait analysis of these models and its connection to the I-V curve characterization discussed in the article.

\sbend

\sbsection{I-V curves and phase portraits}{phase_portraits}

Classically, the mathematical analysis of excitability has centered around the phase portrait analysis of dynamical models. Phase portraits provide a geometrical understanding of excitability thanks to its fundamentally two-timescale nature. The famous FitzHugh-Nagumo model \cite{fitzhugh1961} provides the core illustration of this property through its two state variables: a voltage variable capturing the fast positive feedback process, and a slow refractory variable capturing the slow recovery. The model provides key insight into the generation of an action potential through the analysis of the nullclines and the equilibrium analysis of the system. The inverted N-shape of the fast nullcline reflects the bistable nature of the fast subsystem, while the monotone slow nullcline ensures the monostability of the system in the slow timescale. Remarkably, reduction of more complex spiking models reveals the same excitability picture \cite{sb_rinzel1985a,sb_kepler1992}, showing the generality of the mechanisms discussed by FitzHugh in his seminal paper.

Richer multi-timescale phenomena such as bursting are more difficult to capture with a two-dimensional phase plane \cite{izhikevich2007}. Still, because of their ease of interpretability, two-dimensional phase portrait techniques were also successfully utilized to understand more complex neuronal behavior \cite{hindmarsh1984,sb_rinzel1987,sb_izhikevich2000,franci2014a}. Those models augment the phase plane analysis of spiking with a parameter that deforms the phase portrait in a slower timescale.

Traditionally, phase plane models have been derived from empirical reductions of higher-dimensional conductance-based models. This is not necessarily an easy task, as many distinct currents and parameters control neuronal excitability. Input-output techniques offer an attractive alternative to state-space methods, closer in spirit to the experimental techniques of electrophysiology. Input-output measurements are directly generated through experiments using the classical voltage-clamp technique \cite{sb_hodgkin1952}. The technique utilizes an experimental setup using a high gain negative feedback amplifier that allows fixing the membrane voltage at different levels and measuring the total internal current, thus capturing the step response of the system. Secondly, input-output characterization naturally leads to a multiscale modeling approach as the characterization can be captured at different timescales, corresponding to the timescales of the spiking behavior \cite{drion2015}. This is the idea behind the I-V curve system analysis.

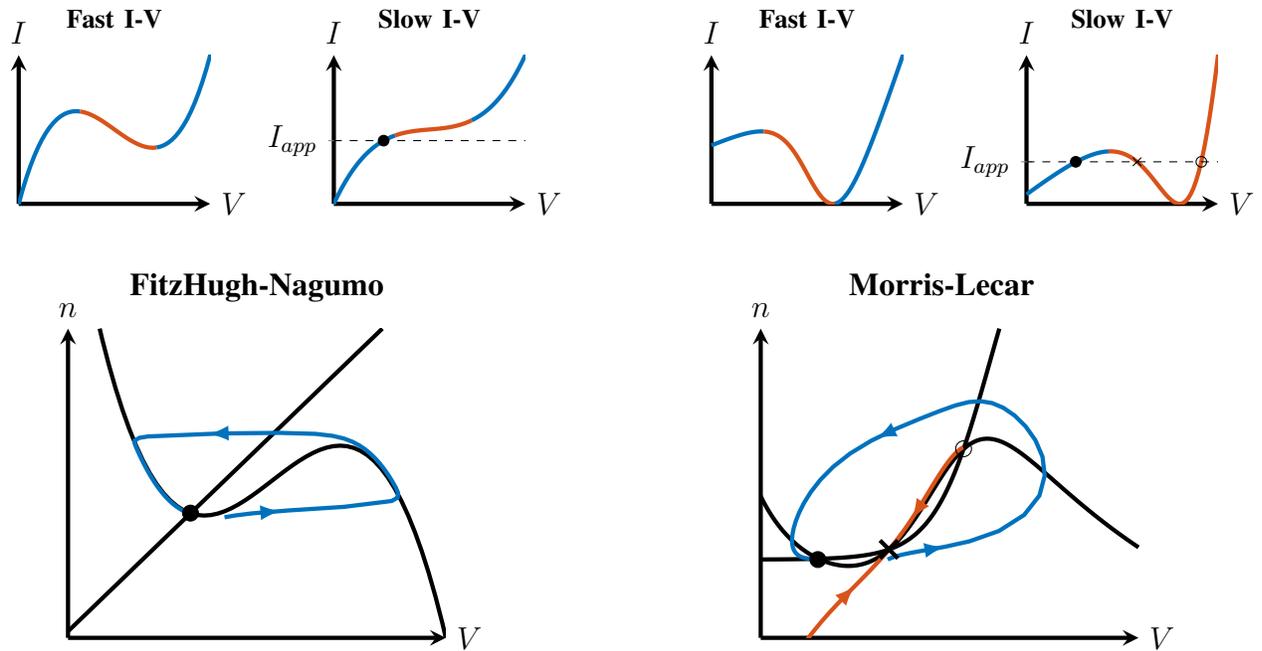
\includetikzfigure{phase_portraits1}{Phase portraits of excitability and their corresponding I-V curves. The FitzHugh-Nagumo phase portrait consists of an inverted N-shaped fast $V$ nullcline and a linear slow $n$ nullcline. The single equilibrium of the system loses stability through a Hopf bifurcation near the local maxima, leading to finite frequency oscillations at the onset. This is reflected in the I-V curves as an N-shaped fast I-V curve, and a monotonic slow I-V curve. On the other hand, Morris-Lecar model can exhibit the loss of stability through a saddle-node on invariant circle (SNIC) bifurcation, observed through the appearance of three nullcline intersections: a stable equilibrium on the left, a saddle point in the middle, and an unstable equilibrium on the right. While the fast I-V curve assumes a similar N-shape, the SNIC bifurcation mechanism is manifested through the non-monotonicity of the slow I-V curve.}

The classical phase portrait pictures can be linked to the input-output characteristic of the system in a simple way. This is shown in \cref{fig:phase_portraits1}. The figure links the standard phase portraits of excitability with their corresponding characteristic fast and slow I-V curves. The two spiking models presented are FitzHugh-Nagumo and Morris-Lecar \cite{sb_morris1981}, representing two different types of common spiking behavior. In the case of FitzHugh-Nagumo, the model undergoes a Hopf bifurcation at the onset of spiking, leading to the oscillations appearing with a non-zero frequency. On the other hand, the parameters of the Morris-Lecar can be set up so that it represents a minimal realization of a spiking neuron with continuous frequency curve \cite{sb_rinzel1989}, characterized by a saddle-node on invariant circle (SNIC) bifurcation at the onset of spiking. Both models share the same N-shaped fast I-V curve, stemming from the existence of bistability in the fast timescale. They differ in the monotonicity of the slow I-V curve: a monotonically increasing slow I-V curve is a signature of the Hopf bifurcation, while the non-monotonic slow I-V curve gives rise to the SNIC bifurcation. 

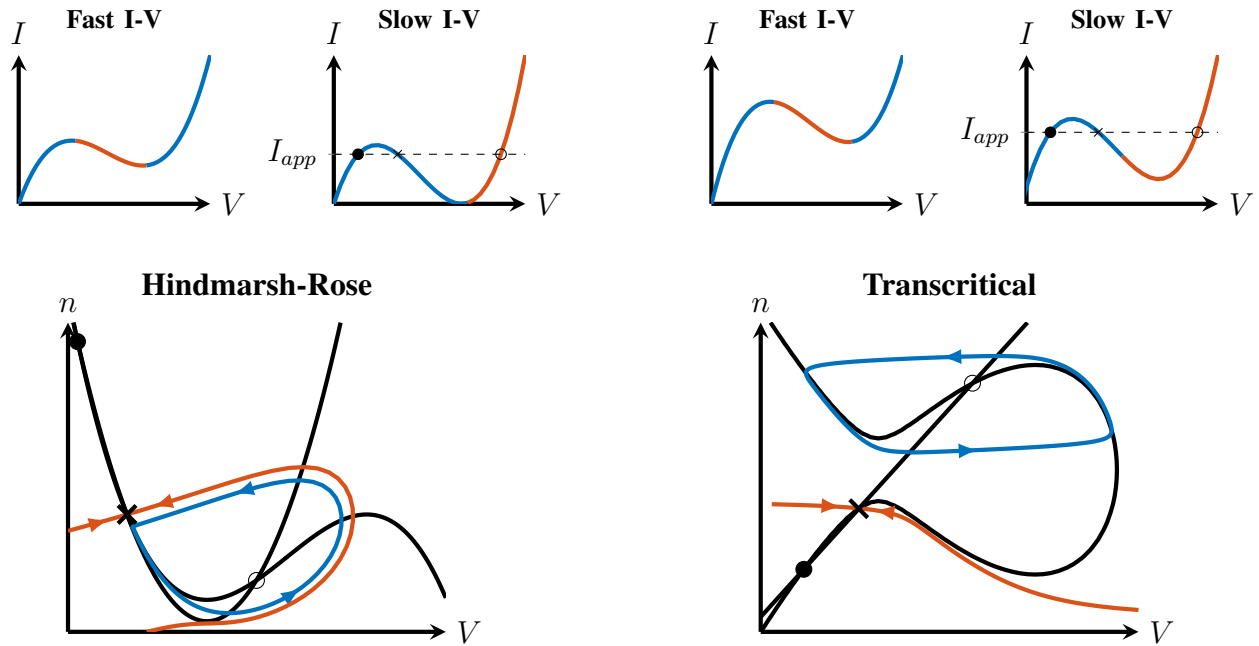
\includetikzfigure{phase_portraits2}{Phase portraits of rest-spike bistability and their corresponding I-V curves. The essential mechanism behind burst excitability is the bistability between a stable rest state and a stable spiking state. In Hindmarsh-Rose model this is achieved through the quadratic form of the slow nullcline, so that three intersection points give rise to a stable equilibrium on the left, a saddle point in the middle, and the unstable equilibrium on the right, surrounded by a stable limit cycle. In the more recent transcritical model, this bistability is achieved instead by mirroring the fast cubic nullcline, giving rise to the same three equilibrium structure. Both mechanisms share qualitatively the same I-V curves, the fast N-shaped I-V curve due to the presence of fast bistability, and the N-shaped slow I-V curve due to the presence of slow bistability. The important characteristic of both models is the presence of slow positive feedback. This is showcased by the appearance of the negative slope region of the slow I-V curves that does not correspond to the negative slope region in the fast timescale, which necessarily comes from a source of slow negative conductance.}

Two representative phase portraits of burst excitability are shown in \cref{fig:phase_portraits2}. The key characteristic of bursting systems is the appearance of bistability between stable rest and spiking states, which allows slower processes to periodically switch the system between the two. The two bursting phase portraits display the different mechanisms of achieving the bistability between these two states. The left phase portrait is the well-known Hindmarsh and Rose model \cite{hindmarsh1984}, where the slow nullcline is non-monotonic. The non-monotonicity of the slow variable reflects the appearance of both positive and negative feedback processes in the slow timescale. At the right-hand side is the phase portrait of the more recent bursting model organized by a transcritical bifurcation \cite{franci2014a}, where bistability is instead achieved by mirroring the fast nullcline, again the consequence of the appearance of both positive and negative slow feedback. Both mechanisms lead to a robust generation of bistability between resting and spiking states of the model. Importantly, both of these models share the same qualitative picture in their input-output I-V curve characterization: the N-shaped fast I-V curved is accompanied by the N-shaped slow I-V curve, a signature of the appearance of slow bistability between the two states. A more extensive analysis of the geometry of rest-spike bistability is provided in \cite{sb_cirillo2020}.

We note that the phase portraits illustrated here do not include classical bistable phase portraits that lack a source of slow positive feedback. Both phase portraits of \cref{fig:phase_portraits1} can exhibit the coexistence of a stable fixed point and a stable limit cycle through the appropriate tuning of the timescales in the models and have often been associated to bursting in neurodynamical studies \cite{sb_rinzel1989}. The models are however bistable only for a precise ratio of timescales and do not capture the robust transitions between slow spiking and bursting. These limitations are further discussed in \cite{franci2013,franci2017}, as well as in terms of the neuromorphic implementation in \cite{ribar2019}.

\sbend

\sbsection{A neuromorphic neuron in hardware}{circuit_implementation}

A neuromorphic neuron obeying the control principles discussed in this article was realized in hardware as part of the PhD dissertation \cite{ribar2019a}. The circuit possess the interconnection architecture presented in "\nameref{sec:synthesis}" and has a simple implementation using the standard neuromorphic building blocks. An individual feedback element is implemented as a series interconnection of a follower-integrator circuit providing the first-order filtering, and a transconductance amplifier, providing the hyperbolic tangent input-output relationship. This interconnection is shown in \cref{fig:filter_amplifier}, and the details of the individual circuits are discussed in the sidebar "\nameref{sb:neuromorphic_building_blocks}".

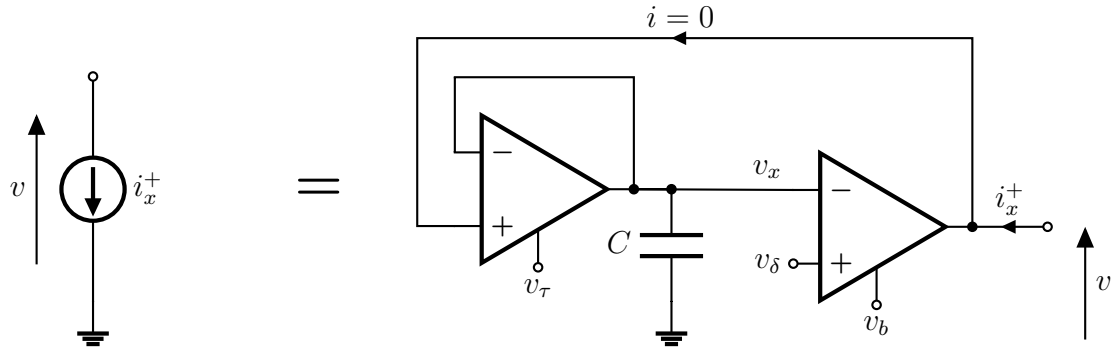
\includetikzfigure{filter_amplifier}{Implementation of a single circuit feedback element. The first transconductance amplifier and a capacitor form a nonlinear first-order filter. The output of the filter is then fed into the second transconductance amplifier which forms the output current $i_x^+$. The figure shows a realization of a positive conductance element, while for a negative conductance element the inputs to the second amplifier are interchanged.}

The main control parameters of the element then correspond to the voltage $v_\tau$ that sets the timescale of the element, the offset voltage variable $v_\delta$, and the gain of the feedback element $i_b$ set through $v_b$. These parameters correspond to the dimensionless parameters introduced in "\nameref{subsec:neuromorphic_architecture}" through simple transformations
\begin{subequations}
\label{eq:parameter_mapping}
\begin{align}
i_b &= \alpha \big(2 G v_0 \big), \\
v_{\delta} &= \delta \big( 2 v_0 \big),
\end{align}
\end{subequations}
where $v_0$ is the voltage constant depending on the integrated process and the circuit temperature, and $G$ defined as the conductance of the passive membrane element around equilibrium
\begin{equation}
G = \frac{di_p(v)}{dv}\bigg|_{v=v_e}.
\end{equation}

The conductance of the passive element and the membrane capacitor define the time constant of the membrane equation
\begin{equation}
T_v = \frac{C}{G},
\end{equation}
and the applied current is simply derived from the dimensionless applied current as
\begin{equation}
i_{app} = I_{app} \big( 2 G v_0 \big).
\end{equation}

A measurement from the hardware realization is shown in \cref{fig:oscilloscope_transition} capturing the continuous modulation of the circuit between spiking and bursting regimes through the sole control of the voltage that defines the gain of the slow negative conductance. This switch drastically changes the input-output behavior of the circuit, enabling the localized control of the network output when the neuron is part of a larger interconnected structure.

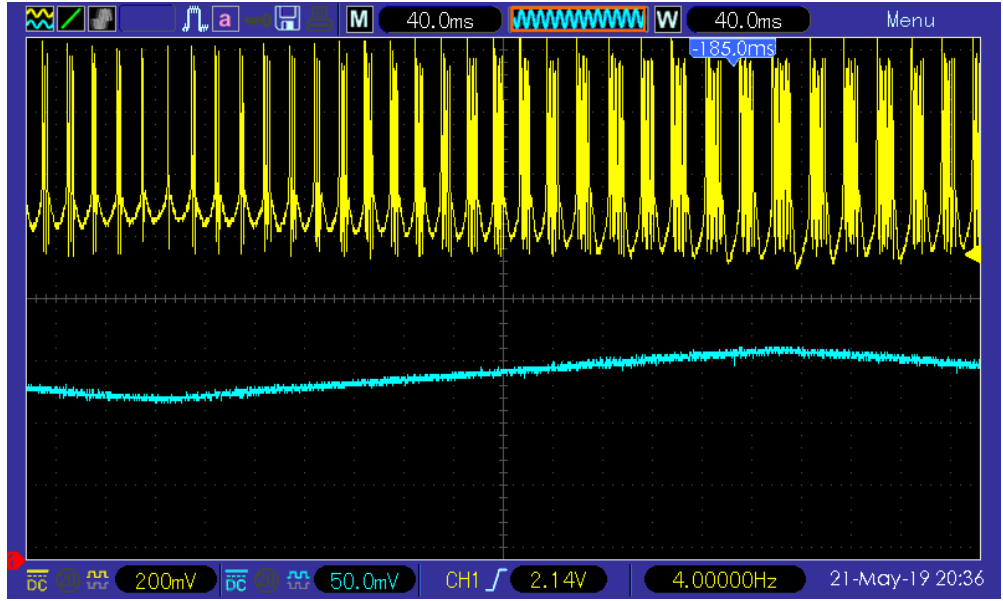
\includefigure{oscilloscope_transition}{0.8}{Transition between slow spiking and bursting in the neuromorphic circuit implementation. A periodic triangle wave is applied to the base voltage of the slow negative conductance (blue trace) so that the circuit is periodically moving between the bursting and slow spiking regimes.}

\sbend

\sbsection{Central pattern generators}{cpg}

Generating robust and flexible rhythmic activity is necessary for a variety of essential biological behaviors such as locomotion and breathing. Special neural networks called central pattern generators (CPGs) are responsible for generating autonomous rhythmic patterns. These patterns then define spatio-temporal sequences of activation that generate synchronized motor behavior. CPG networks vary greatly in structure and complexity \cite{sb_marder1996}, but share a common modular organization. The fundamental module in this organization is a simple two-neuron network commonly known as the \textit{half-center oscillator}.

The half-center oscillator finds its origins in the seminal work of Brown \cite{sb_brown1911}, who investigated the autonomous electrical activity of electro-stimulated muscles of dead animals. The simplest model of a half-center oscillator consists of two neurons mutually interconnected with inhibitory synaptic connections. The individual neurons do not oscillate in isolation but the network rhythm emerges from the interaction. The self-regenerative feedback necessary for sustained oscillatory behavior is provided through the post-inhibitory rebound mechanism discussed in the main article.

The mutual inhibition motif appears to be a fundamental building block shared by all rhythmic neural network.

\sbend
       	
\newpage
\section{Author Biography}
Luka Ribar received the M.Eng. degree in electrical and information sciences and the Ph.D. degree in control engineering from the University of Cambridge in 2015 and 2020, respectively. He is currently a postdoctoral research associate at the University of Cambridge. His main research interests are in dynamics and control of neuronal circuits and neuromorphic engineering.

Email: lr368@cam.ac.uk

Rodolphe Sepulchre (M-96, SM-08, F-10) received the engineering degree and the Ph.D. degree from the Universit\'e catholique de Louvain in 1990 and in 1994, respectively.  He is currently Professor of Engineering at Cambridge University. His research interests are in nonlinear control and optimization.  He co-authored the monographs "Constructive Nonlinear Control" (Springer-Verlag, 1997) and "Optimization on Matrix Manifolds" (Princeton University Press, 2008). From 2009, his research has been increasingly motivated by control questions from neuroscience. A current focus is his ERC advanced grant "Switchlets",  aiming at a multi-scale control theory of excitable systems.  In 2008, he was awarded the IEEE Control Systems Society Antonio Ruberti Young Researcher Prize. He is a fellow of IEEE, SIAM, and IFAC. He has been IEEE CSS distinguished lecturer  between 2010 and 2015. In 2013, he was elected at the Royal Academy of Belgium.

\end{document}